\documentclass[fleqn,usenatbib]{mnras}

\usepackage{newtxtext,newtxmath}

\usepackage[T1]{fontenc}

\DeclareRobustCommand{\VAN}[3]{#2}
\let\VANthebibliography\thebibliography
\def\thebibliography{\DeclareRobustCommand{\VAN}[3]{##3}\VANthebibliography}

\usepackage{graphicx}	
\usepackage{amsmath}	
\usepackage{xcolor}
\usepackage{float}
\usepackage{siunitx}
\usepackage{setspace}
\usepackage{geometry}
\usepackage{graphicx}
\usepackage{natbib}
\usepackage{pdfpages}
\usepackage{subcaption}
\usepackage[toc,page]{appendix}
\usepackage{threeparttable}
\usepackage{xspace}

\newcommand{\Halpha}{H$\rm \alpha$\xspace}

\title{Non-LTE Monte Carlo Radiative Transfer. III. The thermal properties of Tilted and Warped Be Star Discs}

\author[M.W. Suffak et al.]{
M.W. Suffak,$^{1}$\thanks{E-mail: msuffak@uwo.ca}
C.E. Jones,$^{1}$
A.C. Carciofi,$^{2}$
T.H. de Amorim$^{2}$
\\
$^{1}$Department of Physics and Astronomy, Western University, London, ON N6A 3K7, Canada\\
$^{2}$Instituto de Astronomia, Geof\'isica e Ci\'encias Atmosf\'ericas, Universidade de S\~ao Paulo, Brazil\\
}

\date{Accepted 2023 September 07. Received 2023 August 28; in original form 2023 June 19 
}

\pubyear{2023}

\begin{document}
\label{firstpage}
\pagerange{\pageref{firstpage}--\pageref{lastpage}}
\maketitle

\begin{abstract}
We use the three-dimensional Monte Carlo radiative transfer code {\sc hdust} to model Be stars where the disc is tilted from the equatorial plane of the star. We compute 128 models across 4 spectral types, B0, B2, B5 and B8, tilting the disc by \ang{0}, \ang{10}, \ang{20}, and \ang{40}, while varying disc density according to spectral type. We also compute every model for an average and high stellar rotation rate. We first discuss non-tilted disc temperatures and show its non-linear dependence on stellar and disc parameters. We find that tilting the disc minimally affects the density-weighted average disc temperature, but tilting does create a temperature asymmetry in disc cross sections, which is more pronounced for a faster rotation rate. We also investigate the effect tilting has on $V$-band magnitude, polarization, and the \Halpha line. Tilting the disc does affect these observables, but the changes are entirely dependent on the position of the observer relative to the direction of tilt. We find the observables that distinguish tilting from a change in density or geometry are the \Halpha line shape, where it can transition between single-peaked and double-peaked, and the polarization position angle, whose value is dependent on the projected major elongation axis of the disc on the sky. We also present one early and one late-type model with warped discs. We find their temperature structure varies a small amount from the uniformly tilted models, and the different observables correspond to different tilt angles, consistent with their expected volume of origin within the disc.
\end{abstract}



\section{Introduction}
\label{sec:intro}

Classical Be stars are defined as non-supergiant B-type stars that have, or have had, Balmer lines in emission \citep{Collins1987}. These emission lines are known to form in a gaseous circumstellar disc that has developed around the equator of the star. The exact process which leads to the formation of these discs is uncertain, but coupling rapid rotation with non-radial pulsations \citep{Baade2016} is thought to be the most-likely mechanism for stellar mass-loss. In addition to Balmer line emission, Be star discs are also characterised by excess continuum emission, particularly at infrared (IR) and radio wavelengths, and by linear polarization \citep[for recent examples, see][]{Ghoreyshi2021, Marr2022}. The most recent comprehensive review of classical Be stars is given by \cite{rivinius2013classical}.


Observables seen from Be star discs are not only highly dependent on the density structure of the disc, but also on the disc temperature, as the temperature ultimately is what sets the state of the gas through its level populations and ionization state \citep{carciofi2008}. Until the late twentieth century, the temperature of Be star discs was assumed to be constant, or simply fall off with a radial power-law \citep{Waters1986}. The first attempt to self-consistently determine the disc temperature was performed by \cite{Millar1998}, who determined the temperature by equating the rates of energy gain and loss at each point in the disc. They applied this technique to various case studies of both early and late-type Be stars in subsequent publications \citep{Millar1999a, Millar1999b, Millar2000} and found temperature differences of thousands of Kelvin between the midplane and upper edge of the disc. \cite{Jones2004} added to this method by accounting for metals with the inclusion of iron. They found that for the early type star, $\gamma$ Cas, the inclusion of metals lead to an overall cooling of the disc, and a slight heating at the inner most disc, within 3 stellar radii. However for the late type star 1 Del, the most heating occurred on the outer edges of the disc, which was illuminated by light from the poles, and the greatest cooling happened in the middle portion of the disc, not near the dense equatorial plane.

\cite{carciofi2006non} investigated the temperature structure of early-type Be star discs with their 3-dimensional (3D) non-local thermodynamic equilibrium (non-LTE) Monte Carlo code {\sc hdust}. In their models, they found the temperature at the dense midplane of the disc initially drops within 3-5 stellar radii before rising back to the optically thin equilibrium temperature, while the thin upper layers of the disc were approximately isothermal, consistent with \cite{Millar1998}. They also found the disc to be almost completely ionized, except for a small portion in the midplane near the minimum temperature. \cite{carciofi2008} further investigated this non-isothermal structure by presenting a self-consistent solution for the viscous decretion disc (VDD) scenario. They determined that the varying temperature affects the density structure in two ways: 1) the radial temperature gradient changes the radial fall-off of the density, and 2) the reduction in temperature within the midplane results in the collapse of the disc onto itself, thereby causing a decrease in its scale height. They conclude that a non-isothermal disc density model must be used for detailed modelling of Be star disc observables. However, many successful modelling efforts of Be stars using {\sc hdust} have utilized the simpler isothermal density formula while solving for non-isothermal disc temperatures \citep{Silaj2016, Ghoreyshi2018, Suffak2020, Marr2021}.

Over the past decade, the possibility of Be star discs warping, tilting, and precessing, has gained a lot of attention \citep[see][for example]{Martin2011, Brown2019,Suffak2022}. There have been a number of studies using hydrodynamical simulations to predict the nature of warping, tilting, and oscillations of Be star discs in situations where a binary companion's orbit is misaligned to the initial plane of the disc \citep{Martin2014, Cyr2017, Suffak2022}, many of which focus on Be/X-ray binary system parameters \citep[for example]{Martin2014b, Brown2019}. The simulations of \cite{Suffak2022} showed that, under the influence of a misaligned binary companion, a Be star disc can undergo episodes of disc tearing, as well as develop eccentric gaps near the primary star during disc dissipation, in addition to tilting, warping, and precessing. The phenomena of disc precession and disc tearing are the best current explanation for the behaviour of the observables in the Be star Pleione \citep{Marr2022, Martin2022}.

So far, none of the studies that investigated dynamically simulated disc tilting, warping, precession, etc., have investigated the effects this would have on the disc temperature structure, or its observables in a systematic way. As well, late-type Be stars have been dramatically understudied compared to their early-type counterparts. In this paper, we first provide results of static 3D radiative transfer models, showing the temperature structure of non-tilted Be star discs, ranging in spectral type from B0 to B8 (Section \ref{sec:non-tilted_models}). We then show the same discs, uniformly tilted from the equatorial plane, and discuss their temperature structure (Section \ref{sec:tilted_models}), before we present two scenarios where the continuum, Balmer line, and polarization signatures could allow a tilted disc to be detected (Section \ref{sec:observables}). We also briefly discuss how a warped disc may differ from a flat-tilted disc in Section \ref{sec:warp_v_tilt}. Our discussion and conclusions are presented in Section \ref{sec:discussion}.

\section{Non-tilted Disc Temperatures}
\label{sec:non-tilted_models}

We chose a computational grid of four spectral types from B0 to B8, to capture both early and late-type Be star behaviour. The stellar parameters for each spectral type were taken from \cite{Cox2000} and \cite{Silaj2010}, who interpolated their parameters from \cite{Cox2000}. We model our disc density based on the widely-used equation
\begin{equation}
    \rho(r,z) = \rho_0 \bigg(\frac{R_*}{r}\bigg)^n \exp{-\frac{z^2}{2H^2}},
    \label{eq:rho_power_law}
\end{equation}
where $\rho_0$ is the base density, $R_*$ is the equatorial radius of the star, $H$ is the disc scale height, and $r$ and $z$ are respectively the radial and vertical coordinates in the disc. Equation \ref{eq:rho_power_law} is physically motivated by the viscous decretion disc (VDD) model of \cite{Lee1991}, that in its simplest form predicts $n\,=\,3.5$ for an isothermal and geometrically thin disc, and has been used in many studies, such as \cite{Silaj2010, Jones2008, Suffak2022, Marr2021}. The scale height is calculated by
\begin{equation}
    H(r) = \frac{a}{\Omega}\bigg(\frac{r}{R_*}\bigg)^{1.5},
    \label{eq:scale_height}
\end{equation}
where $a$ is the sound speed, calculated assuming a disc temperature 60\% of the star's effective temperature \citep{carciofi2006non}, and $\Omega$ is the Keplerian orbital frequency at the equator of the star. We selected two base density ($\rho_0$) values for each spectral type, based on the limits of base density versus stellar effective temperature shown in figure 8a of \cite{Vieira2017}. Figure 8b of \cite{Vieira2017} also shows there is no bounds on $n$ with effective temperature, so we choose to use values of $n$ of 2 and 3.5 for every spectral type, as these are approximately the lower and upper limits of $n$ for the majority of stars studied in \cite{Vieira2017}. Finally, we compute each model for two different stellar rotation rates, setting the critical fraction, W \citep[defined in equation 6 of][as the ratio of the rotational velocity at the equator to the Keplerian circular orbital velocity at the equator]{rivinius2013classical}, to 0.7 or 0.95. Figure 9 of \cite{rivinius2013classical} shows 0.7 to be about the average $W$ for Be stars, while 0.95 is on the extreme upper end, nearing the critical rotation rate where the outward centrifugal force at the equator would be equal to the inward pull of gravity.

The disc size is held constant for each spectral type at 50 equatorial radii ($R_{eq}$). The equatorial radius was scaled to be consistent with the chosen value of $W$, satisfying the formula \citep{Rimulo2018}
\begin{equation}
    W = \sqrt{2\bigg(\frac{R_{\text{eq}}}{R_\text{p}} - 1\bigg)},
    \label{eq:W_eqn}
\end{equation}
where $R_\text{p}$ is the stellar polar radius. Table \ref{tab:param_table} presents stellar and disc parameters in our models.

\begin{table*}
    \centering
    \caption{Stellar and disc parameters used in our {\sc hdust} grid of models. Left to right is the spectral type, stellar mass, polar radius, fraction of critical rotation velocity, effective temperature, luminosity, disc base density, disc density slope, and model number.}
    \begin{tabular}{cccccccccc}
    \hline\hline
    Sp. Type & M ($\rm M_\odot$) & $\rm R_p$ ($\rm R_\odot$) & $W$ & $T_{\rm eff}$ (K) & $L$ ($\rm L_\odot$) & $\rm \rho_0$ ($\rm g\, cm^{-3}$) & n & Model \# \\ \hline
    B0 & 17.5 & 7.4 & 0.7 & 30000 & 39740 &$1\times 10^{-10}$ & 2/3.5 & 1/2\\
    &&&0.7&&& $1\times10^{-11}$ & 2/3.5 & 3/4 \\
    &&&0.95&&& $1\times10^{-10}$ & 2/3.5 & 5/6 \\
    &&&0.95&&& $1\times10^{-11}$ & 2/3.5 & 7/8 \\
    B2 & 9.11 & 5.33 & 0.7 & 21000 & 4950 &$5\times 10^{-11}$ & 2/3.5 & 9/10 \\
    &&&0.7&&& $5\times10^{-12}$ & 2/3.5 & 11/12 \\
    &&&0.95&&& $5\times10^{-11}$ & 2/3.5 & 13/14 \\
    &&&0.95&&& $5\times10^{-12}$ & 2/3.5 & 15/16 \\
    B5 & 5.9 & 3.9 & 0.7 & 15000 & 690 & $5\times 10^{-12}$ & 2/3.5 & 17/18\\
    &&&0.7&&& $5\times10^{-13}$ & 2/3.5 & 19/20 \\
    &&&0.95&&& $5\times10^{-12}$ & 2/3.5 & 21/22 \\
    &&&0.95&&& $5\times10^{-13}$ & 2/3.5 & 23/24 \\
    B8 & 3.8 & 3.0 & 0.7 & 12000 & 167 & $1\times 10^{-12}$ & 2/3.5 & 25/26\\
    &&&0.7&&& $1\times10^{-13}$ & 2/3.5 & 27/28 \\
    &&&0.95&&& $1\times10^{-12}$ & 2/3.5 & 29/30 \\
    &&&0.95&&& $1\times10^{-13}$ & 2/3.5 & 31/32 \\
    \hline
    \end{tabular}
    \label{tab:param_table}
\end{table*}

\subsection{Azimuthally-Averaged Temperature Slice}
\label{sec:az_avg_slices}

Across all of our models, regardless of spectral type, rotation rate, or disc density parameters, we find the following common traits: (1) the tenuous upper layers of the disc (i.e., far from the midplane) are fully ionized and approximately isothermal; (2) the very dense disc midplane contains the diversity between the models, it can be cooler or hotter than the upper layers depending on model parameters, and is only partially ionized; (3) between these two regions exists a transition layer between the fully ionized outer disc and partly ionized inner disc, where relatively thin, hot sheaths arise. However, when examining the temperature structure in more detail, the behaviour seen from one model to another is very non-linear and is coupled to the disc density structure, spectral type, and stellar rotation rate. We summarize these particularities in appropriate detail below.

\subsubsection{Early Spectral Types}
\label{sec:early-types}

\begin{figure*}
    \centering
    \includegraphics[scale = 0.3]{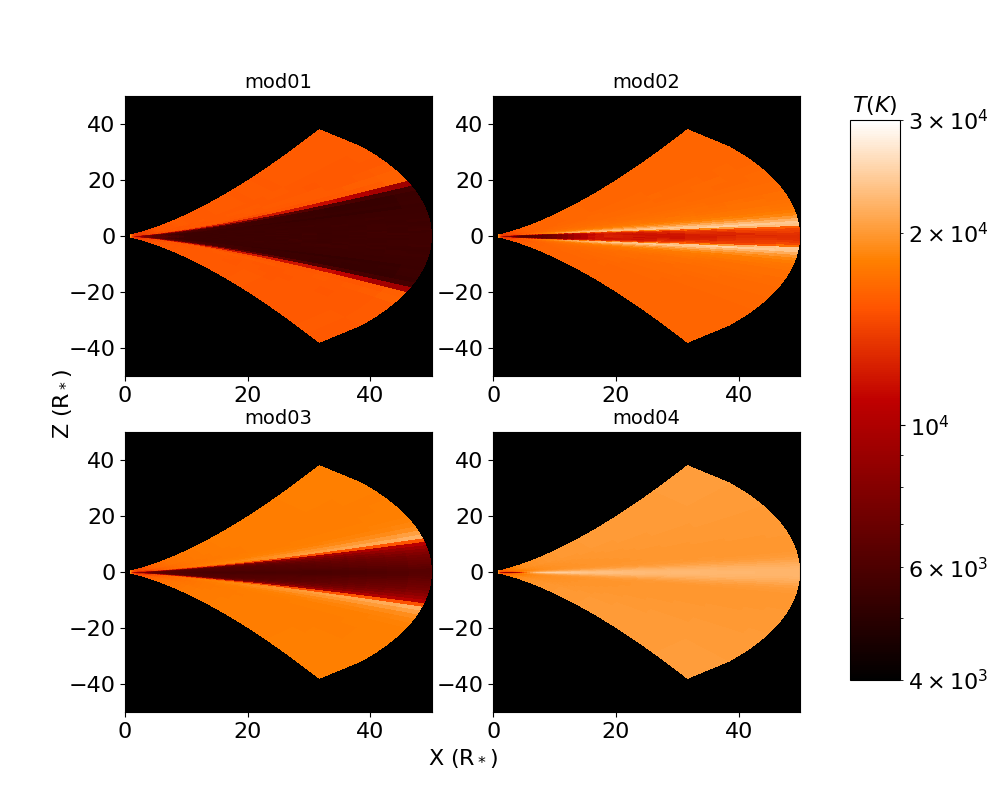}
    \includegraphics[scale = 0.3]{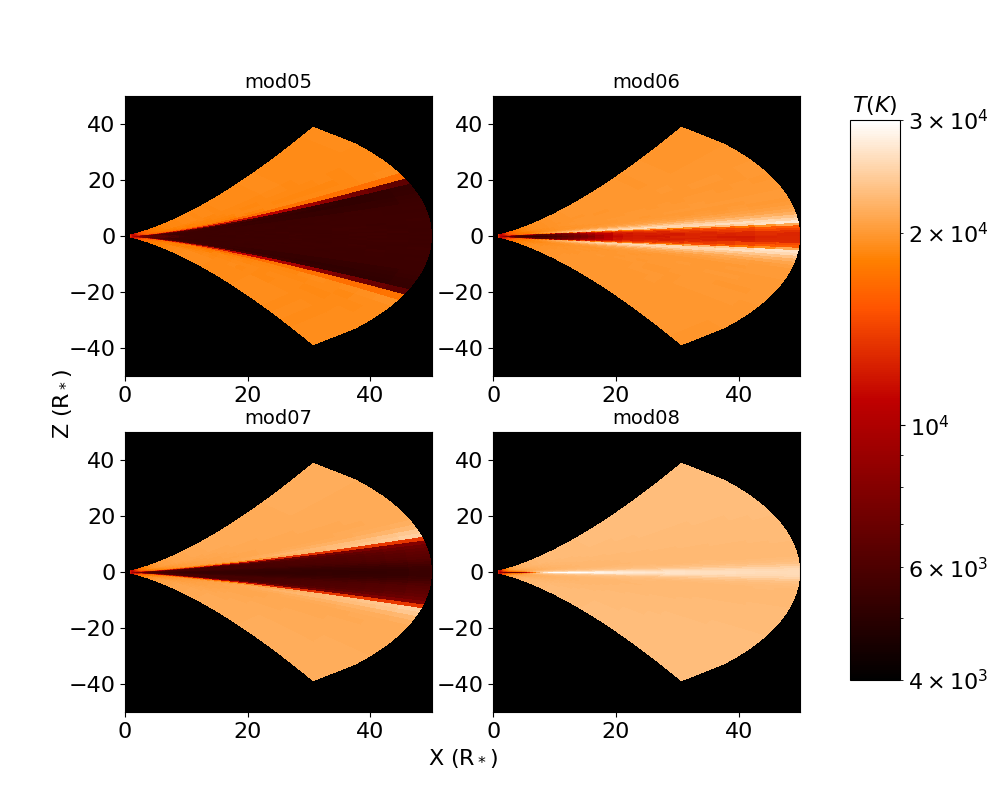}
    \includegraphics[scale = 0.3]{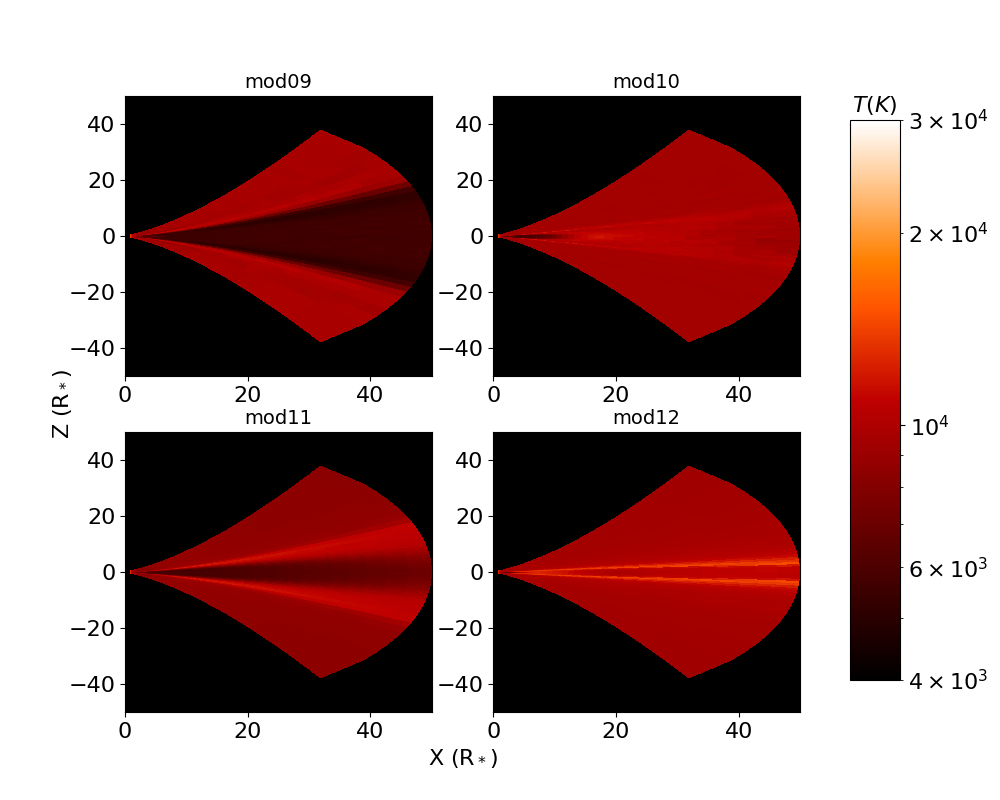}
    \includegraphics[scale = 0.3]{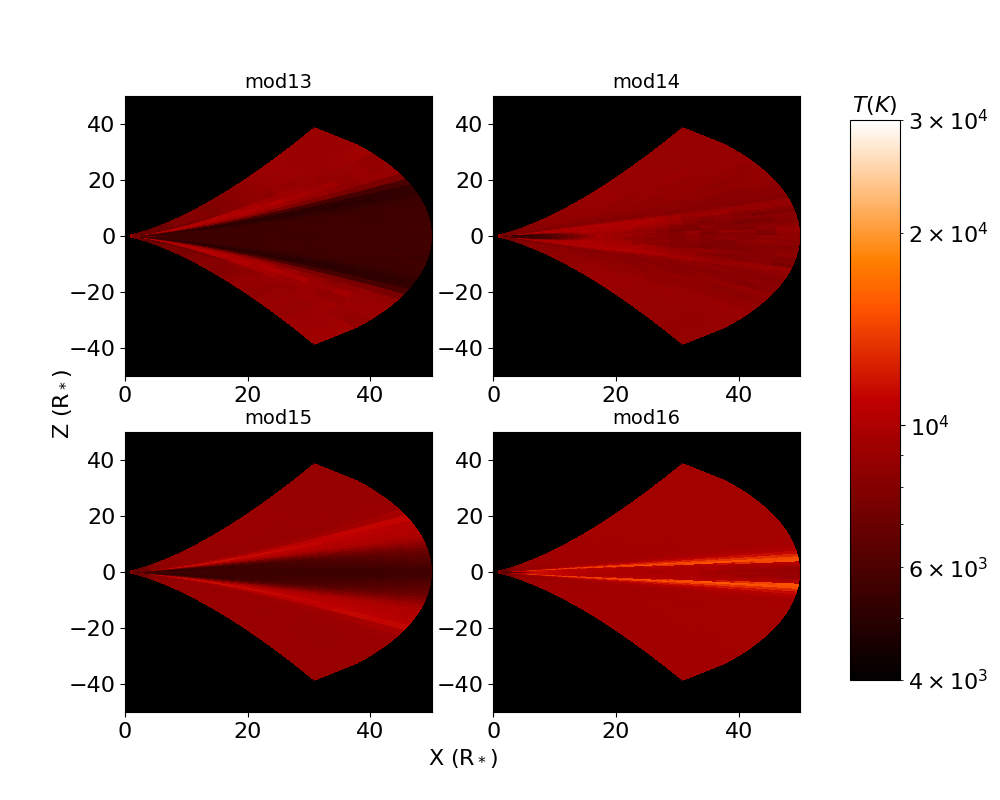}
    \caption{Disc cross sections of the azimuthally-averaged temperatures of models 1-16, as noted in the subplot titles. The colour of each cell in the cross section corresponds to temperature as shown in the colour bar. The axes are in stellar equatorial radii, which is different for each spectral type, consistent with the polar radius listed in Table \ref{tab:param_table}.}
    \label{fig:mod1-16_cross_sec}
\end{figure*}

Figure \ref{fig:mod1-16_cross_sec} shows the azimuthally-averaged cross sections of the non-tilted disc models for the B0 and B2 models of our grid. We see that the models where $n = 2$, whose discs have a slow density fall-off and thus are much more dense than $n=3.5$, have a very large cool, partially ionized region surrounding the midplane of the disc, while the outer regions are much hotter and fully ionized. The inner cool regions are due to the disc being optically thick, while in the outer regions, the density drops and the temperature can reach much higher values. In the models where $n=3.5$, we see that the midplane has a much smaller cool region and then transitions to hotter temperatures with increasing radius as the disc becomes optically thin. The upper hot layers are also larger than in the $n=2$ case due to the densities falling off faster and more of the disc being optically thin. These inner and outer regions are separated by a hot thin sheath, which has also been seen in other publications \citep{carciofi2006non, Sigut2007}. 

\begin{figure}
    \centering
    \includegraphics[scale=0.3]{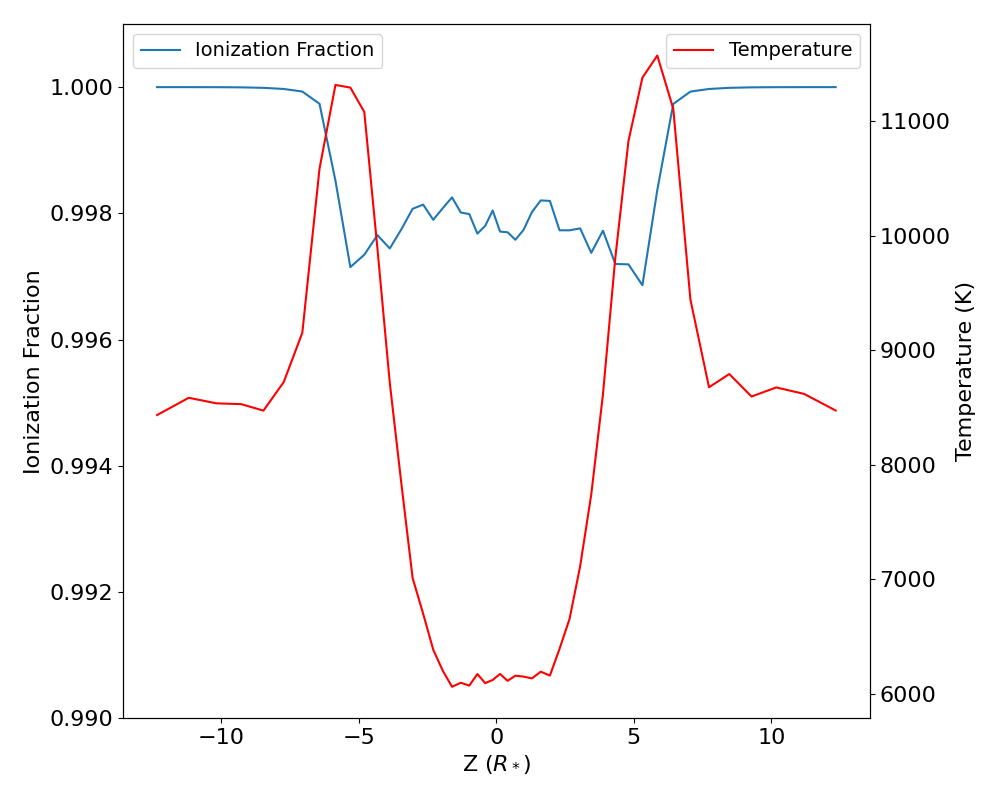}
    \caption{Ionization fraction (blue, left $y$-axis), and temperature (orange, right $y$-axis) versus $z$ height in the non-tilted disc of model 11. The radial distance of the measurements from the central star is about 30 $R_*$.}
    \label{fig:ion_frac_mod11}
\end{figure}

In Figure \ref{fig:ion_frac_mod11}, we have plotted the disc temperature and ionization fraction (fraction of hydrogen in the disc that is ionized) for a column at a radial position 30 $R_*$ for model 11, which prominently displays these hot sheaths. We can see that the spike in temperature (i.e., the hot sheath) occurs right at the boundary between the cooler inner portion where the disc is partially ionized, and the hotter outer layers where the disc is fully ionized. This can be explained by the inner cool region being optically thick and locally trapping the UV radiation, so as the vertical direction offers the largest escape probability due to lower opacity, the UV radiation travels vertically and further heats the gas directly above and below the inner cool region. When we compute the bound-free and bound-bound optical depths, as well as the hydrogen level populations, they show inverse profiles to the ionization fraction in Figure \ref{fig:ion_frac_mod11}, being highest around the midplane of the disc and trending towards zero as height in the disc increases. These same trends occur for all models that have these hot sheaths in their cross sections, including both early- and late- spectral types.

The position of these hot sheaths also noticeably changes between models, as they move closer together as the $n$ value rises, or as the disc base density decreases. Both of these changes to the density structure make the upper disc layers more tenuous, which allows the disc to be fully ionized to a greater vertical depth, making the inner partially ionized disc portion thinner, and thus the transition regions closer to the midplane. This can easily be seen by comparing the cross sections of model 1 to model 2 and 3, respectively. There are also cases where the disc is so tenuous that it is nearly entirely ionized, even in the midplane, seen for example in model 4. Here there is no cooler section in the midplane. Instead the midplane is hotter than the upper disc layers, due to the denser midplane being able to reprocess UV radiation and increase the role of diffuse radiation in that area.

In models 5 to 8 and 13 to 16, we see that increasing the stellar rotation rate to $W=0.95$ does not change the qualitative temperature patterns in the disc. However the temperatures in the upper disc are notably higher than in the slower rotating case and the hot sheaths and disc midplane can be slightly warmer than with slower rotation. This indicates that the hotter stellar poles caused by increased gravity darkening at this high rotation is able to ``carve" into the disc deeper, penetrating the disc midplane with more UV radiation and raising its temperature.

\subsubsection{Late Spectral Types}
\label{sec:late-types}

\begin{figure*}
    \centering
    \includegraphics[scale = 0.3]{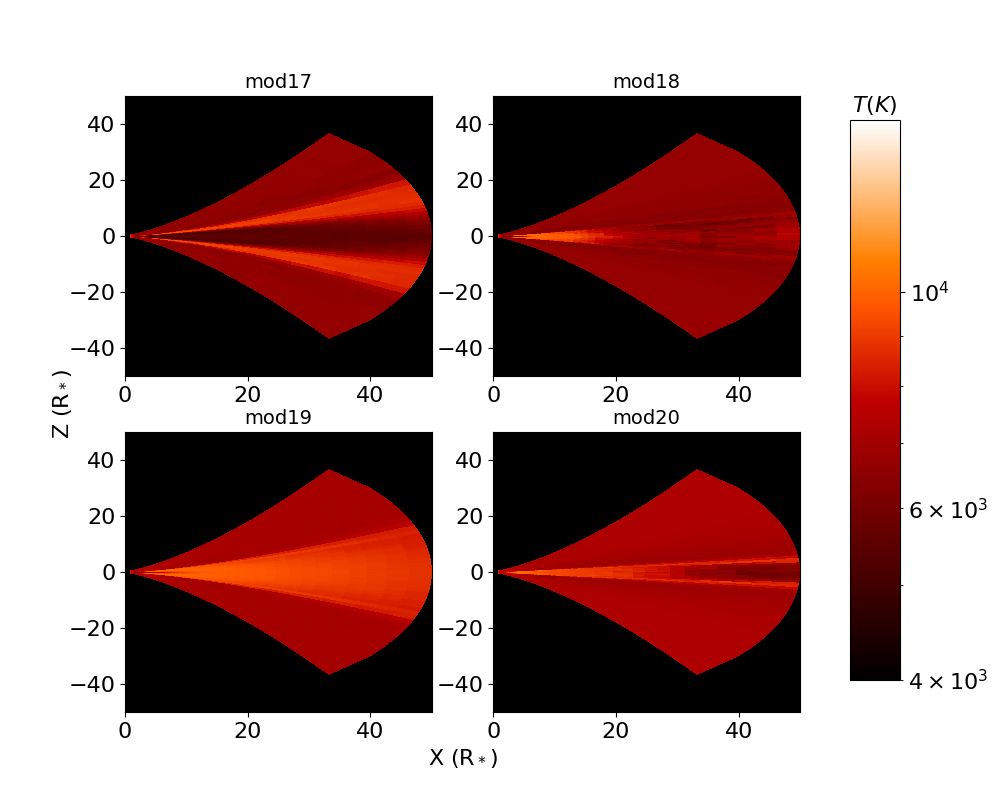}
    \includegraphics[scale = 0.3]{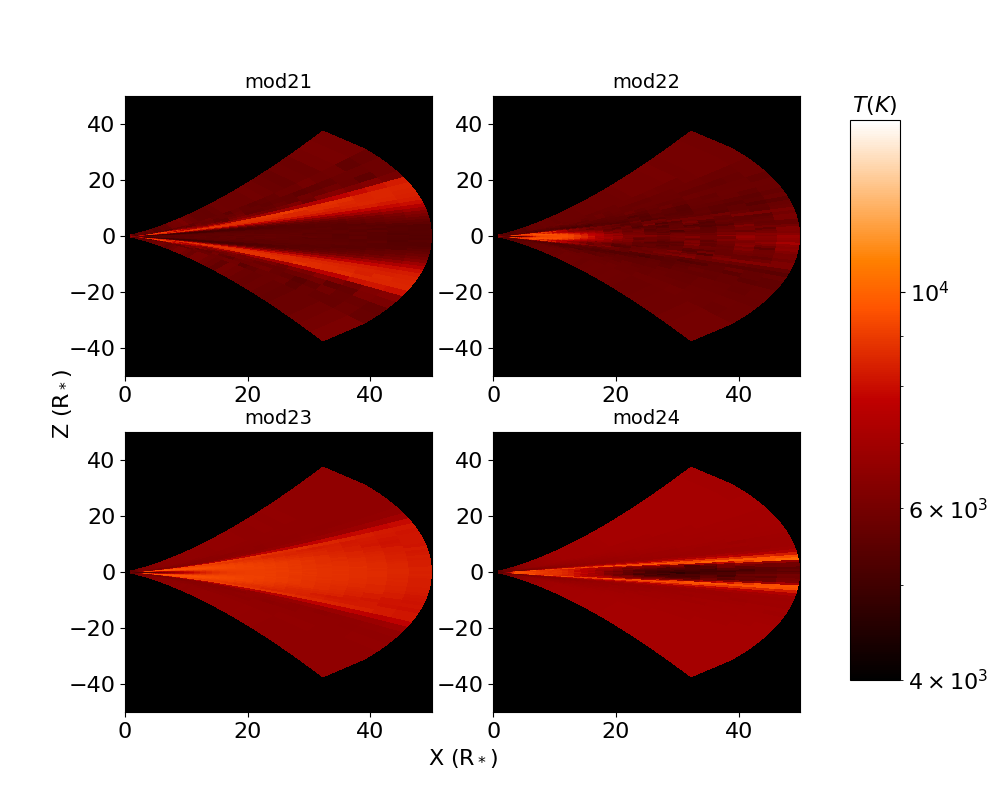}
    \includegraphics[scale = 0.3]{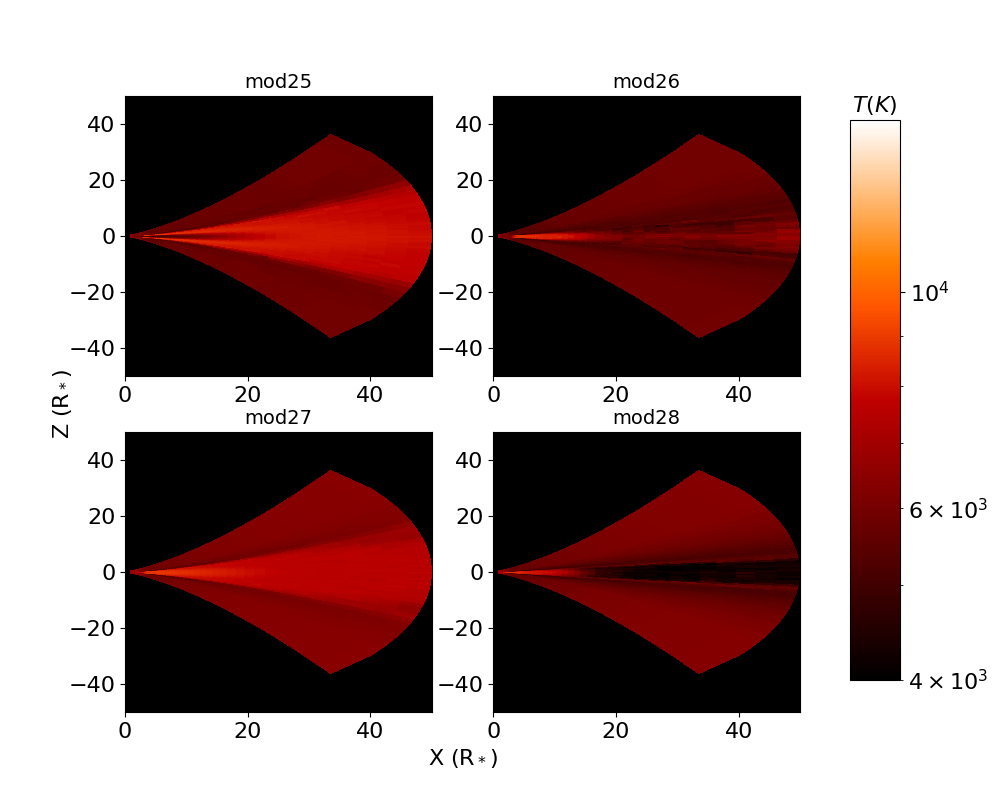}
    \includegraphics[scale = 0.3]{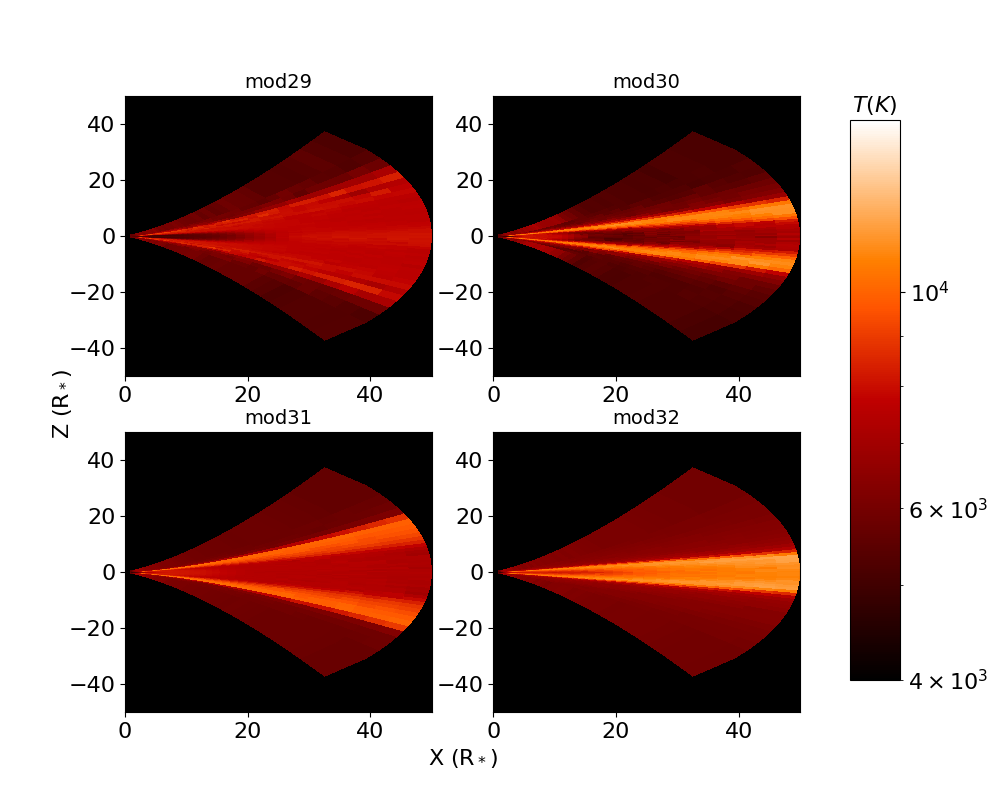}
    \caption{Same as Figure \ref{fig:mod1-16_cross_sec}, but for models 17-32. Note the change in the maximum temperature of the colour bar scale.}
    \label{fig:mod17-32_cross_sec}
\end{figure*}

The cross sections for our B5 and B8 non-tilted models are shown in Figure \ref{fig:mod17-32_cross_sec}. In these later spectral types, we see some different behaviour in the temperature structure than the early B0 and B2 type stars. The highest density, lowest $n$ model for the B5 spectral type is similar to the analogues for B0 and B2 stars. However, at lower densities, for $n\,=\,2$ in both B5 and B8 type stars, we see the midplane is hotter than the outer disc, which is opposite the early type stars. This is the same as \cite{Millar1999a} found in their work for the late-type star 1 Del. They explain that this temperature inversion is due to collisions populating the upper levels, and thus photoionization from these upper levels is able to heat the gas, while the disc remains optically thick to Lyman continuum photons. However, as these hot midplane sections are radially extended in our discs when the density has already fallen off exponentially, collisions are not going to be a major factor, and thus this hotter midplane would be due to the discs ability to reprocess the UV radiation (as mentioned for model 4 in Section \ref{sec:early-types}) and locally heat the denser midplane through diffuse radiation. Conversely, we see in models where $n=3.5$, that the midplane temperature drops off in the outermost disc. Due to the much faster drop off of density compared to the $n=2$ model, there is not enough diffuse radiation contributed from the disc itself to make up for the lack of UV radiation reaching the outer midplane of the disc from the late-type star.

With a higher rotation rate, we see the B5 models largely retain the same structure as their slower-rotating counterparts, however the rapidly rotating B8 models, particularly models 30, 31, and 32, display dramatically hotter disc midplanes, as well as hot sheaths which did not appear in the slower rotating case. We interpret this again as the hotter poles being able to ``carve" farther into the disc and cause greater heating in the midplane. Here the high stellar rotation gives qualities of both an early-type star from the hot poles, and a late-type star from the very cool equator causing this large change in temperature cross section.

\section{Tilted-Disc Temperatures}
\label{sec:tilted_models}

\begin{figure}
    \centering
    \begin{subfigure}[b]{0.39\textwidth}
        \centering
        \includegraphics[width=\textwidth]{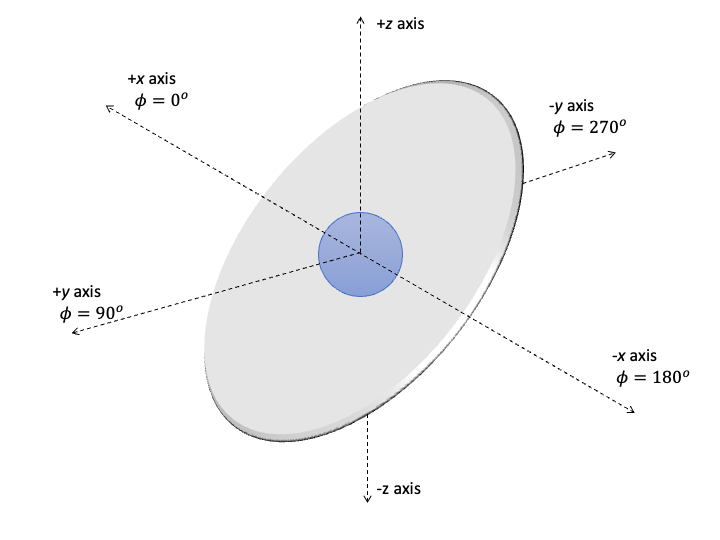}
        \caption{}
        \label{fig:schem_view1}
    \end{subfigure}
    \begin{subfigure}[b]{0.39\textwidth}
        \centering
        \includegraphics[width=\textwidth]{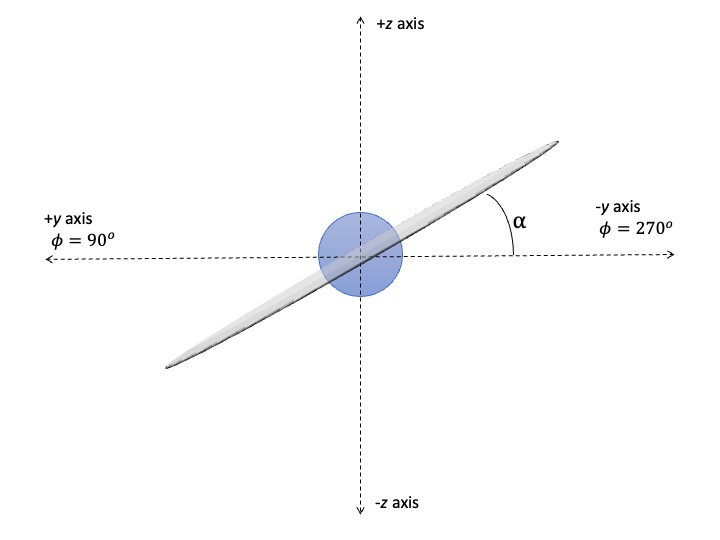}
        \caption{}
        \label{fig:schem_view2}
    \end{subfigure}
    \caption{Two schematics showing the orientation of the tilted disc with respect to the $x$, $y$, and $z$ axes. The azimuthal angle ($\phi$) is defined with $\phi=\ang{0}$ along the positive $x$-axis as shown. The disc is tilted about the $x$-axis, and thus the tilt angle $\alpha$ is defined from the $y$-axis. The central star is represented by the blue circle at the origin of both diagrams. The disc and star sizes are not to scale.}
    \label{fig:tilt_schematic}
\end{figure}

To expand our work on our non-tilted grid, we tilted all of our models listed in Table \ref{tab:param_table} by $\alpha\,=\,$ \ang{10}, \ang{20}, and \ang{40} away from the equatorial plane. Figure \ref{fig:tilt_schematic} shows a schematic of the orientation of our tilted disc models in cartesian coordinates, where the star lies at the origin. The disc is tilted about the $x$-axis, so the disc tilt angle is measured from the $y$-axis. When referring to azimuthal angles ($\phi$), we denote the positive $x$-axis as having $\phi = \ang{0}$, and the positive $y$-axis having $\phi = \ang{90}$. Thus $\phi$ values of $\ang{180}$ and $\ang{270}$ correspond to the negative $x$ and $y$-axes, respectively. The height of the disc midplane can thus be given as 
\begin{equation}
    Z(r_i,\phi_j) = -r_i\sin\bigg\{\arctan\bigg[\sin(\phi_j)\tan\big(\alpha\big)\bigg]\bigg\},
    \label{eq:z_heights}
\end{equation}
where $r_i$ and $\phi_j$ are the radial and azimuthal coordinates on the midplane, and $\alpha$ is the disc tilt angle about the $x$-axis (either \ang{10}, \ang{20}, or \ang{40}).

\begin{figure*}
    \centering
    \begin{subfigure}[b]{0.39\textwidth}
        \centering
        \includegraphics[width=\textwidth]{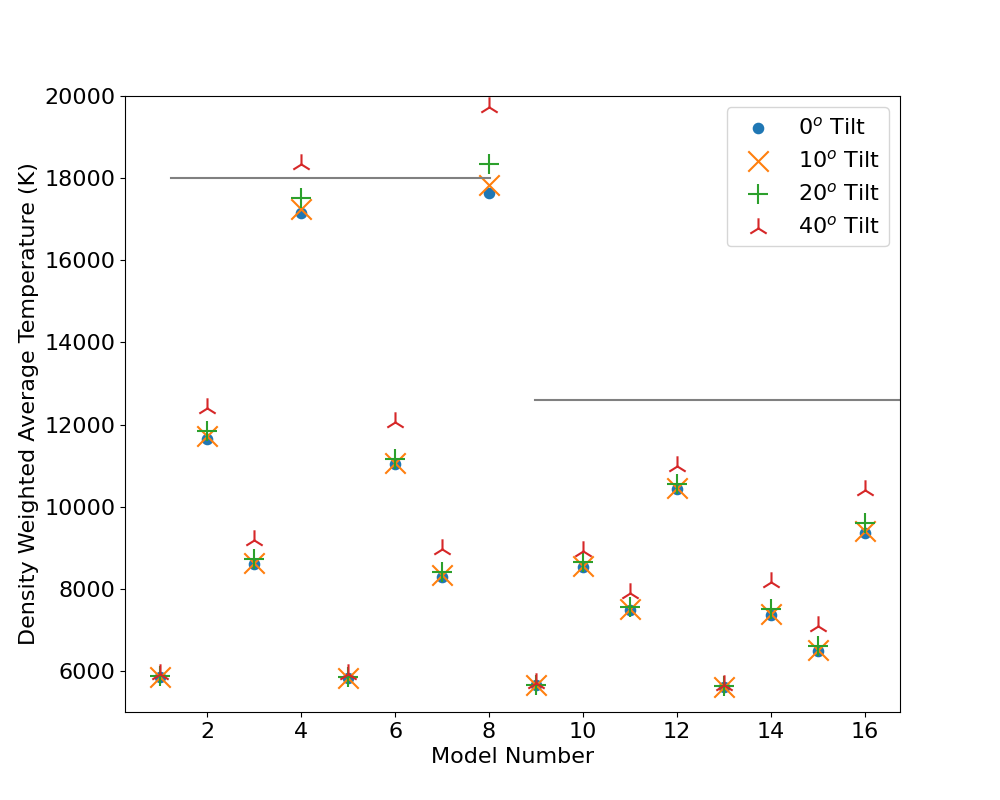}
        \caption{}
        \label{fig:avg_temps_early}
    \end{subfigure}
    \begin{subfigure}[b]{0.39\textwidth}
        \centering
        \includegraphics[width=\textwidth]{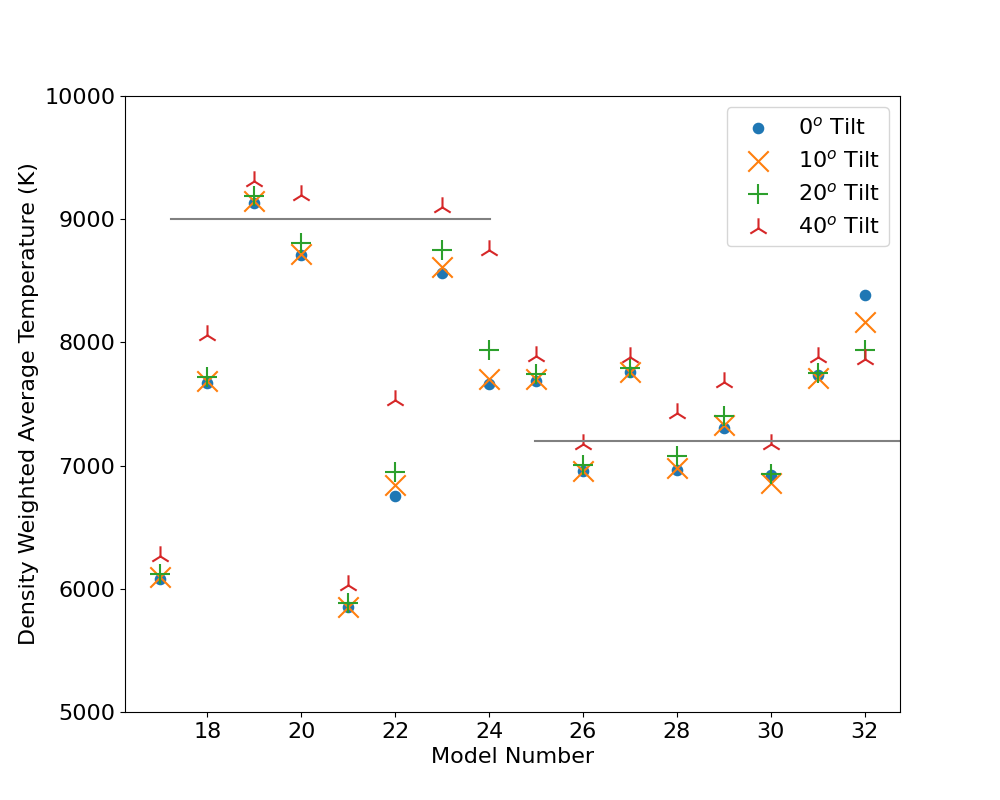}
        \caption{}
        \label{fig:avg_temps_late}
    \end{subfigure}
    \caption{Density weighted average disc temperatures for our early-type (a) and late-type (b) models. The model number, listed in Table \ref{tab:param_table} is on the x-axis. The tilt angle of each disc model is indicated by the legend. The grey lines indicate 60\% of the star's $T_{\rm eff}$ for each model.}
    \label{fig:avg_tilted_disc_temps}
\end{figure*}

To assess any global changes in disc temperature due to disc tilting, we calculate the mass-averaged disc temperature, $\Bar{T}_M$, using the formula
\begin{equation}
    \Bar{T}_M = \frac{1}{M_{\rm disc}}\sum_{i=0}^N T_i \rho_i V_i,
    \label{eq:temp_approx}
\end{equation}
where $M_{disc}$ is the total mass of the disc, and $T_i$, $\rho_i$ and $V_i$ are the temperature, density, and volume of the $i$-th cell. The sum is performed over all $N$ cells in the disc. This formula was adapted from the equation for density-weighted average temperature of \cite{McGill2013}. The results of these calculations are presented in Figure \ref{fig:avg_tilted_disc_temps}. We find that there is a clear relationship between $\Bar{T}_M$ and disc tilt angle: the greater the tilt angle, the greater $\Bar{T}_M$. While this trend is clear, the changes are not large, with most discs varying in temperature by less than 1000 K. The change is especially small in our densest models, where the change in average temperature cannot be distinguished between being caused by the tilt or simply the error in the nature of Monte-Carlo simulations. We also note that most of these density-weighted average temperatures are below $60\% \, T_{\rm eff}$, which \cite{carciofi2006non} found to be a good isothermal approximation of Be star discs. Since the density-weighted average weights the most dense, and hence most optically thick regions the highest, it is not surprising that our least dense models come closest to, or sometimes end up greater than, the $60\% \, T_{\rm eff}$ mark. This is consistent with the results of our non-tilted discs seen in Figures \ref{fig:mod1-16_cross_sec} and \ref{fig:mod17-32_cross_sec} where the lower density discs have a higher temperature. It is worth mentioning that by definition, our tilted models are not anchored at the equator of the star, thus the innermost disc may have slightly inflated temperatures due to directly seeing a hotter part of the star than if it were aligned at the equator.

\subsection{Detailed Temperature Structure}
\label{sec:detailed_structure}

\begin{figure}
    \centering
    \includegraphics[scale=0.3]{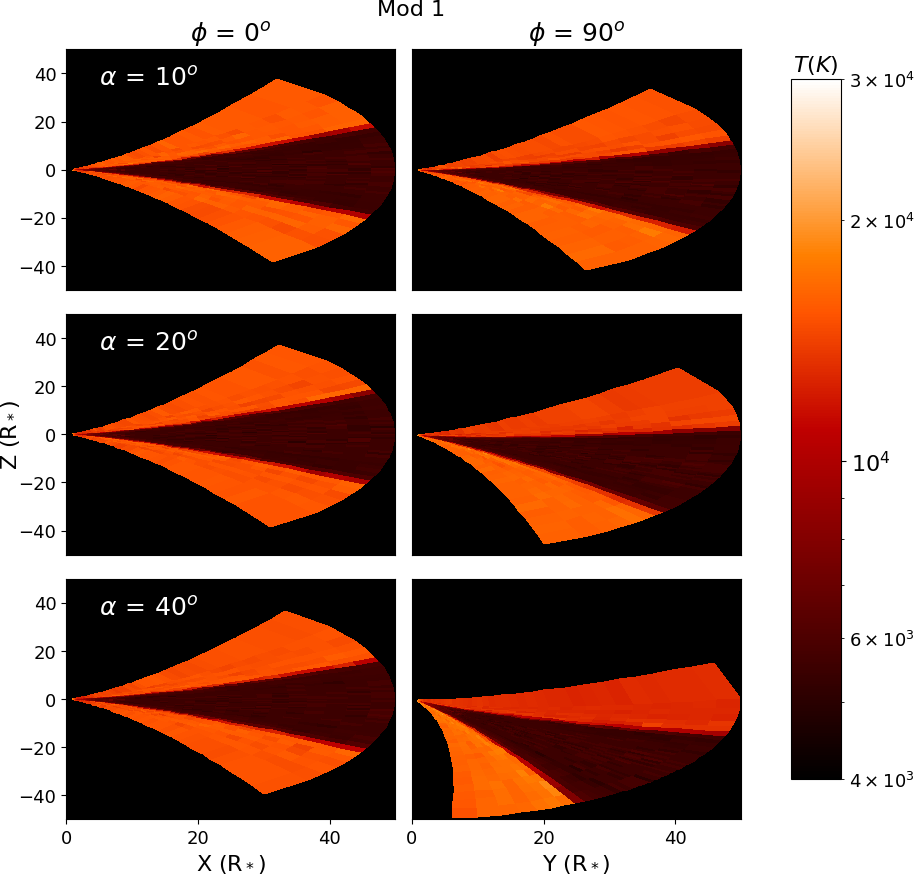}
    \caption{Temperature structure of the line of nodes cross section (left column) and the cross section farthest from the equator (right column) of the tilted discs with parameters of model 1 (see Table \ref{tab:param_table}). The top row is for a \ang{10} tilt, middle row for \ang{20} tilt, and bottom row for a \ang{40} tilt, as indicated by the $\alpha$ value in the leftmost plot of each row. The colour corresponds to the temperature of a given cell, as indicated by the colour bar on the right.}
    \label{fig:mod33_tilt_temps}
\end{figure}

Although the density-weighted average disc temperature doesn't change appreciably with tilt angle, we find that a tilted disc can have significant changes in the temperature structure of certain areas of the disc. An example is shown in Figure \ref{fig:mod33_tilt_temps}, where we show cross sections of discs with parameters of model 1, tilted by $\ang{10}$, $\ang{20}$, and $\ang{40}$. In the line of nodes cross section of the disc ($\phi\,=\,\ang{0}$, left column) we see that changes in the temperature structure from the non-tilted model (top left panel of Figure \ref{fig:mod1-16_cross_sec}) are hard to detect. However, in the cross section farthest from the equator ($\phi\,=\,\ang{90}$, right column of Figure \ref{fig:mod33_tilt_temps}), we see noticeable temperature difference between the top and bottom of the disc, with the top of the disc becoming cooler and the bottom of the disc becoming hotter. This is due to the effect of gravity darkening arising from rapid rotation, resulting in the stellar equator being cooler than the pole. Thus, the part of the disc that moves closer to the equator when tilted, ends up being cooler than the non-tilted solution, and the part that moves closer to the stellar pole ends up hotter than the non-tilted case. We note that the disc midplane does not significantly change temperature when tilted because its high density compared to the rest of the disc makes it insensitive to the change in radiation input caused by the tilting of the disc.

\begin{figure}
    \centering
    \includegraphics[scale = 0.3]{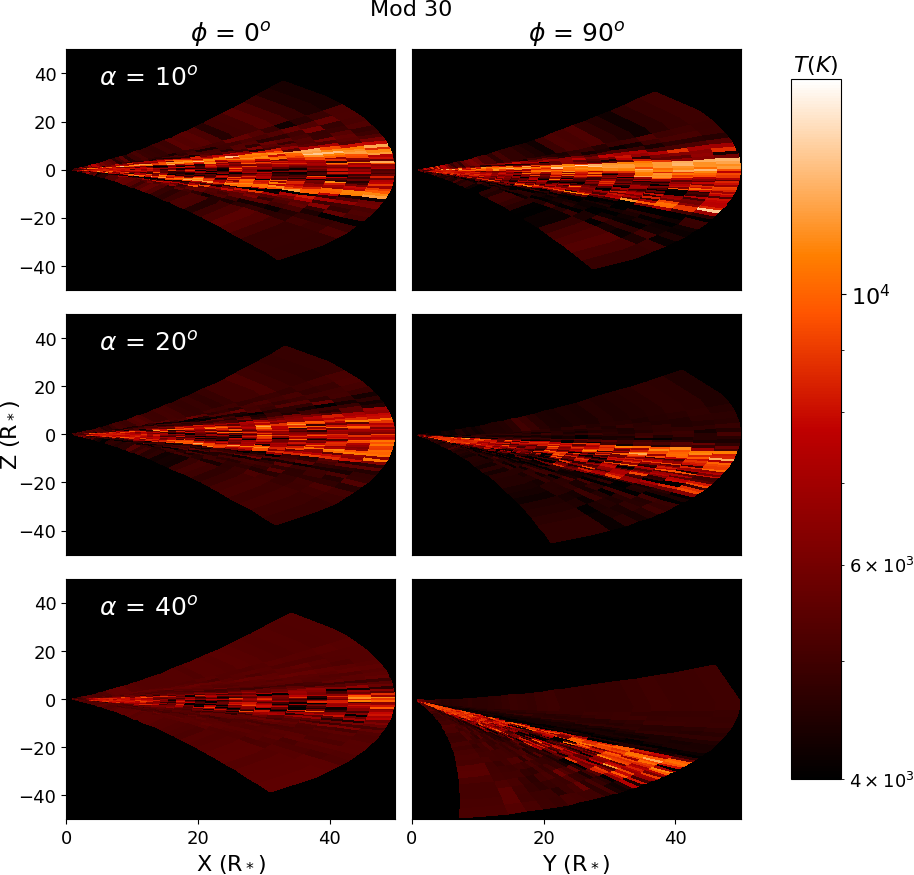}
    \caption{Same as Figure \ref{fig:mod33_tilt_temps}, but for the parameters of model 30. Note the change in the maximum temperature of the colour bar scale.}
    \label{fig:mod62_tilt_temps}
\end{figure}

In Figure \ref{fig:mod62_tilt_temps}, we show the same plots as Figure \ref{fig:mod33_tilt_temps}, but for one of our late-type models. We see that the temperature trends of the $\phi\,=\,\ang{90}$ cross section of the disc are the same, with the top of the disc cooling and the bottom heating as it is oriented closer to the pole of the star. In the line of nodes of the disc, however, we see a change in the temperature structure: the hot bands that were on either side of the midplane in the non-tilted models, greatly lessen in temperature as the overall disc tilt angle increases. Since the central star does not change, this change in temperature structure is due to the diffuse radiation field of the disc affecting the disc temperature significantly. Plots similar to Figures \ref{fig:mod33_tilt_temps} and \ref{fig:mod62_tilt_temps} for all other models are presented in Appendix \ref{sec:tilted_temps}. 

\begin{figure*}
    \centering
    \begin{subfigure}[b]{0.39\textwidth}
        \centering
        \includegraphics[width=\textwidth]{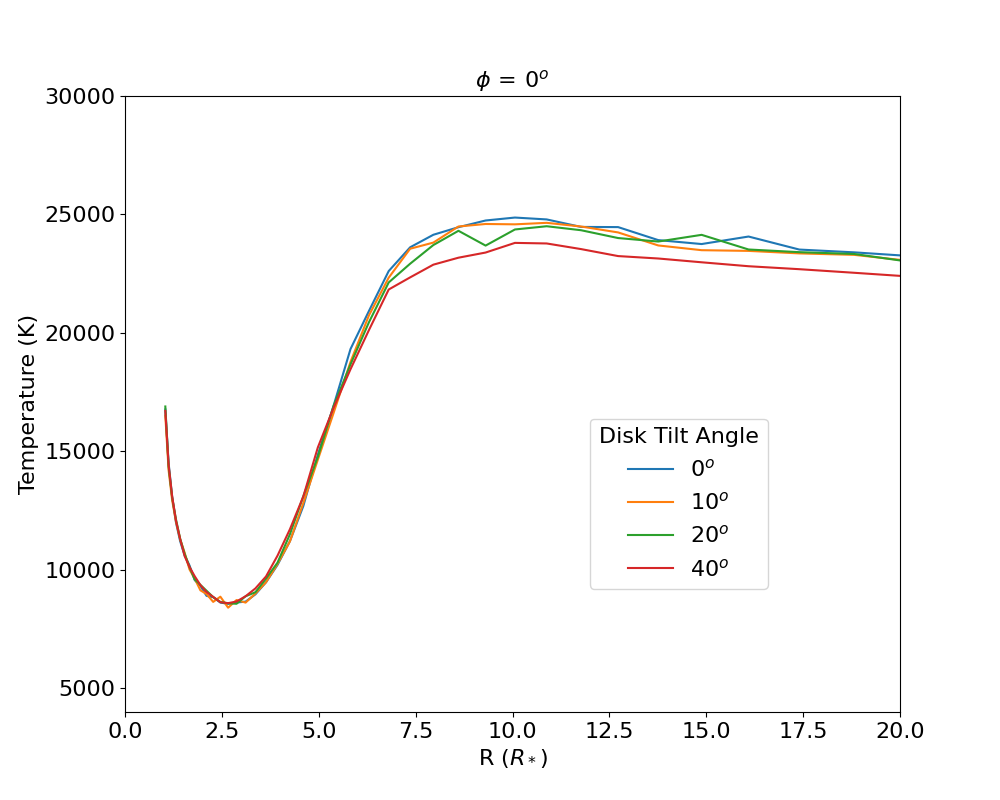}
        \caption{}
        \label{fig:temp_rad_1}
    \end{subfigure}
    \begin{subfigure}[b]{0.39\textwidth}
        \centering
        \includegraphics[width=\textwidth]{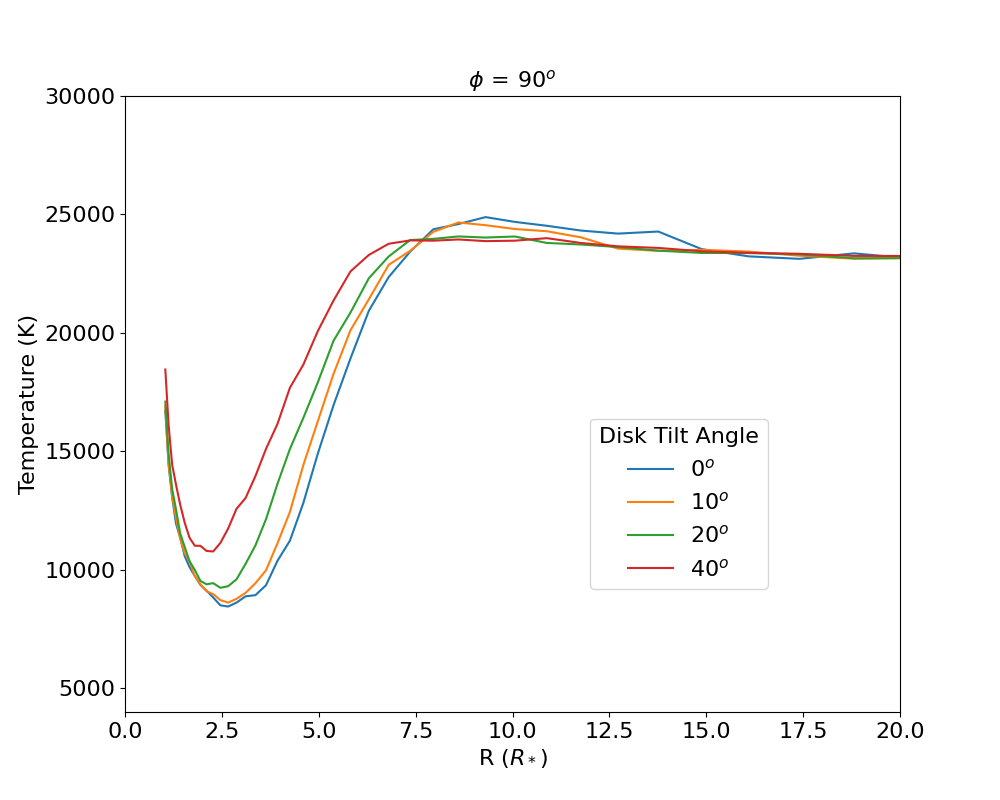}
        \caption{}
        \label{fig:temp_rad_2}
    \end{subfigure}
    \caption{Plots of temperature vs. radius of model 4 at the midplane of the disc in two different azimuthal directions; $\phi\,=\,\ang{0}$ (a), and $\phi\,=\,\ang{90}$ (b). The four different lines are for different disc tilt angles as indicated by the legend.}
    \label{fig:midplane_radial_temps}
\end{figure*}

Overall we see that the average cell temperature of the $\phi\,=\,\ang{90}$ cross sections of the disc can differ by as much as 30\% when compared to the non-tilted models. The large majority of this difference however, is in the optically thin outer disc. As shown in Figure \ref{fig:midplane_radial_temps}, while the midplane temperature does differ in the $\phi\,=\,\ang{90}$ cross sections of the disc from the non-tilted model, the difference is quite small. Appendix \ref{sec:midplane_temps} contains similar plots of the midplane for all models. As the tilt angle increases, we see the temperature of the innermost disc increase as well. This is due to the midplane being oriented closer to the pole, and therefore directly seeing a hotter area of the star. We also see that the midplane temperature profile of the tilted disc still has the same structure as past publications \citep{Millar1998, carciofi2006non}, with the temperature reaching a minimum within the first few stellar radii before increasing to an approximately isothermal temperature in the outer disc. It is important to note that for some of our densest models, the midplane temperature does not increase from its minimum due to the high density of the disc. It is also noteworthy that this structure becomes less pronounced as one moves towards the top and bottom of the disc, away from the midplane.

\section{Tilted Disc Observables}
\label{sec:observables}

Due to the 3-dimensional nature of our simulations, we specify our observer position by two spherical coordinates, the polar angle $\theta$ and the azimuthal angle $\phi$. $\theta$ ranges from $\ang{0}$ when looking pole-on with the star, to $\ang{90}$ when looking at the equatorial plane, while $\phi$ is defined in the range [\ang{0},\ang{360}) in the same manner as shown in Figure \ref{fig:tilt_schematic}. 

Observables of Be stars are highly dependent on the orientation of the disc with respect to the observer. It is well known that a non-tilted disc seen edge-on will have dimmer photometric magnitudes, higher polarization levels, and shell emission lines, compared to a disc viewed pole-on, which will have brighter photometric magnitudes, close to zero polarization, and single-peaked emission lines. However, in a tilted disc scenario, looking edge-on with the disc may not imply one is looking equator-on with the star, and a face-on view of the disc would not be pole-on with the star. Thus, the changes in observables that occur due to disc tilting would be expected to be highly dependent on the orientation of the tilt with respect to the observer.

In their hydrodynamic simulations, \cite{Suffak2022} showed that a Be star disc can tilt and then precess under the influence of a misaligned binary companion. We now examine these two scenarios where a tilted disc may be able to be detected through a change in their observables.

\subsection{Case I: Viewing a Disc with Varying Tilt Angle}

In this first scenario, a misaligned binary companion torques the disc above and below the equatorial plane of the primary star, meaning the axis of the disc's tilt is aligned with the companion's line of nodes. Here, the orientation between the disc tilt axis and the stationary observer is constant, thus the observer merely sees the disc tilt in a constant direction over time. 

\begin{figure}
    \centering
    \includegraphics[scale = 0.35]{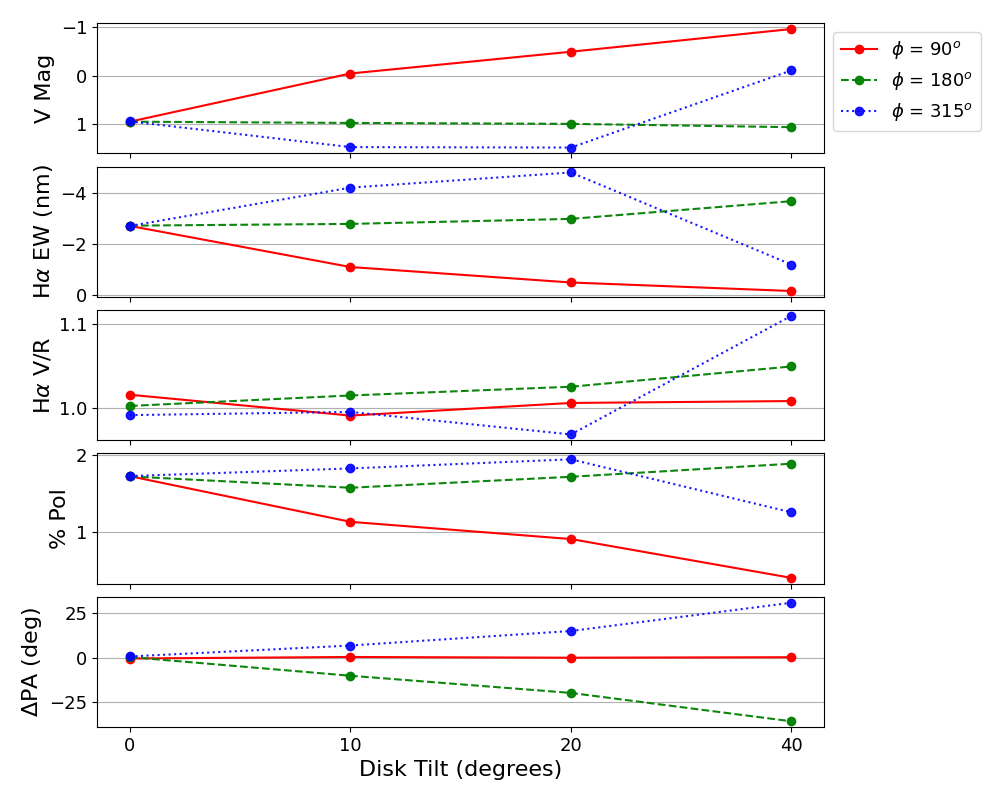}
    \caption{Plots showing (top to bottom) $V$-band magnitude, H$\alpha$ equivalent width, H$\alpha$ V/R ratio, polarization percentage, and polarization position angle (PA) in the $V$-band, versus disc tilt angle, for systems with parameters of model 1 (Table \ref{tab:param_table}). All points are for a $\theta$ observing angle of \ang{80}, and the different coloured lines indicate different $\phi$ observing angles as indicated by the legend. Some $\phi$ directions may be degenerate and thus not every line will show on every plot.}
    \label{fig:bigplot_tilt_mod1}
\end{figure}

Since our observing coordinates are defined with respect to the central star and tilt axis of the disc, we can simulate a disc tilting over time by plotting any observable viewed from the same ($\theta$, $\phi$) observer position over four simulations where a model disc is tilted by \ang{0}, \ang{10}, \ang{20}, and \ang{40}. An example of this is shown in Figure \ref{fig:bigplot_tilt_mod1}, which shows, for a constant $\theta$ of $\ang{80}$ and $\phi$ of $\ang{90}$, $\ang{180}$, and $\ang{315}$, $V$-band magnitude, H$\alpha$ equivalent width (EW), H$\alpha$ violet-to-red (V/R) ratio, polarization percentage and position angle (PA) in the $V$-band, versus disc tilt angle for systems with parameters of model 1. Here we see the change in an observable as the disc tilts is very dependent on the observer's position. 


For example, at $\phi\,=\,\ang{0}$ or $\ang{180}$, the observer will be aligned with the tilt axis of the disc, thus the $V$ magnitude, EW, and percent polarization do not vary as much as at other $\phi$ angles, however the polarization position angle will vary greatly with increasing tilt angle. This is in contrast to $\phi\,=\,\ang{90}$, where the disc is tilted to be more face-on with the observer. As the tilt angle increases, we see $V$ magnitude increase, while EW and polarization decrease and the position angle stays constant. Not plotted is $\phi\,=\,\ang{270}$, where the disc tilts to be more edge-on with the observer, and the trends would be generally reversed from the $\phi\,=\,\ang{90}$ case. Finally, for an intermediate azimuthal angle of $\ang{315}$, all of the observables vary greatly, with the only clear trend being the polarization position angle. Note that these results are degenerate with $\phi\,=\,\ang{225}$, while the degenerate pair of $\ang{45}$ and $\ang{135}$ will give similar results. The polarization position angle is thus the one observable which shows constant change as the disc tilts (except when the observer is exactly perpendicular to the tilt axis). We also see for this simulation, large V/R ratios of about 10\% at certain azimuthal observing angles when the disc is tilted by $\ang{40}$. This is an extreme example however, since most of our models don't reach a V/R ratio of more than 2 to 5\%.


\begin{figure}
    \centering
    \includegraphics[scale = 0.35]{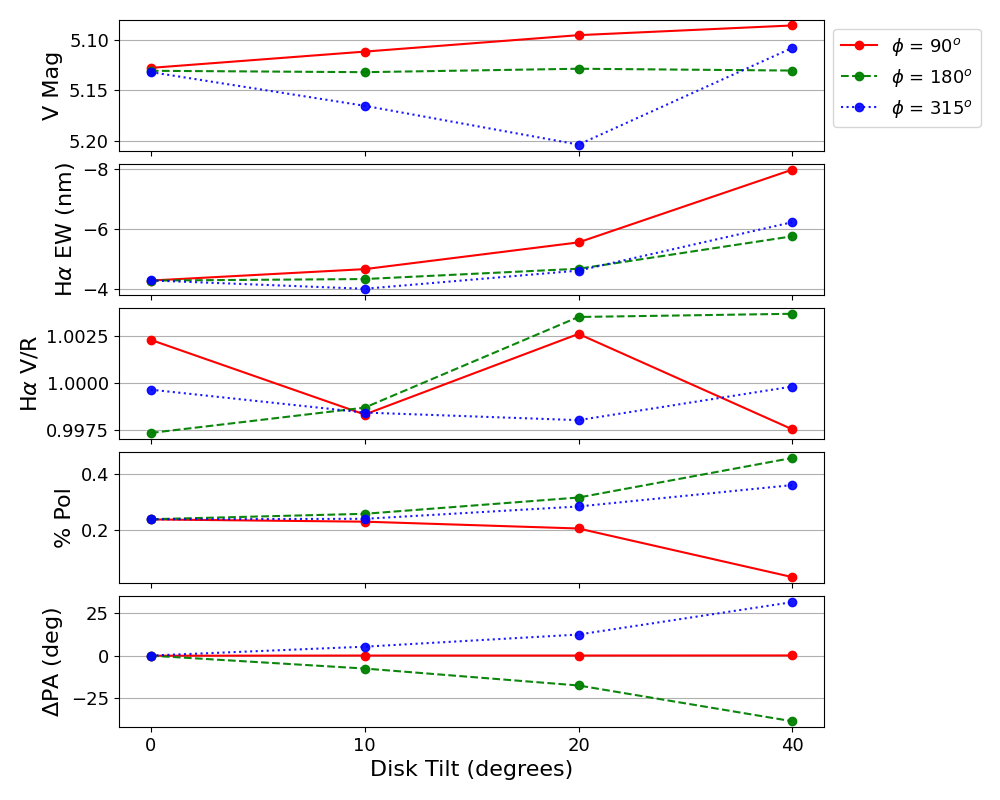}
    \caption{Same format as Figure \ref{fig:bigplot_tilt_mod1}, but for parameters of model 29.}
    \label{fig:bigplot_tilt_mod29}
\end{figure}

Figure \ref{fig:bigplot_tilt_mod29} shows the same plot as Figure \ref{fig:bigplot_tilt_mod1}, but for a late-type B8 star, model 29. We see another example of the variability of the change of observables depending on observer position, and that the magnitude of these changes for late-type stars are much less than those of the early-type example shown previously, particularly the $V$ magnitude and polarization, which is expected due to the lower disc density.

To briefly investigate disc obscuration effects, we took two models, 10 and 25, at a tilt angle of $\ang{40}$ and computed them with a disc radius of 20 $R_*$ instead of 50. These models presented no difference in observables aside from a slightly lower \Halpha EW. This is expected, as the \Halpha emission comes from a large portion of the disc, while visible continuum emission and polarization come from the inner few stellar radii. This shows that the material in the outer disc is largely optically thin, and thus the size of a tilted disc would not affect observables other than \Halpha until the disc radius is reduced to a few stellar radii.




\subsubsection{\Halpha Line Shapes}

The shape of the \Halpha line in Be stars is largely due to Doppler shift caused by the relative velocity between the disc material and the observer. In a non-tilted disc, the \Halpha line in Be stars is seen as single-peaked when looked pole-on with the star and face-on with the disc ($\theta\,=\,\ang{0}$), and is a double-peaked line when observed at other inclination angles due to disc material moving both toward and away from the observer. As the \Halpha emission line is a defining characteristic of Be stars, we also looked at the shapes of the emission lines in our simulations to see if the changing shape might show indications of disc tilting.

\begin{figure}
    \centering
    \includegraphics[scale=0.35]{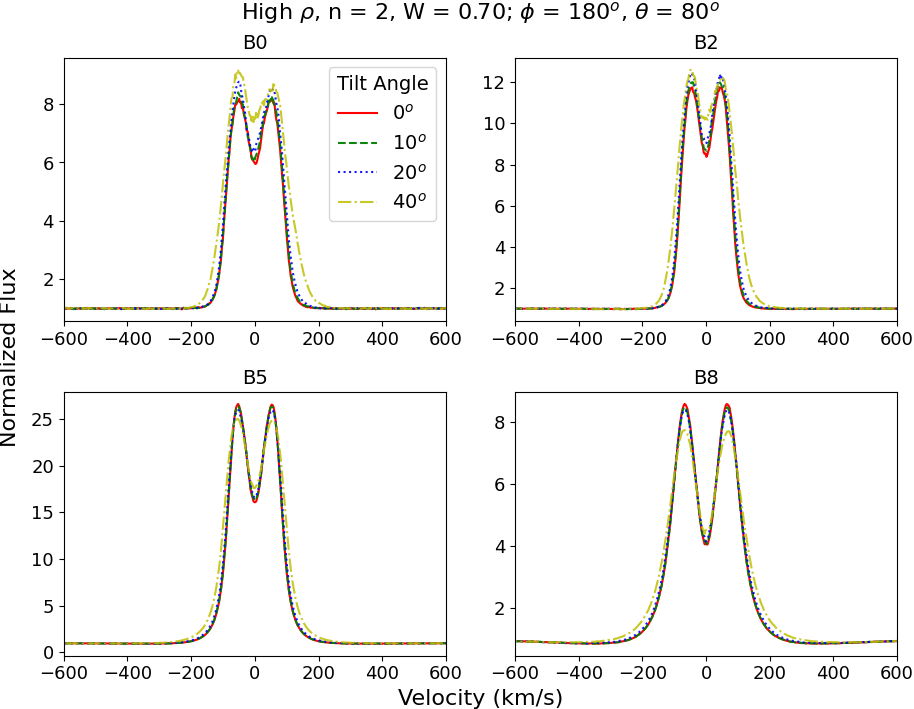}
    \caption{Simulated \Halpha lines of models 1 (top left), 9 (top right), 17 (bottom left) and 25 (bottom right), for four different disc tilt angles as indicated by the legend. The model spectra are seen from an observer at position $\phi = \ang{180}$, $\theta = \ang{80}$.}
    \label{fig:Halpha_mod01}
\end{figure}

Overall we find there are three different patterns in which the tilt of the disc can affect the \Halpha line. The first is shown in Figure \ref{fig:Halpha_mod01}, where $\phi\,=\,\ang{180}$ and the observer is aligned with the tilt axis of the disc, so the disc is seen as rotating either clockwise or counter-clockwise from the observers perspective. In this case, we see the line strength stay approximately the same, but the V/R ratio increases with increasing tilt angle. This effect is particularly noticeable in the early-type stars at high ($\theta\, \geq \, \ang{60}$) inclinations, where the projected area of the disc does not change very much with increasing tilt angle. The lines plotted here are one of the most extreme V/R ratios that we have obtained from our models.

\begin{figure}
    \centering
    \includegraphics[scale=0.35]{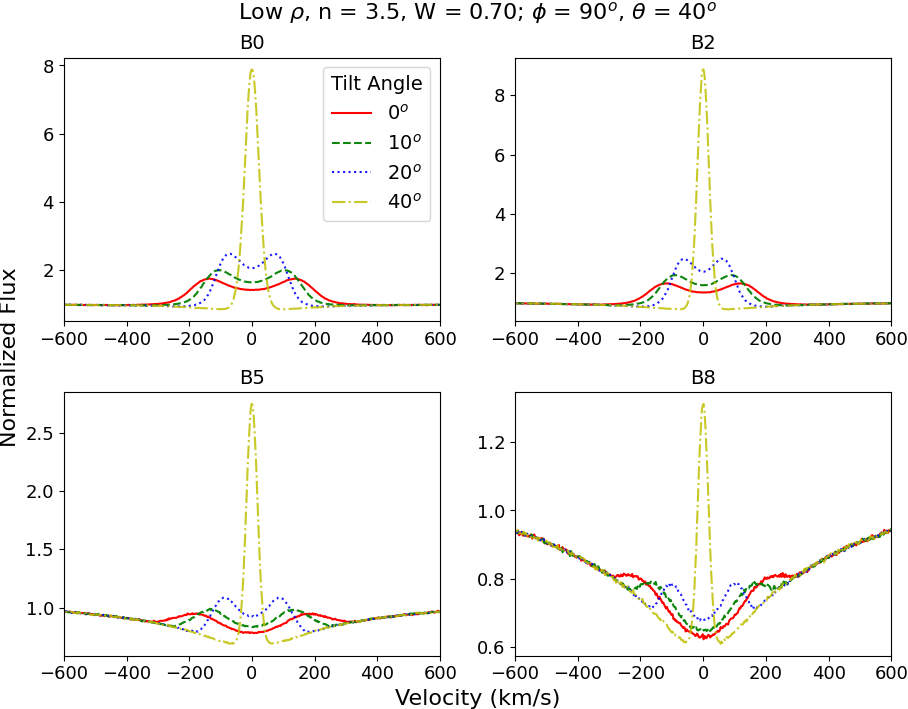}
    \caption{Simulated \Halpha lines of models 4 (top left), 12 (top right), 20 (bottom left) and 28 (bottom right), for four different disc tilt angles as indicated by the legend. The model spectra are seen from an observer at position $\phi = \ang{90}$, $\theta = \ang{40}$.}
    \label{fig:Halpha_mod04}
\end{figure}

The second pattern is shown in Figure \ref{fig:Halpha_mod04}, where $\phi\,=\,\ang{90}$ and the disc tilts to be more face-on or edge-on with the observer. Here we see that the line starts out as double-peaked for $\ang{0}$ tilt, and transitions to a single-peaked line when the tilt is $\ang{40}$. This behaviour occurs for all spectral types and densities. The reverse of this process is also seen in our models, with the lines shifting from single-peaked to double-peaked with increasing tilt angle for pole-on observing angles. It is worth noting that the equivalent width of the line is approximately constant in these scenarios as well.

\begin{figure}
    \centering
    \includegraphics[scale=0.35]{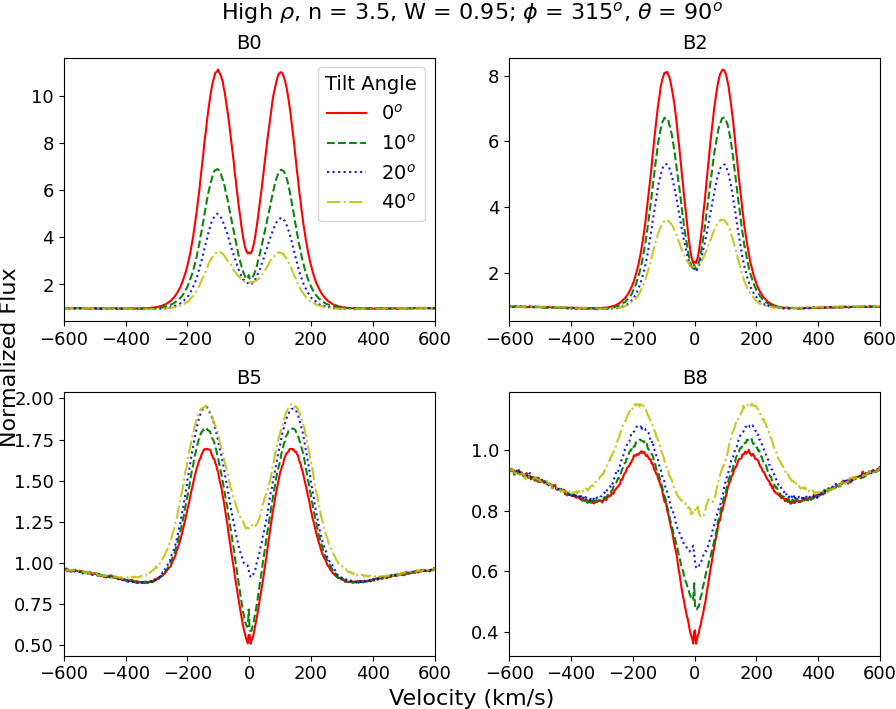}
    \caption{Simulated \Halpha lines of models 6 (top left), 14 (top right), 22 (bottom left) and 30 (bottom right), for four different disc tilt angles as indicated by the legend. The model spectra are seen from an observer at position $\phi = \ang{315}$, $\theta = \ang{90}$.}
    \label{fig:Halpha_mod06}
\end{figure}

Finally when $\phi\,=\,\ang{315}$, the motion of the disc relative to the observer is a combination of a rotation clockwise or counterclockwise in the plane of the sky, and a tilt to be more edge-on or face-on. Figure \ref{fig:Halpha_mod06} shows this case, where the line shape and peak separation stay the same, but the line decreases in strength as tilt angle increases for the B0 and B2 type stars, and slightly increases in strength for the late type stars. The reverse is also seen for some observing angles. These changes in the normalized lines are largely due to a change in continuum level rather than a change in the emission itself.

\subsection{Case II: Watching a Tilted Disc Precess}

The second scenario where a tilted disc may be able to be detected is where an already tilted disc precesses under the influence of a misaligned binary companion. This occurs in many simulations of \cite{Suffak2022}, particularly after mass-loss from the disc is turned off, the disc is no longer anchored to the equator of the star, and the line of nodes of the disc is free to rotate about the primary star.

\begin{figure}
    \centering
    \includegraphics[scale = 0.35]{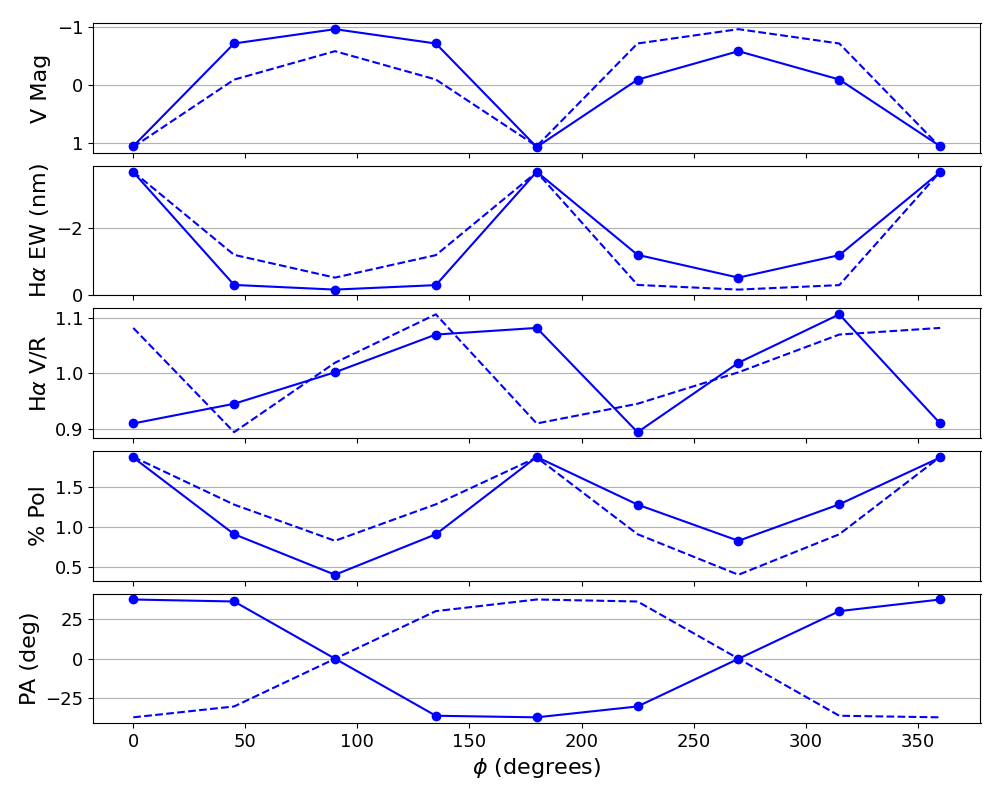}
    \caption{Top to bottom, the $V$-band magnitude, H$\alpha$ equivalent width, H$\alpha$ V/R ratio, polarization percentage in the $V$-band, and position angle, versus azimuthal viewing angle $\phi$, for model 1 with the disc tilted $\ang{40}$. The system is viewed at a $\theta$ of $\ang{80}$. The dashed line is shifted by $\ang{180}$ to facilitate comparison between periods.}
    \label{fig:bigplot_phi_mod97}
\end{figure}

By holding the polar viewing angle ($\theta$) constant, and moving around the star and disc in $\phi$, we can see what observational signature a disc may present if it were precessing about the pole of the star. This is shown in Figure \ref{fig:bigplot_phi_mod97}, where we've plotted the same quantities as Figure \ref{fig:bigplot_tilt_mod1} versus $\phi$, for model 1 with a disc tilt of $\ang{40}$. From this Figure we can see that, as the observational viewpoint moves around the star/disc system at a constant $\theta$, the observables oscillate quite significantly as the disc moves from being edge-on with the observer at $\ang{0}$ and \ang{180}, to being more face-on at $\ang{90}$ and $\ang{270}$. Moving the observer like this is exactly the same as if the disc were rigidly precessing about the pole of the star and the observer was stationary, but this is accomplished here without the need for computationally expensive hydrodynamical simulations.

Unlike the case of tilting a disc, precession shows signatures in all observables. The percent polarization and $V$ magnitude will oscillate at half the precession period as more or less of the inner disc becomes visible as the disc precesses about the star. The \Halpha EW also oscillates at half the precession period and can increase with $V$ magnitude as more of the disc is visible, or decrease as $V$ magnitude increases due to the changing continuum level. The position angle will oscillate about zero at the same period as the precession. These period/half-period trends can be seen easily with the dashed line in Figure \ref{fig:bigplot_phi_mod97} shifted $\ang{180}$. We note here that the half-period trends of $V$ magnitude, \Halpha EW, and percent polarization are not perfectly symmetric because the $\theta$ of $\ang{80}$ means the observer will see slightly more or less of the central star from opposite sides of the disc. If $\theta$ were $\ang{90}$, the half-period trends would ideally have perfect symmetry. The V/R trend is more complex, as the trend from $\phi\,=\,\ang{90}$ to $\ang{270}$ is the reverse of the trend from \ang{270} to \ang{90}. This is due to the relative velocities ``reversing" as the observer moves to the opposite side of the disc, thus the trend in this case is antisymmetric, with \ang{90} and \ang{270} being the nodes of the oscillation.

\section{Warped vs. Tilted Discs}
\label{sec:warp_v_tilt}

We recognize our flat tilted models are limited, particularly at the star-disc boundary, where we have the inner disc tilting the same amount as the outer disc. In reality, it is more likely that the Be star disc would be anchored to the equator of the star, and the rest of the disc would be warped away from the equator by some degree.

To test the difference this may cause, we chose to apply a warp to models 10 and 25, instead of a flat tilt. To warp the computational grid, we fix the first radial bin at the equator, and then linearly increment the degree of tilt with each subsequent radial bin, up to a maximum tilt of $\ang{40}$ at the furthest radius. The height of the disc midplane is then given by
\begin{equation}
    Z(r_i,\phi_j) = -r_i\sin\bigg\{\arctan\bigg[\sin(\phi_j)\tan\big(\alpha\frac{i}{49}\big)\bigg]\bigg\},
    \label{eq:z_heights_warped}
\end{equation}
where the only difference from Equation \ref{eq:z_heights} is that now we have altered the disc tilt angle $\alpha$, such that it gets bigger with increasing radius ($i$ denotes the radial cell index, which goes from 0-49 in our models), so the disc becomes warped instead of a flat tilt.
These models also have a disc radius of 50 $R_*$, and thus the most highly inclined outer parts of the disc do not contribute much to the observables in question here as they are known to originate in the inner disc, which is only moderately tilted.

\begin{figure*}
    \centering
    \begin{subfigure}[b]{0.39\textwidth}
        \centering
        \includegraphics[width=\textwidth]{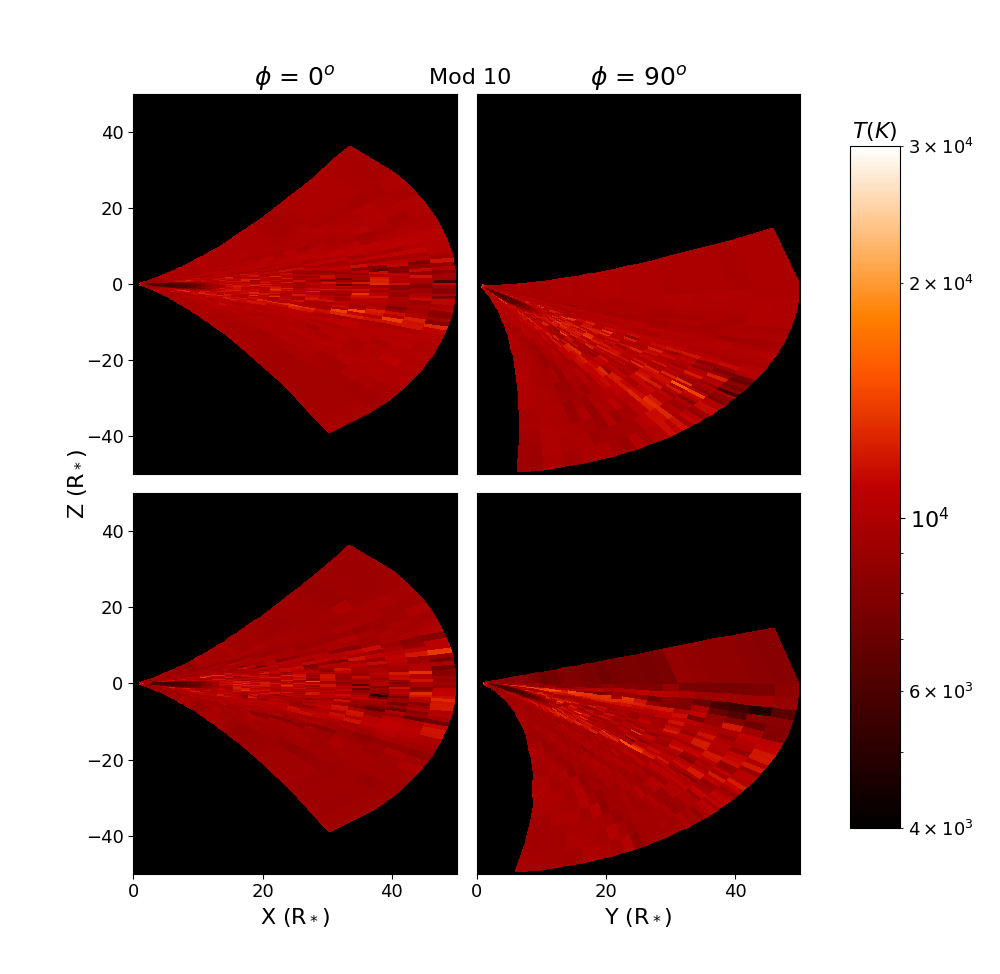}
        \caption{Model 10, full view}
        \label{fig:warped_temp_compare}
    \end{subfigure}
    \begin{subfigure}[b]{0.39\textwidth}
        \centering
        \includegraphics[width=\textwidth]{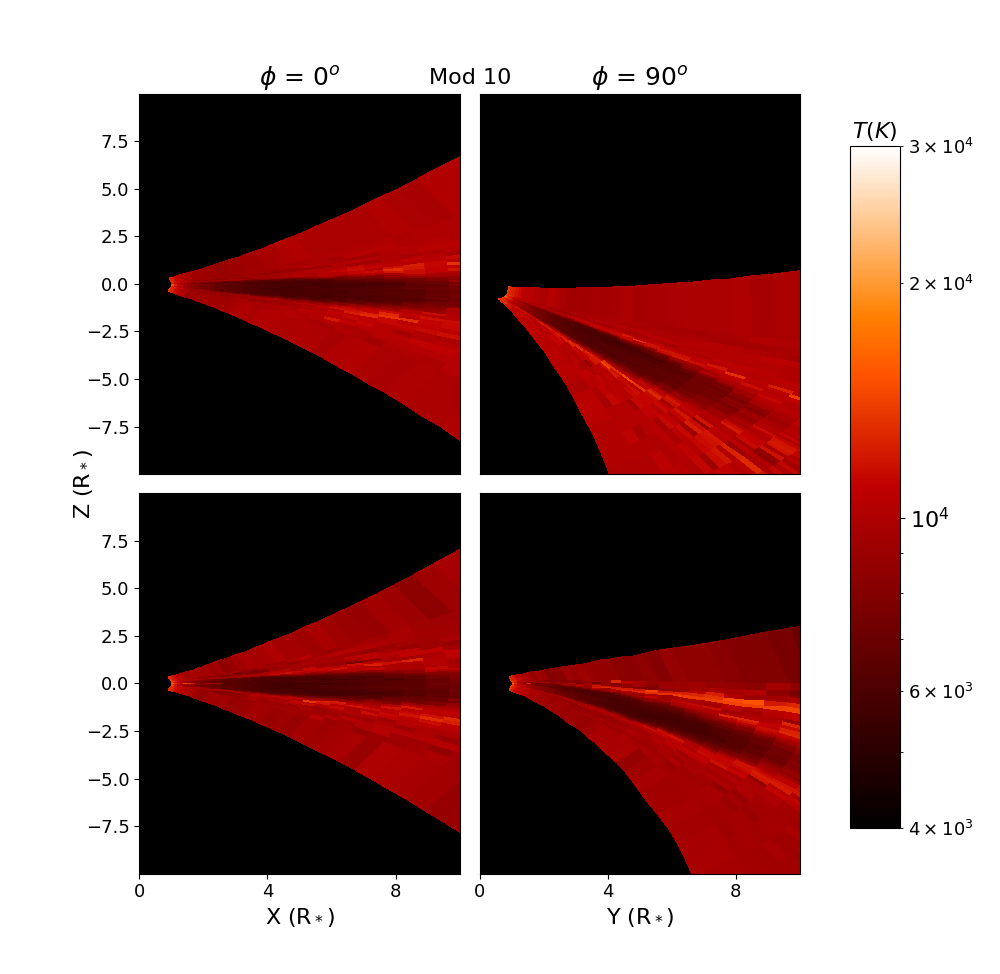}
        \caption{Model 10, zoomed in to 10 $R_*$.}
        \label{fig:warped_temp_compare_zoom}
    \end{subfigure}
    \begin{subfigure}[b]{0.39\textwidth}
        \centering
        \includegraphics[width=\textwidth]{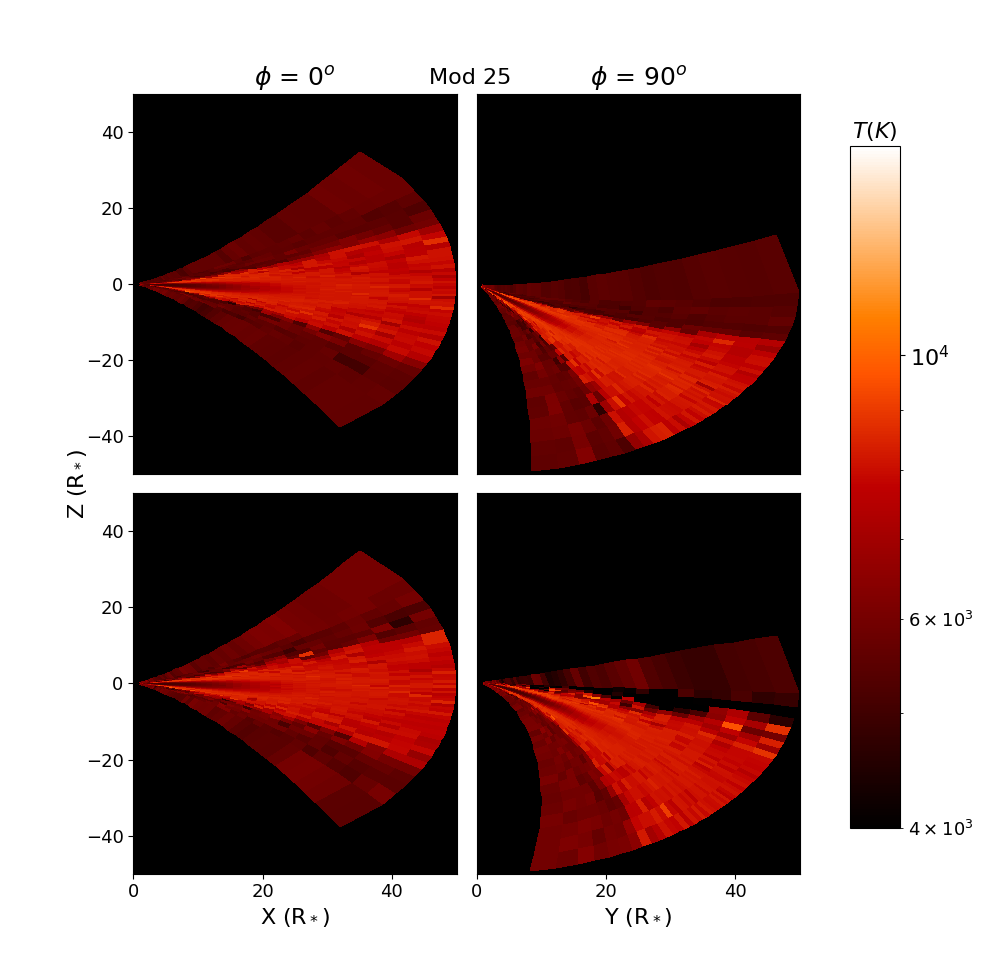}
        \caption{Model 25, full view}
        \label{fig:warped_temp_compare_late}
    \end{subfigure}
    \begin{subfigure}[b]{0.39\textwidth}
        \centering
        \includegraphics[width=\textwidth]{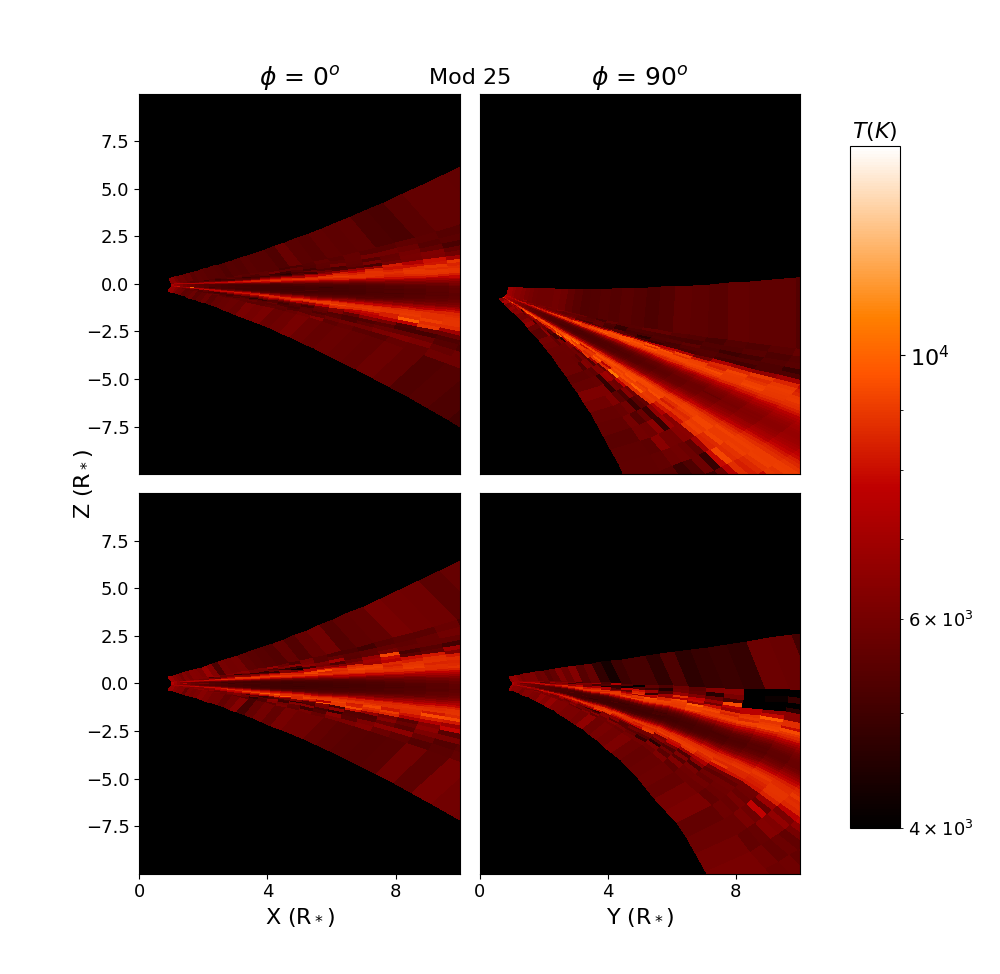}
        \caption{Model 25, zoomed in to 10 $R_*$.}
        \label{fig:warped_temp_compare_zoom_late}
    \end{subfigure}
    \caption{Each panel is a similar to Figure \ref{fig:mod33_tilt_temps}; (a) and (b) are for model 10, while (c) and (d) are for model 25. The top row in each subplot is for the model with a flat $\ang{40}$ tilt, while the bottom row is for the model warped to a maximum of $\ang{40}$.}
    \label{fig:warped_temp}
\end{figure*}


Figure \ref{fig:warped_temp} shows temperature cross sections of the warped disc compared to its flat tilted counterpart. We see that in the inner warped disc, the upper and lower edges of the disc are slightly cooler than the flat tilted model. This is due to the inner disc still being anchored at the equator in the warped case, and thus the inner disc is seeing less radiation than when it is tilted at $\ang{40}$ with the rest of the disc. This effect is seen even in the non-warped slice of the disc at $\phi\,=\,\ang{0}$, due to less diffuse radiation within the disc being able to heat this cross section. In the outer warped disc, the upper and lower edges are respectively cooler and warmer than the flat-tilted case due to the upper edge being shielded by the inner disc at the equator, and the lower edge being freely exposed to radiation from higher stellar latitudes. Figure \ref{fig:warped_temp} also shows how, not surprisingly, the temperature structure warps with the warped density of the disc, and that the interior structure is the same as the tilted model aside from this warp.

With respect to observables, comparison of the warped disc to the different tilted simulations highlights how different areas of the disc are responsible for different observables. To compare these observables between the tilted models and warped models, we held the $\theta$ observing angle constant, and calculated a chi-squared value over all $\phi$ observing angles between each tilted model and the warped model. For the early-type warped model, we find that the $V$ magnitude, percent polarization, and polarization PA of the warped model are best matched by the non-tilted and $\ang{10}$ tilted models, and occasionally the $\ang{20}$ tilted model. This is expected, as these observables originate in the inner disc, which is the least tilted part of the warped disc. On the other hand, \Halpha EW is best matched by the $\ang{40}$ tilt for near pole-on $\theta$ angles, and by $\ang{0}$ and $\ang{10}$ tilt angles for $\theta$ values greater than $\ang{50}$. This is due to a large increase in the continuum emission at certain $\phi$ angles for the $\ang{40}$ flat tilted model, which causes the \Halpha EW to drop significantly. The V/R ratio of the \Halpha line also follows the same trends as the EW. The same trends are seen for the late-type (model 25) warped disc, except that the \Halpha line best matches the flat $\ang{40}$ tilted disc for all viewing inclinations due to the optically thin disc.

\begin{figure}
    \centering
    \includegraphics[scale=0.35]{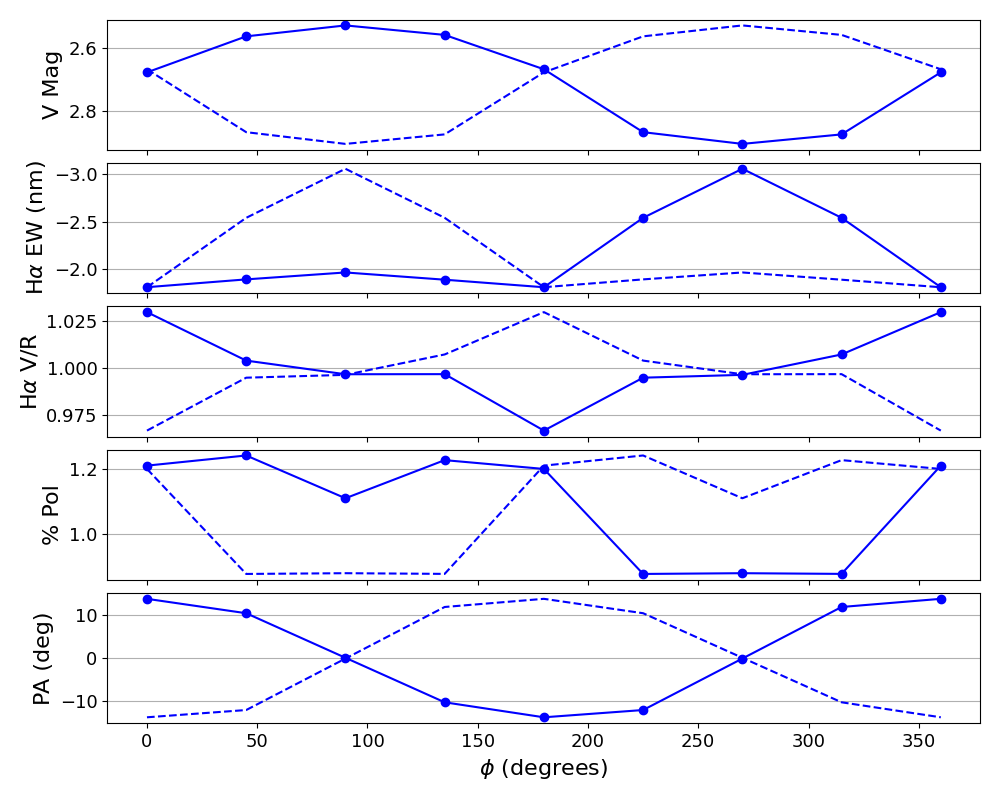}
    \caption{Same as Figure \ref{fig:bigplot_phi_mod97}, but for the warped disc of model 10. The system is viewed at $\theta\,=\,\ang{80}$.}
    \label{fig:bigplot_warped_mod139}
\end{figure}

For comparison purposes, in Figure \ref{fig:bigplot_warped_mod139} we show the observables of a precessing warped disc, similar to what was shown in Figure \ref{fig:bigplot_phi_mod97} for the flat tilted model. The stellar and disc parameters of both Figures are different, so they cannot be directly compared, however we do see that a warped disc, if it were precessing, produces the same period and half-period trends as the flat tilted model, with some asymmetry in the photometry, polarization, and \Halpha line.

\section{Discussion and Conclusions}
\label{sec:discussion}

In this paper, we have shown how tilting a Be star disc out of the equatorial plane of the primary star can affect the disc's temperature structure, as well as it's observables. We modelled B0, B2, B5, and B8 type stars, with different densities, two different rotation rates, and disc tilt angles of $\ang{0}$, $\ang{10}$, $\ang{20}$ and $\ang{40}$. 

We find that the temperature structure between non-tilted early and late type stars can differ greatly, and the behaviour from model to model is highly non-linear. The exact temperature structure is dependent on the disc density configuration, the spectral type, and the stellar rotation rate, which means depending on the model parameters, we see particular trends in the temperature behaviour.

In our non-tilted models we see that all discs have an inner cool region, the extent of which dramatically depends on the density exponent $n$. This can be of significance, as low excitation lines, particularly Fe~{\sc ii}, are known to originate in these cool inner disc volumes \citep{carciofi2006non}. Fe~{\sc ii} emission lines have been well documented in Be stars \citep{Hanuschik1996}, and their line-cooling effects have also been explored \citep{Jones2004}. Since Fe~{\sc ii} emission lines originate in these cool areas, their shape could be used as a tracer of the radial extent of these regions, assuming the width of the line is largely due to Doppler broadening. In this sense, if the Fe~{\sc ii} line had large peak separation and a sharp drop-off in the wings, the center cool region would be relatively small, however a large cool region would mean a large formation loci for Fe~{\sc ii} and its line shape could be similar to Balmer emission lines, albeit with lower peak intensity. Thus, Fe~{\sc ii} lines may be a valuable constraint on the value of $n$ in Be star disc models, and shows the great importance of having a non-isothermal disc model \citep[which was attempted but not conclusively shown by][]{Klement2015}.

We find presence of hot bands above and below the midplane in nearly all disc density configurations, consistent with findings in other studies \citep{Millar1998, carciofi2006non, Sigut2007, McGill2011}. We offer the first concrete explanation of these sheaths, showing in Figure \ref{fig:ion_frac_mod11} that the sheaths occur right at the boundary between where the disc is partially and fully ionized. This strongly indicates that these sheaths are the result of UV radiation that has been trapped in the optically thick, cold inner disc, escaping vertically through the disc and adding excess heat to the optically thin outer disc right at the boundary of this partially ionized region. We also predict that if the disc is dense enough, diffuse radiation near the midplane of the disc can play a large role in heating the disc midplane, sometimes causing the midplane to be warmer than the upper disc layers despite not being fully ionized, particularly seen in our models with late-type stars.

We also investigated the difference between the star having an average rotation rate of 70\% the critical velocity versus a high rotation rate, 95\% of the critical velocity. In our B0, B2, and B5 models, this increase in rotation had marginal effects, only heating the outer disc and midplane slightly, but keeping the overall temperature structure the same. This is different when looking at our B8 models, where an increase in rotation caused a large increase in midplane temperature, as well as the appearance of prominent hot sheaths. In this case, the combination of the higher rotation giving hotter poles, along with the lower densities used for our B8 models, allow the radiation from hot poles to ``carve" farther into the disc, causing substantial heating in and around the midplane.

Overall the temperature structure of our non-tilted models is remarkably similar to the works of \cite{Millar1998, Millar1999, Millar1999a, Millar1999b, Millar2000}, who did detailed work on both the early-type star $\gamma$ Cas, and the late-type star 1 Del, using an escape probability method and by balancing the energy contributions to calculate the disc temperature structures of these stars. This similarity comes despite their code using a different density prescription, and only having five hydrogen energy levels included. {\sc hdust}, used here, uses a Monte Carlo technique to solve the radiative transfer,  has 12 non-LTE and 25 LTE hydrogen energy levels, and also accounts for line radiation and bound-bound transitions. We argue that the strong agreement of the temperature distributions between previous work, including \cite{carciofi2006non} for B3 spectral types also using {\sc hdust}, and the new work presented here, infers that the temperature and ionization levels are primarily controlled by photoionization — recombination equilibrium. This agreement also provides strong evidence of the broad applicability of our work to gaseous discs.

In tilting the disc, we see modest large-scale changes to the disc's average temperature, with it increasing slightly as the disc tilt angle increases. On a smaller scale, we find that with increasing tilt angle, the part of the disc that moves towards the equator becomes cooler, while the portion of the disc moving towards the stellar pole becomes hotter. These changes are explained by gravity darkening of the rapidly rotating star, making the stellar poles hotter than the stellar equator. This anti-symmetric change of disc temperatures is why the overall average disc temperature does not change appreciably when the disc is tilted. The temperature in the midplane of the disc is largely unchanged by the disc tilting due to it's higher density than the rest of the disc, although for already optically thin discs, the midplane can vary in temperature as well. This behaviour is seen across all spectral types.

Examining the observables of our tilted disc simulations, we offer two scenarios where a disc tilt may be detected. 

The first case is where the disc is actively observed to be tilting. In this scenario, the change in observables is entirely dependent on the direction of tilt relative to the observer. A disc may appear completely differently with a $\ang{90}$ or $\ang{180}$ change of relative orientation as the disc either moves to be more face-on or more edge-on with the observer as it tilts. This variability would make it difficult to interpret whether changes in observables of a Be star would be due to a disc tilting or simple changes in disc density or size. The strongest evidence of disc tilting would appear in the polarization PA where, if one were looking along the axis of the disc's tilt, the position angle should exactly match the tilt angle of the disc, however the position angle will change some amount as long as the observers line of sight is not exactly perpendicular to the tilt axis. No other change in geometry would cause a change of tens of degrees in the position angle, making it a key measurement to look at for proof of disc tilting. The shape of the \Halpha line would also be a clear indication of disc tilting as the shape changes from single-peaked to double-peaked and vice-versa. This change could not be brought about by a simple change of density structure or a larger/smaller disc, and could only occur with a major change of disc geometry such as tilting of the \Halpha emitting region. This would also be seen in other emission lines as well, not just \Halpha. The advantage of these two observables being the leading indicators of tilting is that one of them should appear no matter the orientation of the disc, given a large enough disc tilt. If the disc is tilting to be more face-on or edge-on with the observer, the \Halpha line shape would change in shape while the position angle would not. On the other hand, if the observer was looking more aligned with the tilt axis of the disc, the polarization position angle would change, while the \Halpha line shape would be approximately constant. Thus, both the emission line shape and polarization position angle are key signatures of disc tilting. Another difference that could set a tilted disc apart from a non-tilted disc is the V/R ratio in the H$\alpha$ emission lines, however this ratio is not particularly strong in our models apart from a few cases of early-type stars where the disc density and tilt angle is highest. There is no clear pattern to why those few models show stronger V/R ratios than others, so it would be difficult to discern, without further constraints by other observables, whether V/R ratios in actual observations of Be stars are due to a tilted disc or a density enhancement in the disc, like those produced by spiral enhancements in $\zeta$ Tau \citep{Stefl2009, Escolano2015}, and 48 Lib \citep{Silaj2016}.

The second case is where the disc is already tilted, and undergoing precession due to the influence of a misaligned binary companion. We are able to simulate the disc precessing about the stellar pole by holding the $\theta$ observing angle constant and changing $\phi$ only. Here we find that the percent polarization, $V$ magnitude, and \Halpha EW oscillate at half the precession period, although the oscillation will not be perfectly symmetric unless the observer is directly aligned with the stellar equator. The position angle, on the other hand, will oscillate in sync with disc precession. The V/R ratio undergoes an antisymmetric half-period oscillation, with nodes at $\phi\,=\,\ang{90}$ and $\ang{270}$ due to the violet and red sides of the disc reversing when the observer moves to the other side of the disc.

We then investigated two other scenarios. First, we computed two truncated disc models out to a radius of 20 $R_*$, to see any possible obscuring effects that may happen compared to a disc 50 $R_*$ in size. These simulations revealed that the outer disc from 20 to 50 stellar radii only marginally increased \Halpha emission, while not changing other examined observables. The temperature structure was also unchanged out to 20 stellar radii, as expected.

Second, and more importantly, we computed two models that were linearly warped up to a maximum angle of $\ang{40}$. This model revealed a cooler outer disc temperature versus its flat tilted counterpart, and an inner temperature structure that followed the warp of the disc. The observables of this model are essentially a mix of all the tilted models together, with some observables better matching the non-tilted or $\ang{10}$ models, while others matched the higher tilt models. This shows how important it is to recognize that Be star discs emit some wavelengths from dense inner volumes while other wavelengths come from larger radial positions in the disc. These warped models are merely an initial test to see what effects a warped disc may introduce. A proper warped disc study is beyond the scope of this paper, though certainly merits its own study.

These simulations are a vital step towards simulations of more complex disc configurations, such as ones containing warped discs, or those presented by \cite{Suffak2022}, which contain holes and tearing of the disc. The flat-tilted models here will be a good benchmark for analysis of these discs that are tilted, warped, and have asymmetric density distributions. With the fundamentals presented here, we will be able to tackle more complicated Be star systems such as Pleione, which is suspected to have a periodic tearing disc \citep{Marr2022, Martin2022}.

\section*{Acknowledgements}

We thank the anonymous referee for their thorough comments which improved the paper. We greatly acknowledge the work of \cite{MarrThesis}, whose preliminary work on the temperature of tilted discs inspired and aided this work. C.E.J. acknowledges support through the National Science and Engineering Research Council of Canada. A. C. C. acknowledges support from CNPq (grant 311446/2019-1) and FAPESP (grants 2018/04055-8 and 2019/13354-1). THA acknowledges support from FAPESP (grant 2021/01891-2 ). This work was made possible through the use of the Shared Hierarchical Academic Research Computing Network (SHARCNET).

\section*{Data Availability}

Although there is no observational data, the {\sc hdust} models computed for this work can be made available upon request.

\bibliographystyle{mnras}
\bibliography{aastexBeStarbib}
\begin{appendices}

\section{Tilted Temperature cross sections}
\label{sec:tilted_temps}
\begin{figure*}
    \centering
    \begin{subfigure}[b]{0.4\textwidth}
        \centering
        \includegraphics[width=\textwidth]{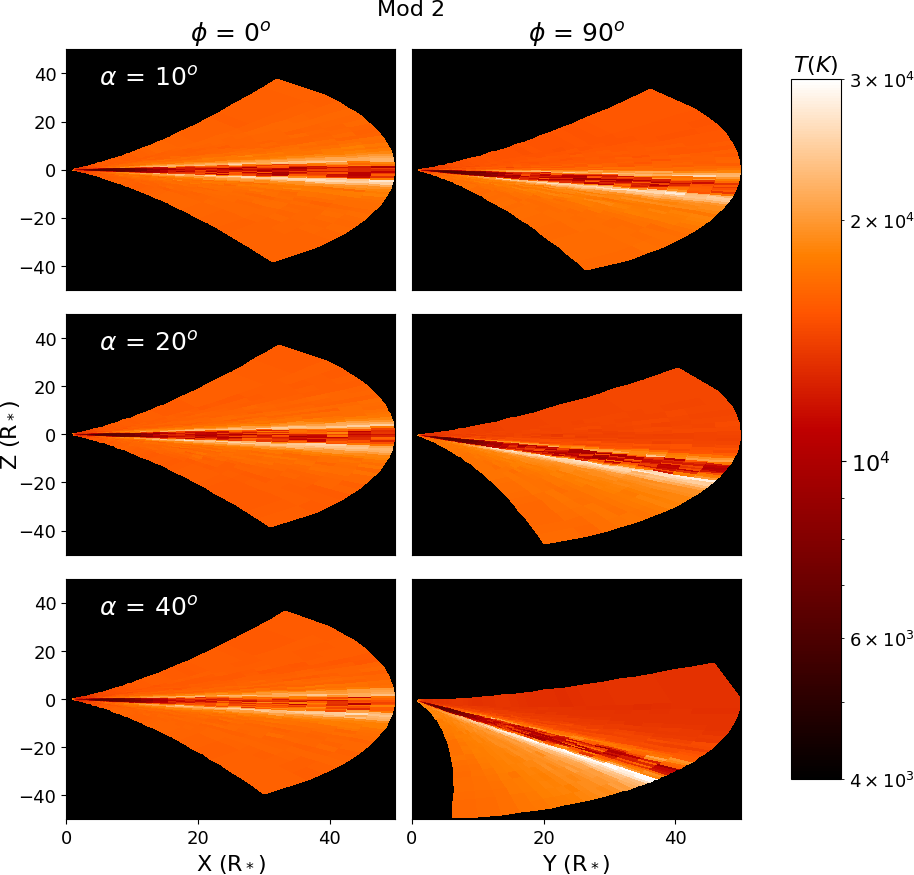}
        \vspace{3pt}
    \end{subfigure} 
    \hspace{5pt}
    \begin{subfigure}[b]{0.4\textwidth}
        \centering
        \includegraphics[width=\textwidth]{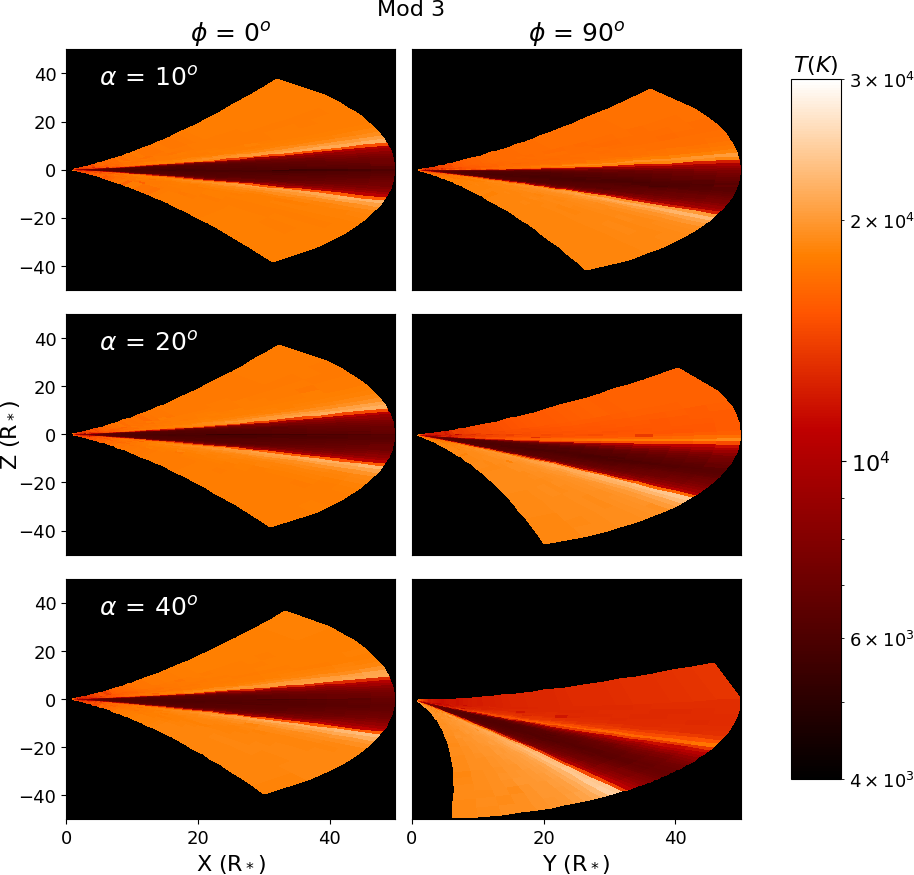}
        \vspace{3pt}
    \end{subfigure} 
    \begin{subfigure}[b]{0.4\textwidth}
        \centering
        \includegraphics[width=\textwidth]{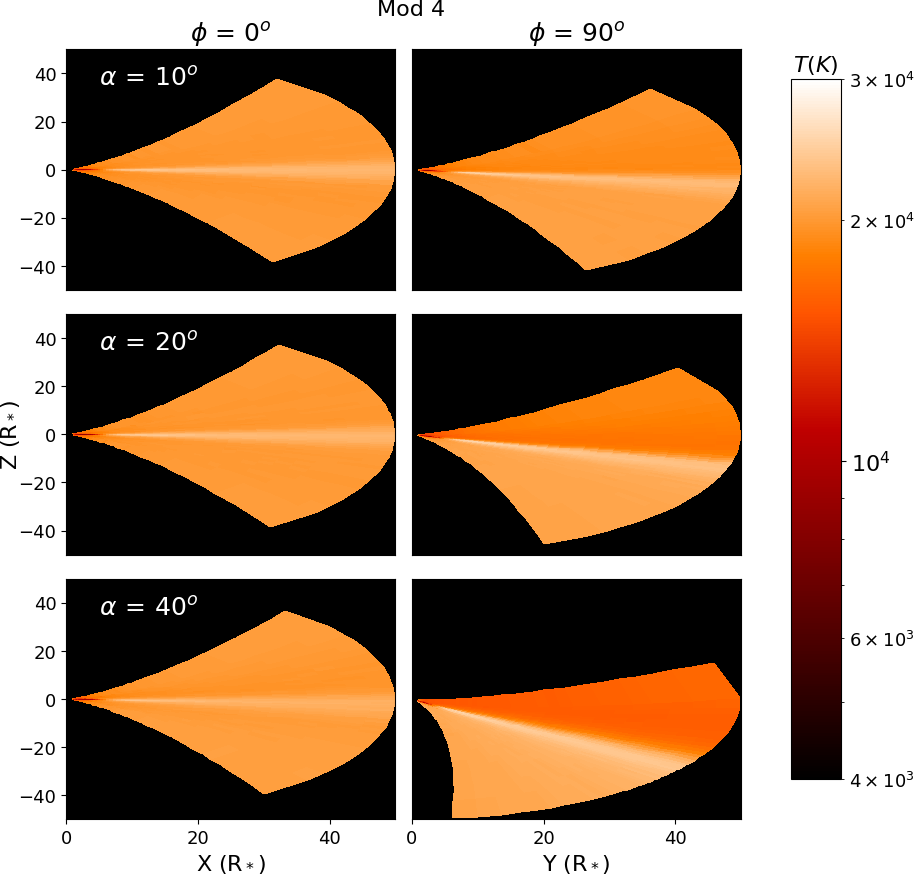}
        \vspace{3pt}
    \end{subfigure} 
    \hspace{5pt}
    \begin{subfigure}[b]{0.4\textwidth}
        \centering
        \includegraphics[width=\textwidth]{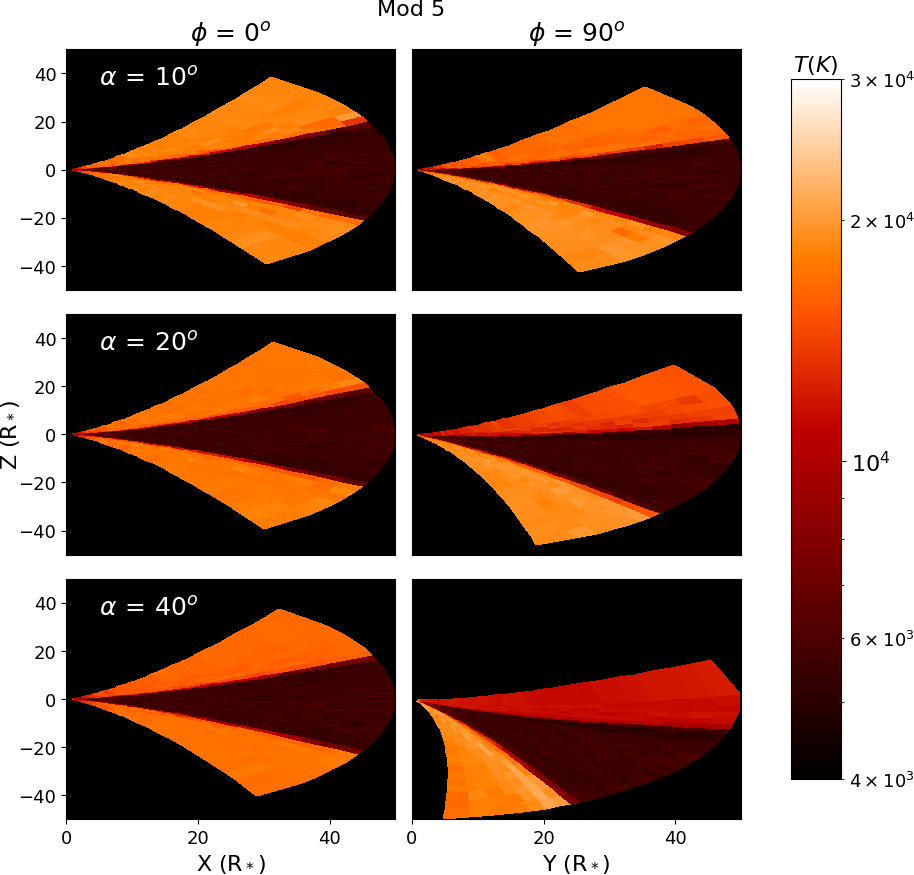}
        \vspace{3pt}
    \end{subfigure} 
    \begin{subfigure}[b]{0.4\textwidth}
        \centering
        \includegraphics[width=\textwidth]{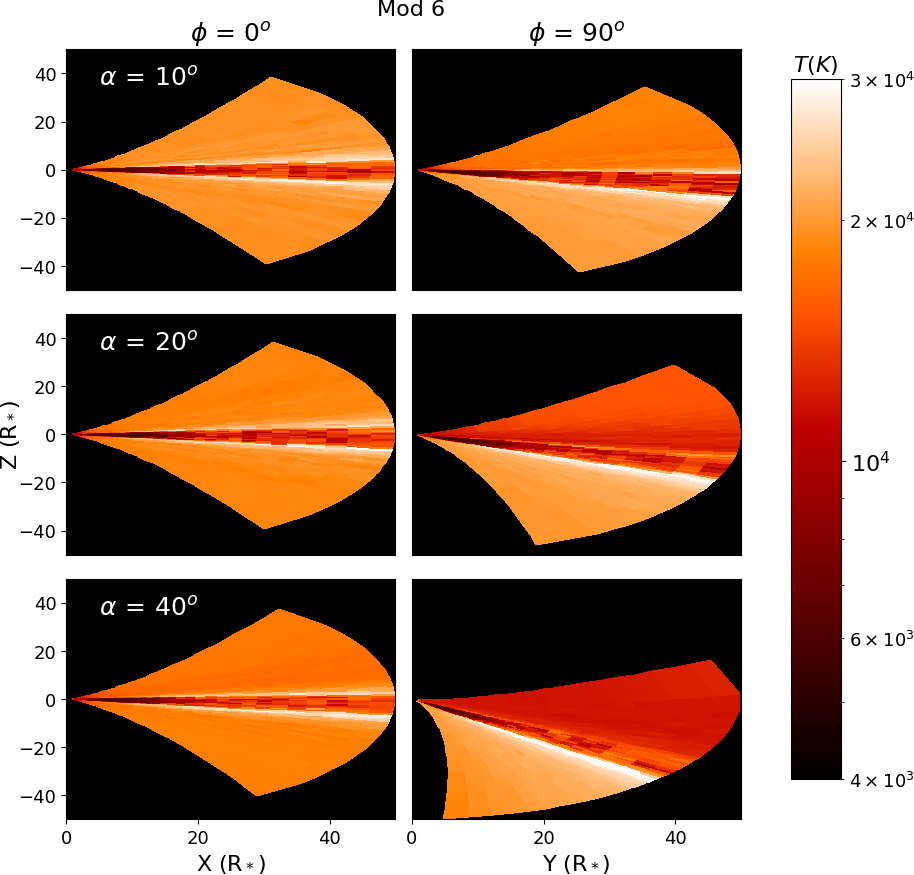}
    \end{subfigure} 
    \hspace{5pt}
    \begin{subfigure}[b]{0.4\textwidth}
        \centering
        \includegraphics[width=\textwidth]{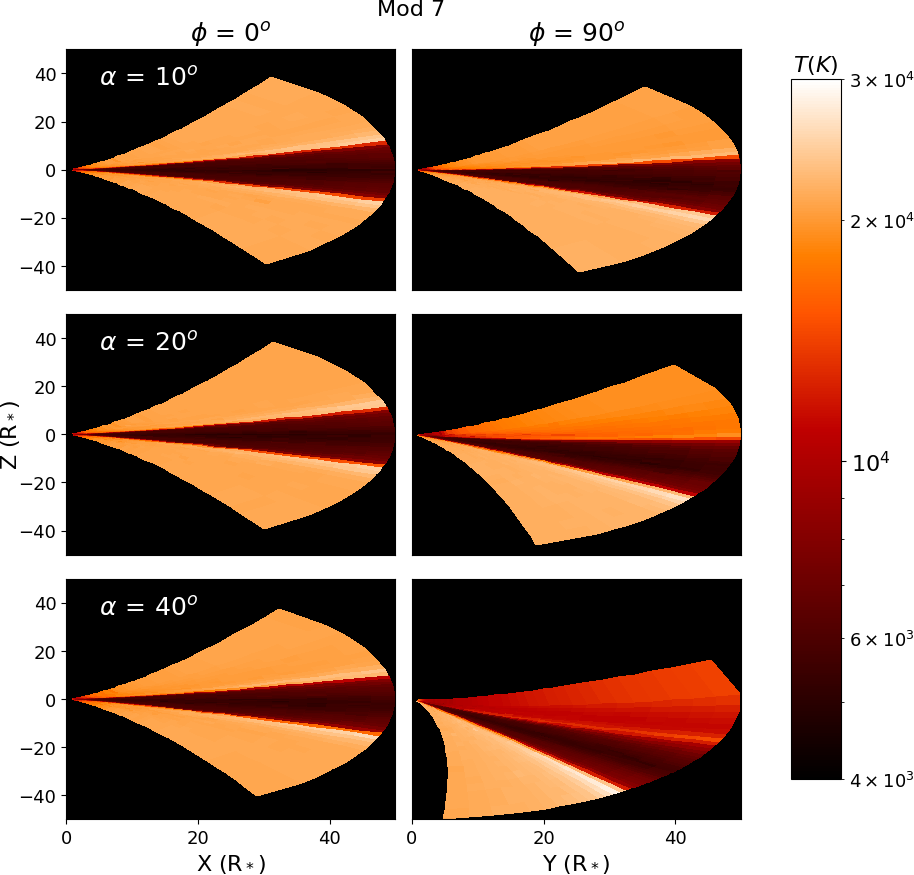}
    \end{subfigure} 
    \caption{Same as Figure \ref{fig:mod33_tilt_temps}, but for models 2-7.}
    \label{fig:mod34-39_tilt_temps}
\end{figure*}

\begin{figure*}
    \centering
    \begin{subfigure}[b]{0.4\textwidth}
        \centering
        \includegraphics[width=\textwidth]{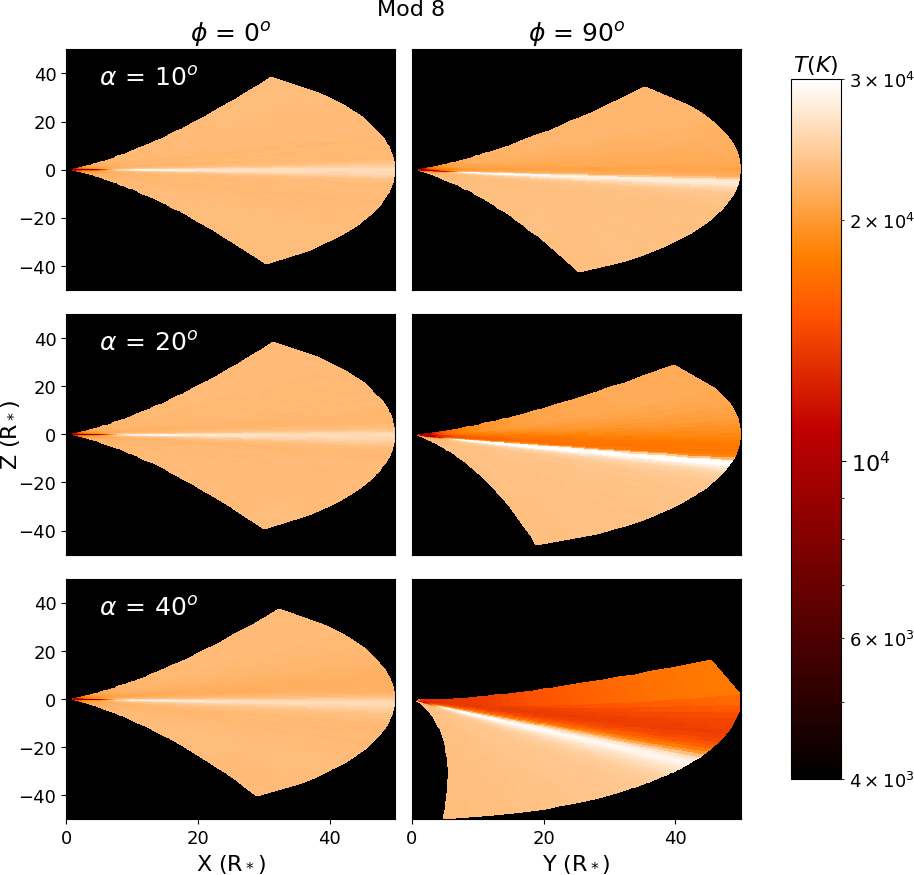}
        \vspace{3pt}
    \end{subfigure} 
    \hspace{5pt}
    \begin{subfigure}[b]{0.4\textwidth}
        \centering
        \includegraphics[width=\textwidth]{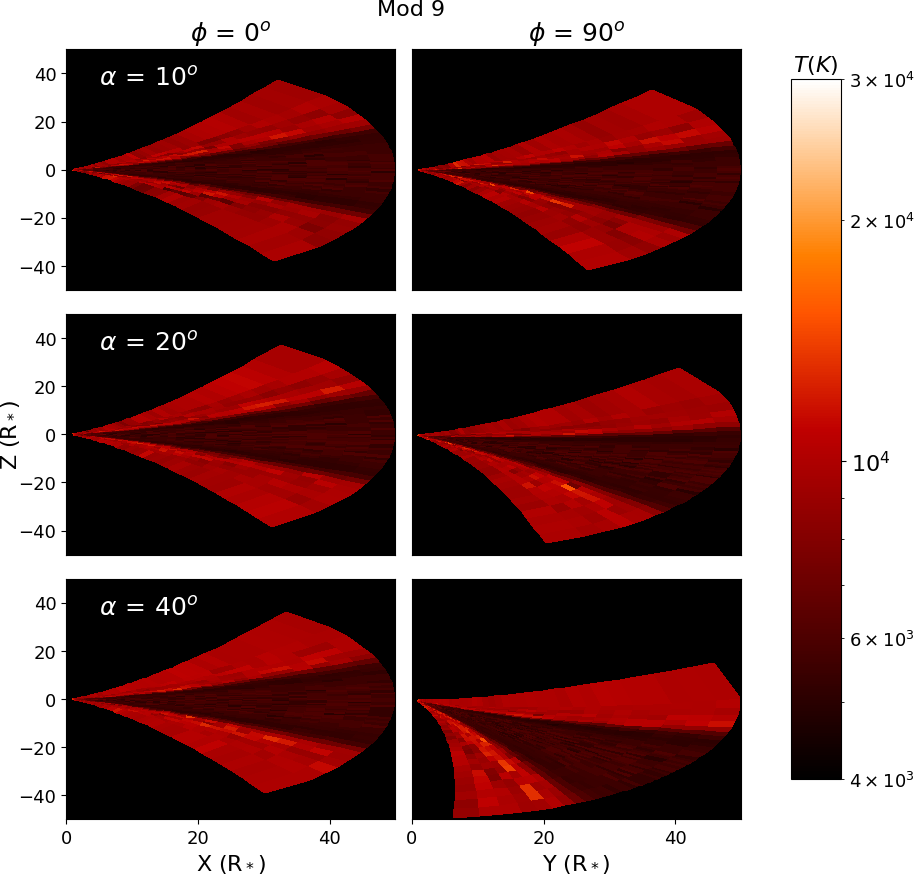}
        \vspace{3pt}
    \end{subfigure} 
    \begin{subfigure}[b]{0.4\textwidth}
        \centering
        \includegraphics[width=\textwidth]{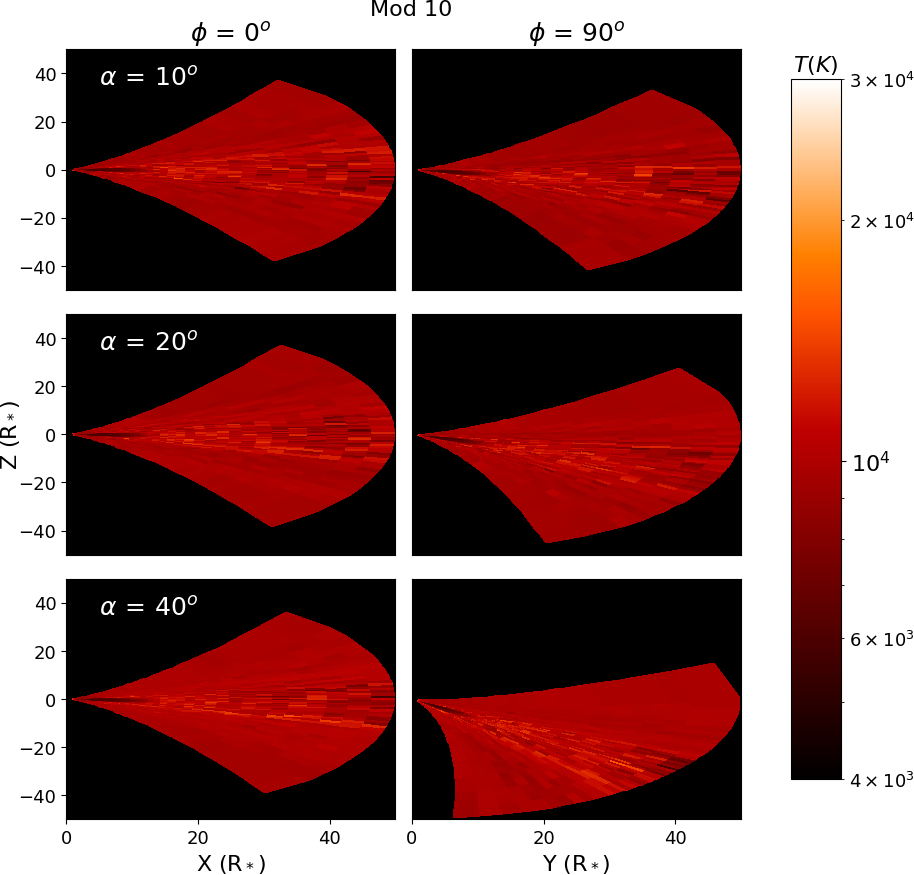}
        \vspace{3pt}
    \end{subfigure} 
    \hspace{5pt}
    \begin{subfigure}[b]{0.4\textwidth}
        \centering
        \includegraphics[width=\textwidth]{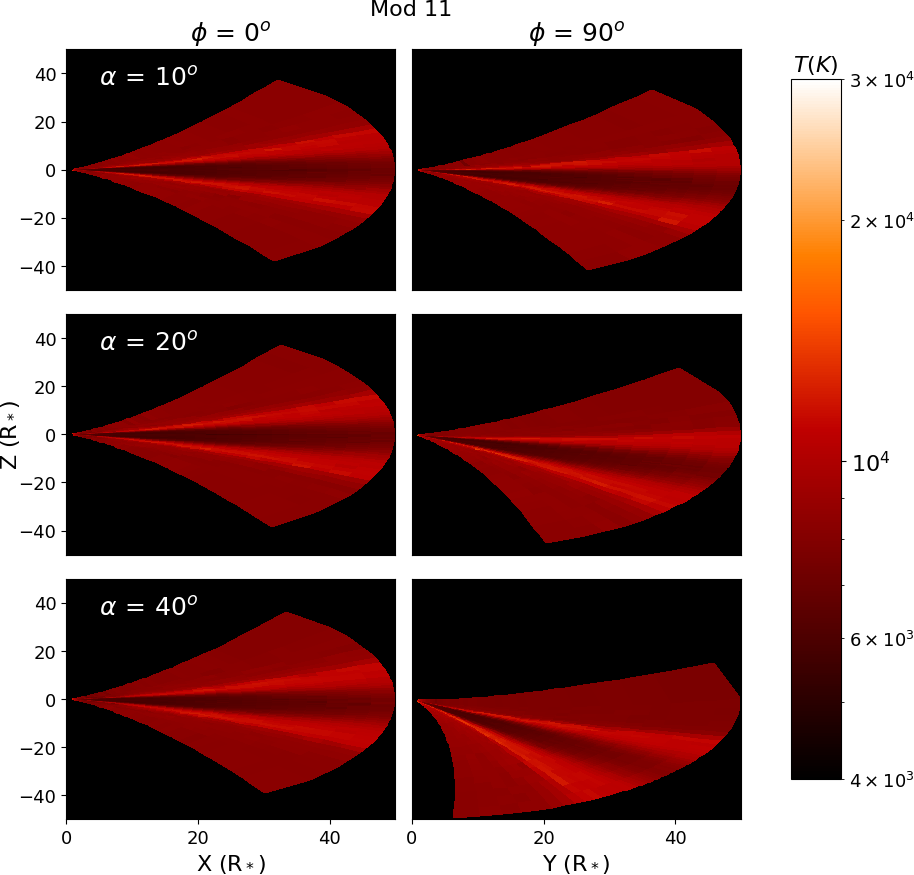}
        \vspace{3pt}
    \end{subfigure} 
    \begin{subfigure}[b]{0.4\textwidth}
        \centering
        \includegraphics[width=\textwidth]{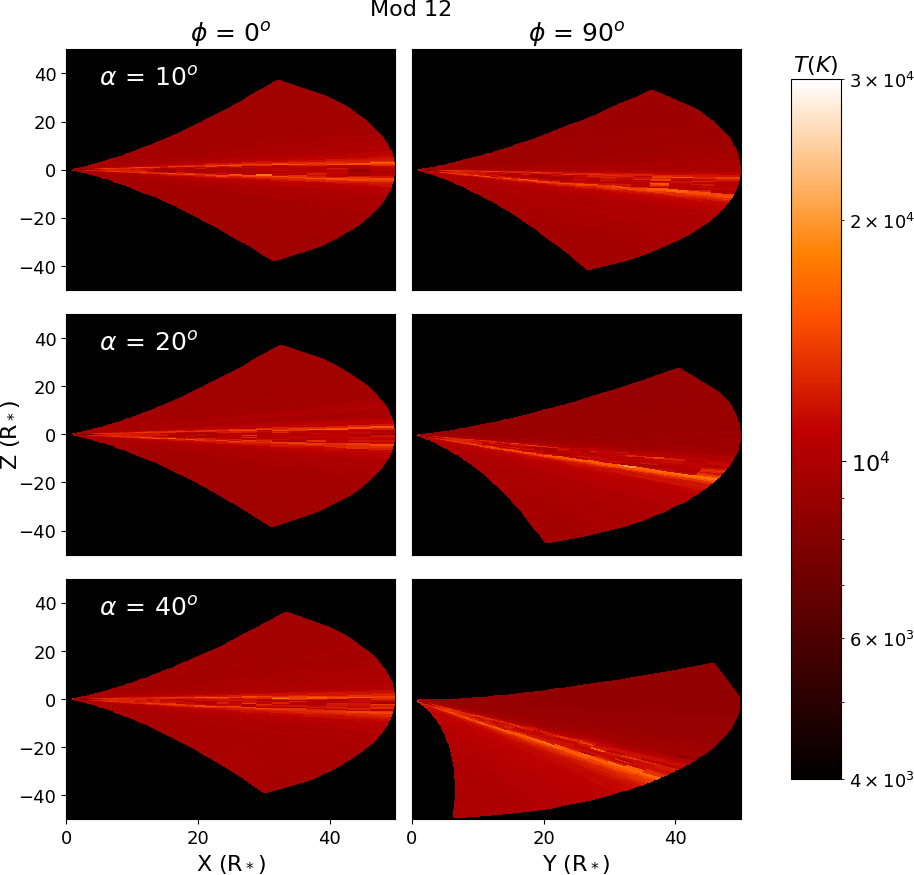}
    \end{subfigure} 
    \hspace{5pt}
    \begin{subfigure}[b]{0.4\textwidth}
        \centering
        \includegraphics[width=\textwidth]{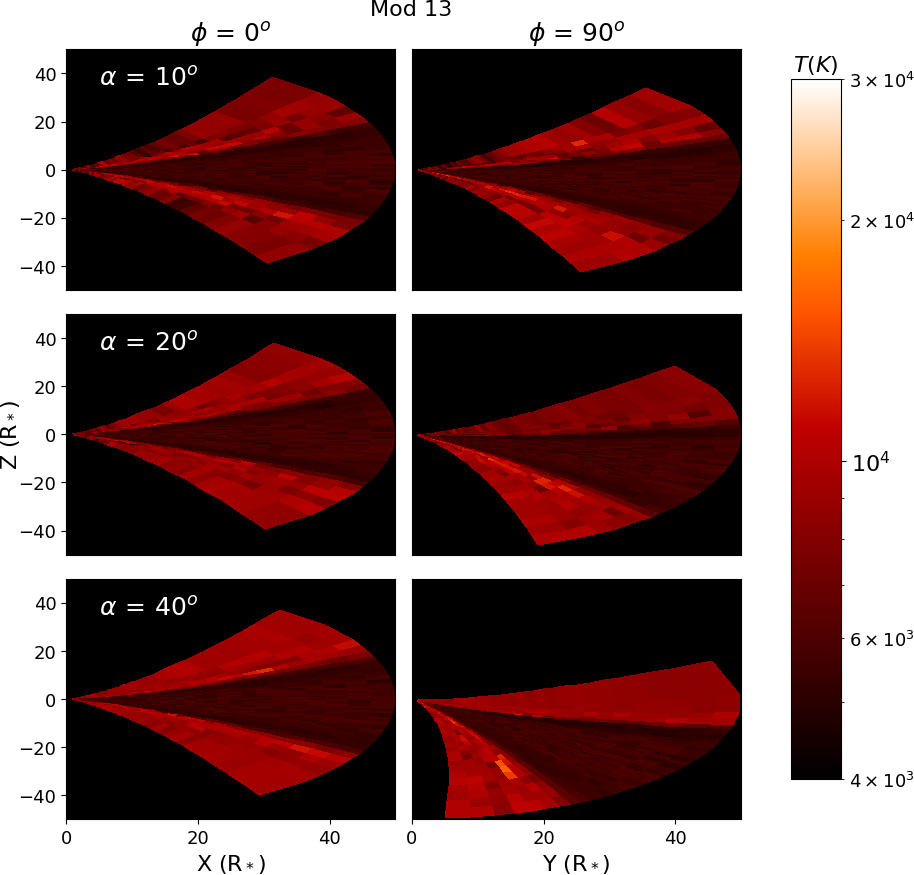}
    \end{subfigure} 
    \caption{Same as Figure \ref{fig:mod33_tilt_temps}, but for models 8-13.}
    \label{fig:mod40-45_tilt_temps}
\end{figure*}

\begin{figure*}
    \centering
    \begin{subfigure}[b]{0.4\textwidth}
        \centering
        \includegraphics[width=\textwidth]{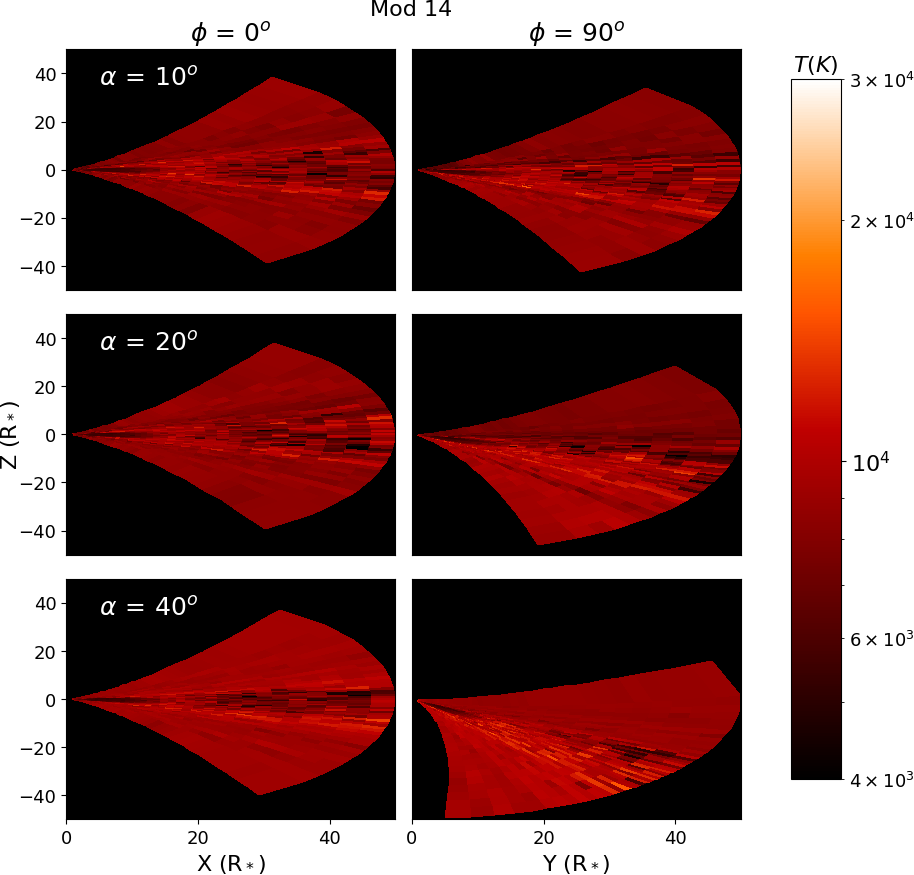}
        \vspace{3pt}
    \end{subfigure} 
    \hspace{5pt}
    \begin{subfigure}[b]{0.4\textwidth}
        \centering
        \includegraphics[width=\textwidth]{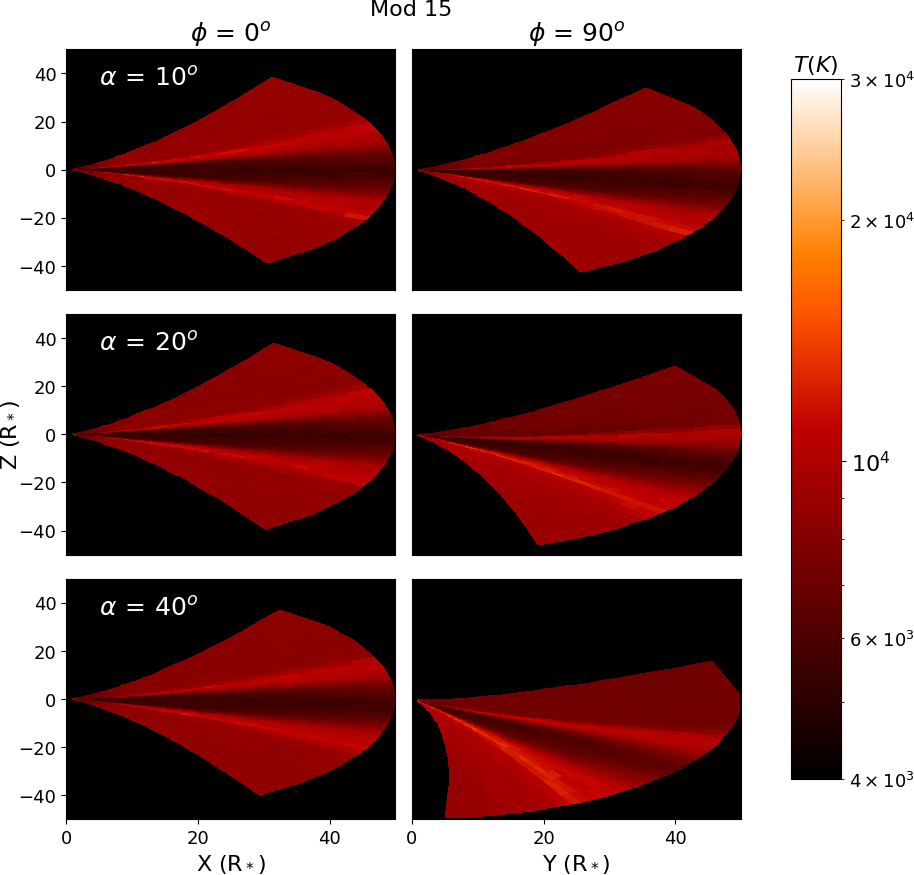}
        \vspace{3pt}
    \end{subfigure} 
    \begin{subfigure}[b]{0.4\textwidth}
        \centering
        \includegraphics[width=\textwidth]{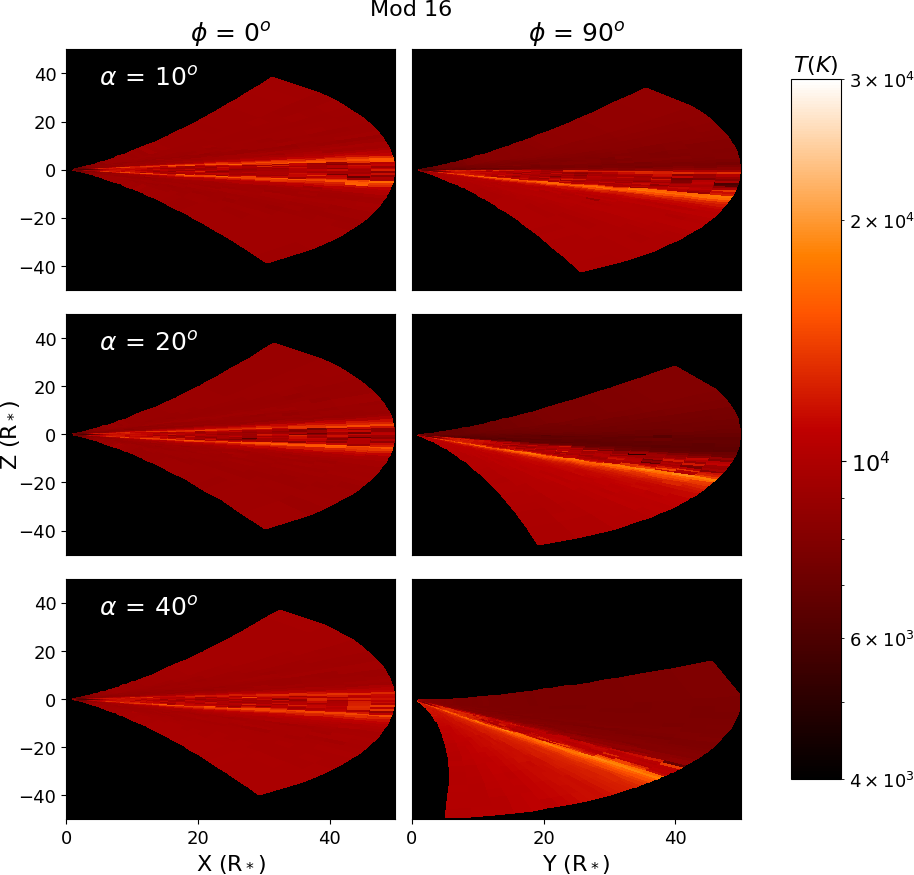}
        \vspace{3pt}
    \end{subfigure} 
    \hspace{5pt}
    \begin{subfigure}[b]{0.4\textwidth}
        \centering
        \includegraphics[width=\textwidth]{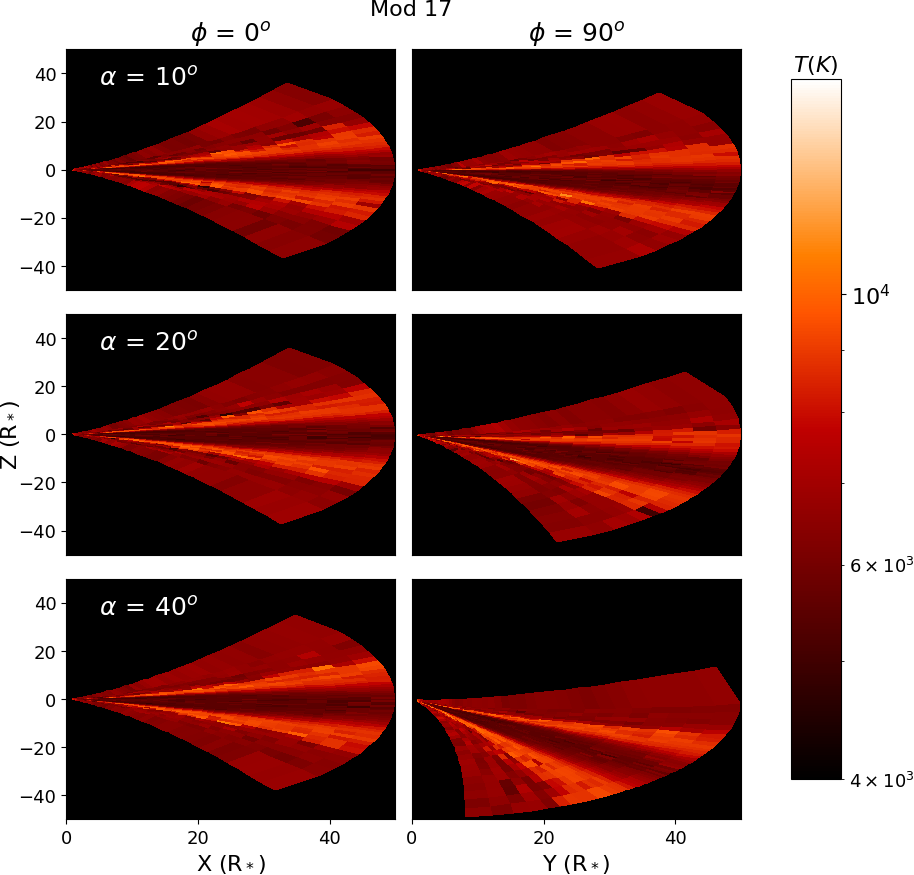}
        \vspace{3pt}
    \end{subfigure} 
    \begin{subfigure}[b]{0.4\textwidth}
        \centering
        \includegraphics[width=\textwidth]{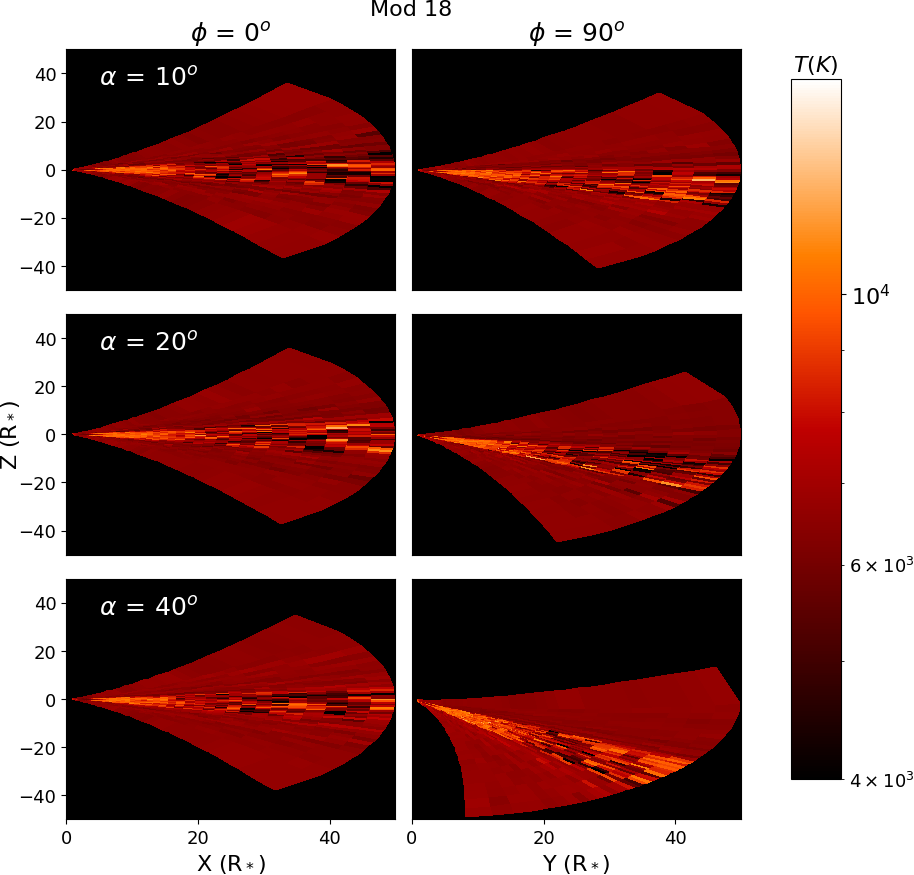}
    \end{subfigure} 
    \hspace{5pt}
    \begin{subfigure}[b]{0.4\textwidth}
        \centering
        \includegraphics[width=\textwidth]{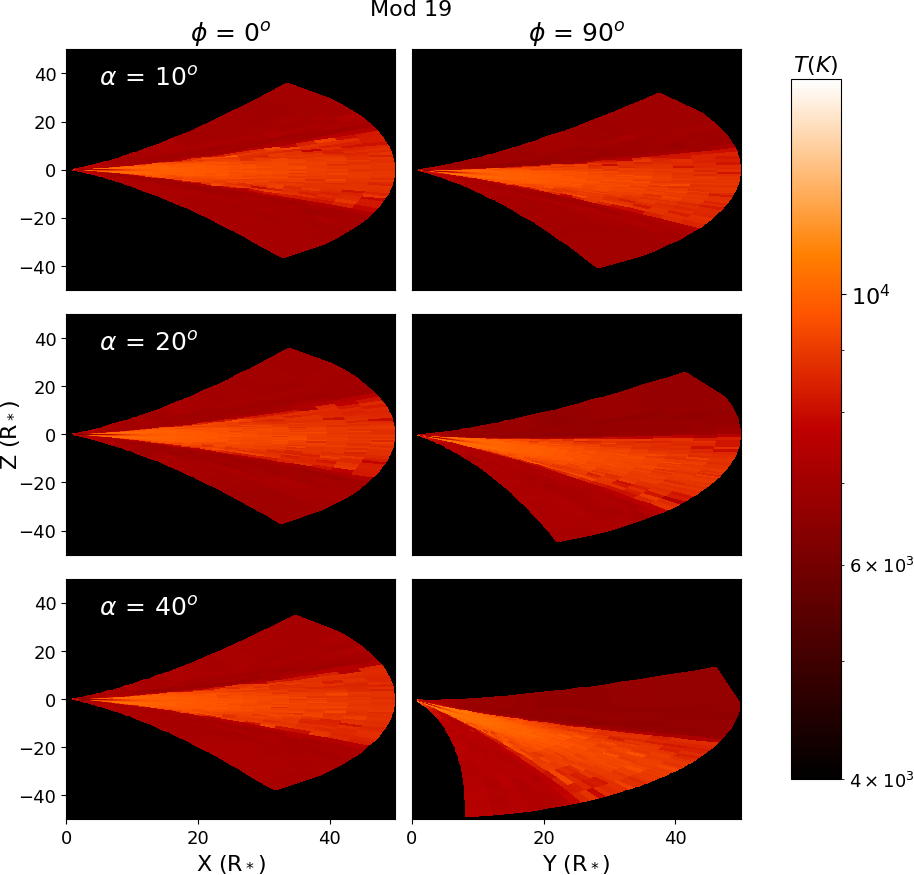}
    \end{subfigure} 
    \caption{Same as Figure \ref{fig:mod33_tilt_temps}, but for models 14-19.}
    \label{fig:mod46-51_tilt_temps}
\end{figure*}

\begin{figure*}
    \centering
    \begin{subfigure}[b]{0.4\textwidth}
        \centering
        \includegraphics[width=\textwidth]{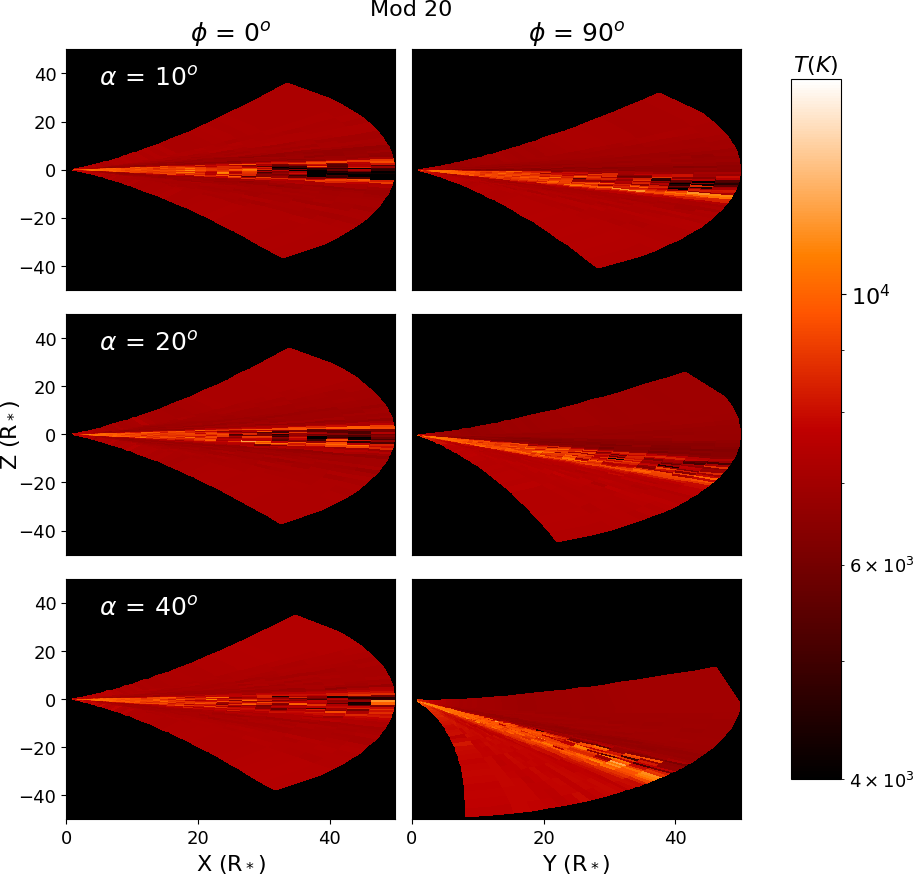}
        \vspace{3pt}
    \end{subfigure} 
    \hspace{5pt}
    \begin{subfigure}[b]{0.4\textwidth}
        \centering
        \includegraphics[width=\textwidth]{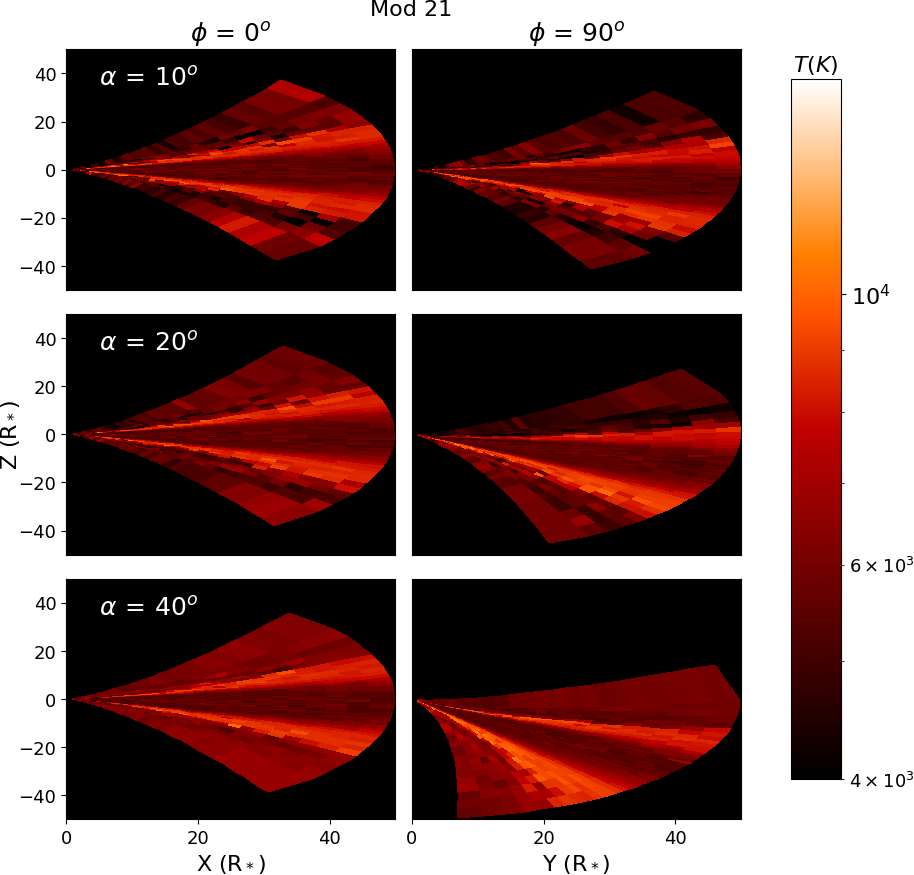}
        \vspace{3pt}
    \end{subfigure} 
    \begin{subfigure}[b]{0.4\textwidth}
        \centering
        \includegraphics[width=\textwidth]{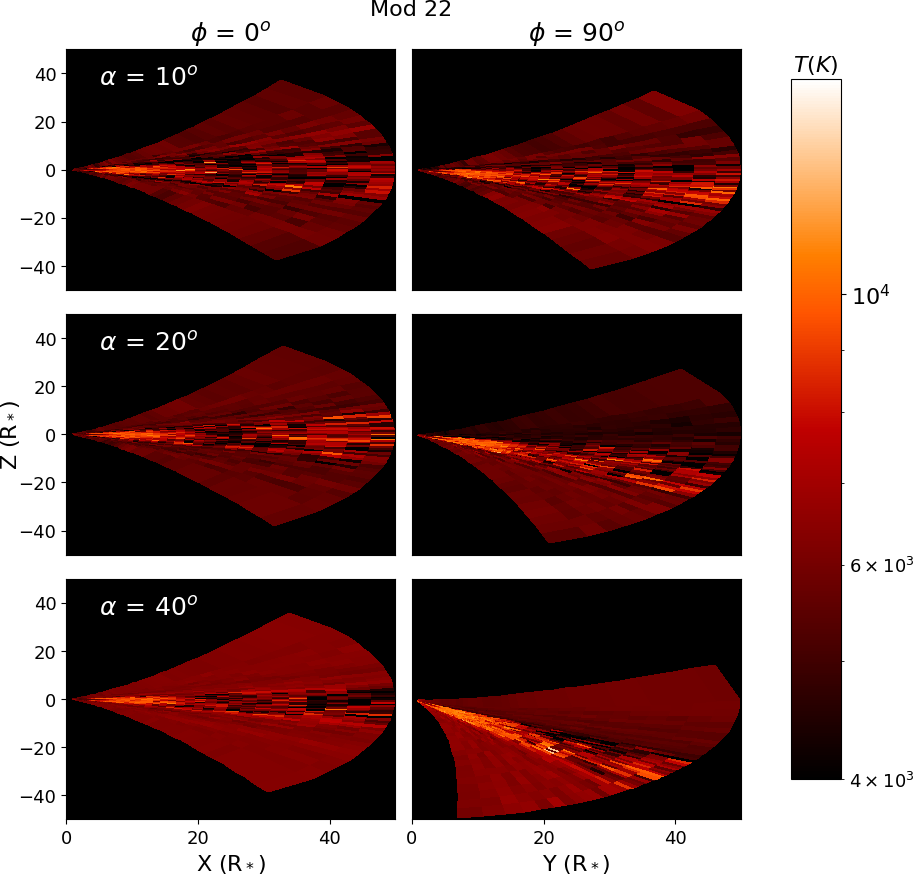}
        \vspace{3pt}
    \end{subfigure} 
    \hspace{5pt}
    \begin{subfigure}[b]{0.4\textwidth}
        \centering
        \includegraphics[width=\textwidth]{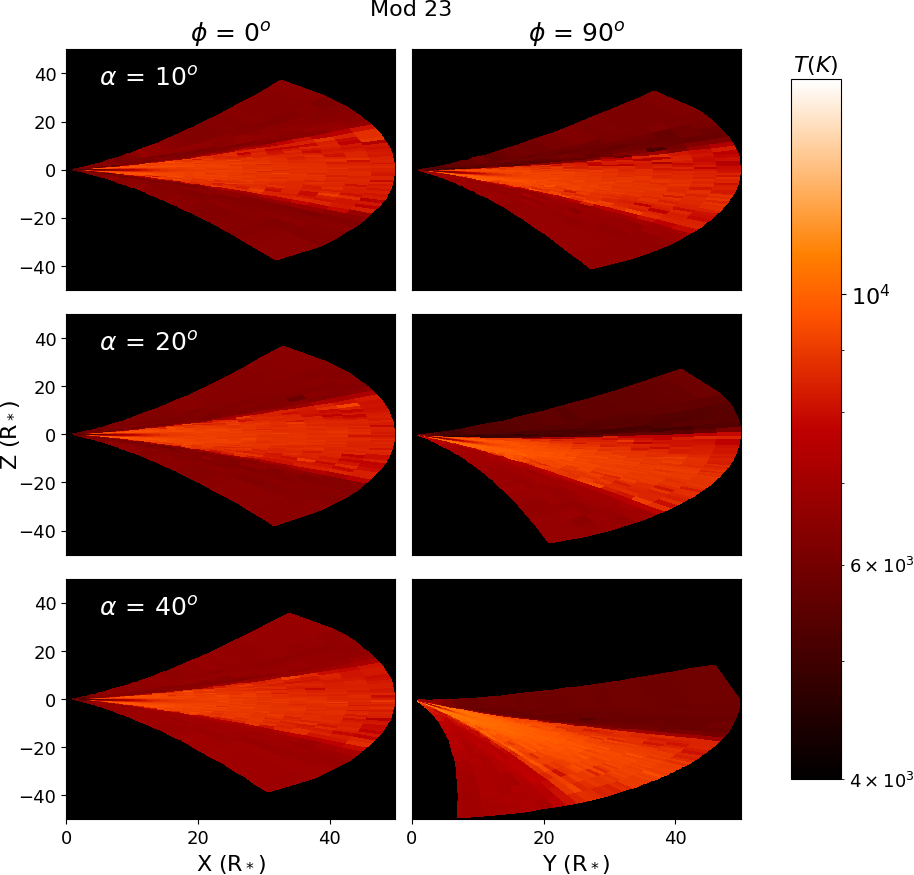}
        \vspace{3pt}
    \end{subfigure} 
    \begin{subfigure}[b]{0.4\textwidth}
        \centering
        \includegraphics[width=\textwidth]{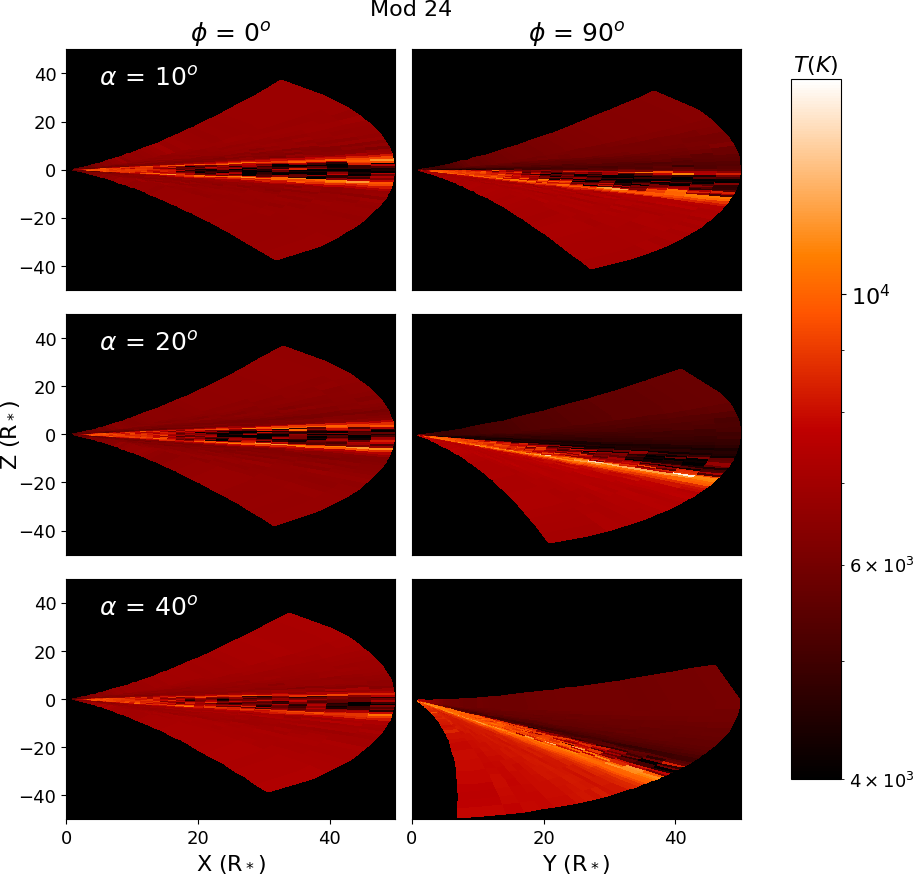}
    \end{subfigure} 
    \hspace{5pt}
    \begin{subfigure}[b]{0.4\textwidth}
        \centering
        \includegraphics[width=\textwidth]{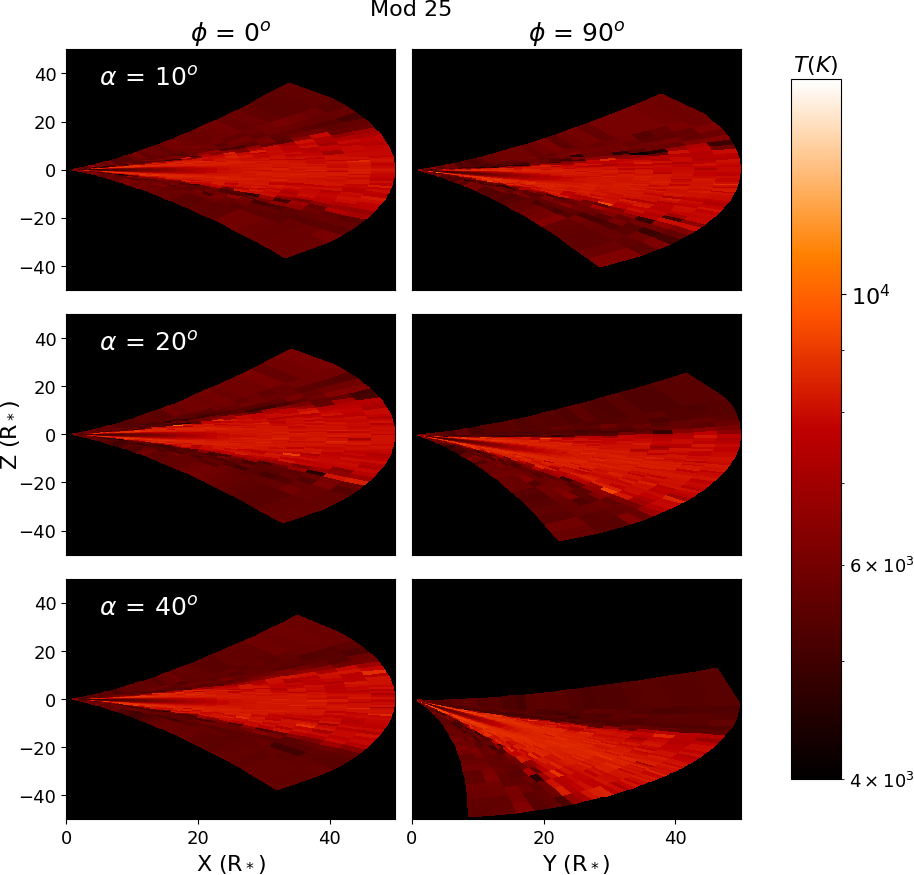}
    \end{subfigure} 
    \caption{Same as Figure \ref{fig:mod33_tilt_temps}, but for models 20-25.}
    \label{fig:mod52-57_tilt_temps}
\end{figure*}

\begin{figure*}
    \centering
    \begin{subfigure}[b]{0.4\textwidth}
        \centering
        \includegraphics[width=\textwidth]{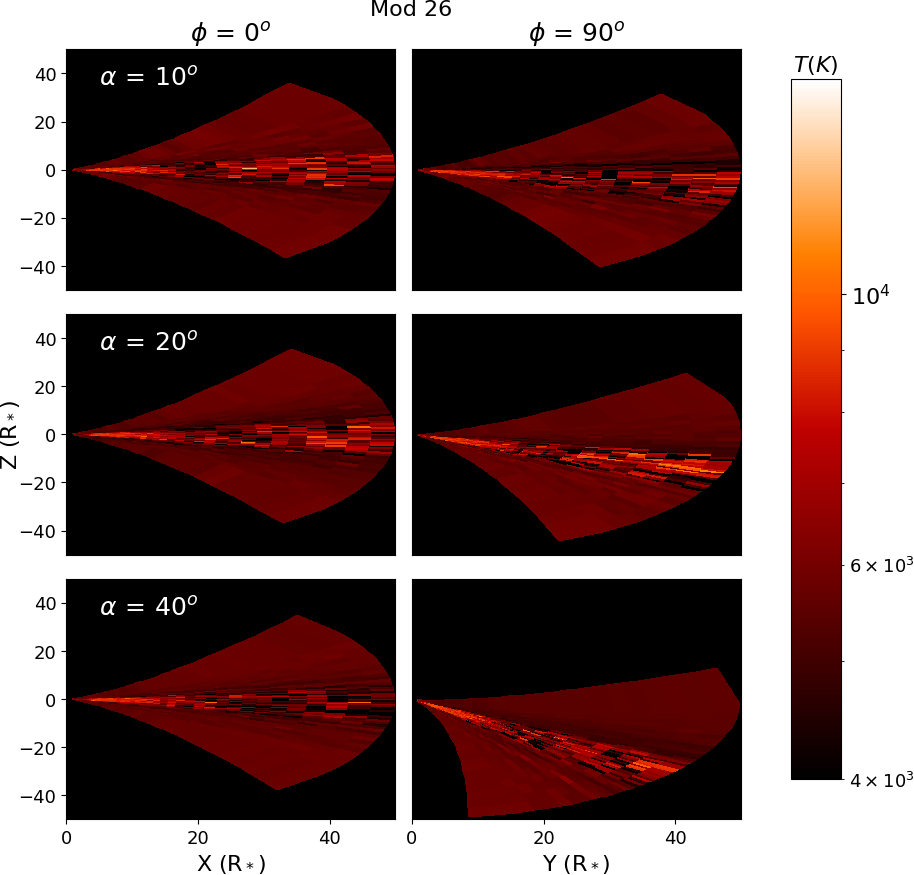}
        \vspace{3pt}
    \end{subfigure} 
    \hspace{5pt}
    \begin{subfigure}[b]{0.4\textwidth}
        \centering
        \includegraphics[width=\textwidth]{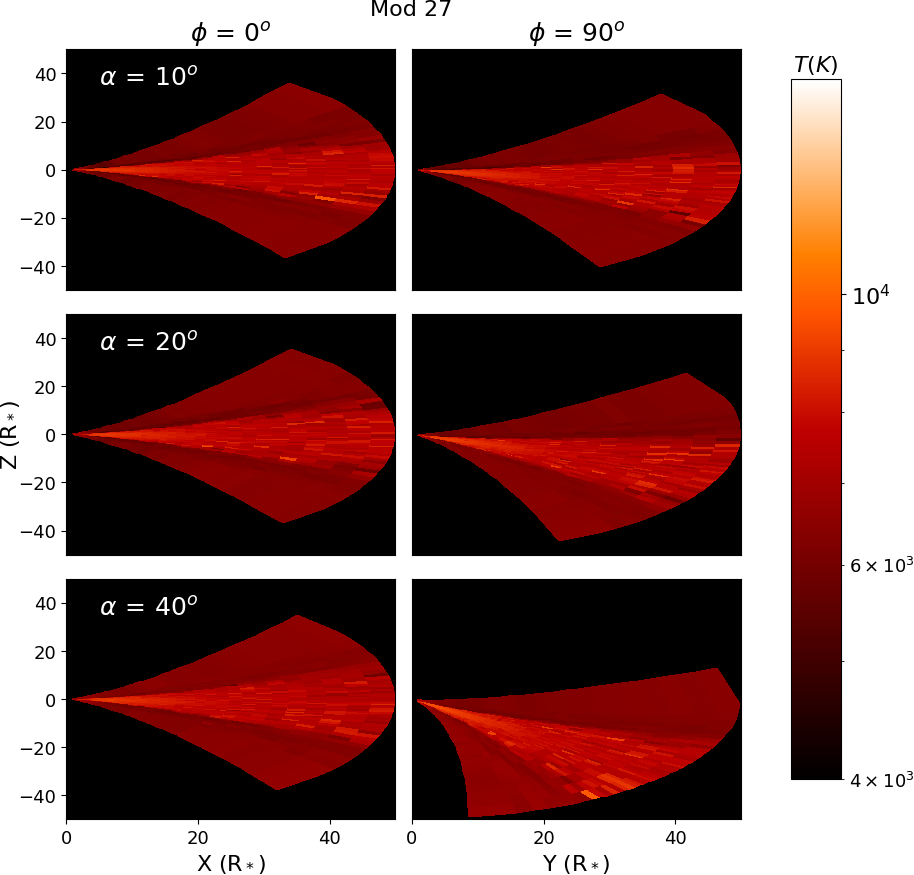}
        \vspace{3pt}
    \end{subfigure} 
    \begin{subfigure}[b]{0.4\textwidth}
        \centering
        \includegraphics[width=\textwidth]{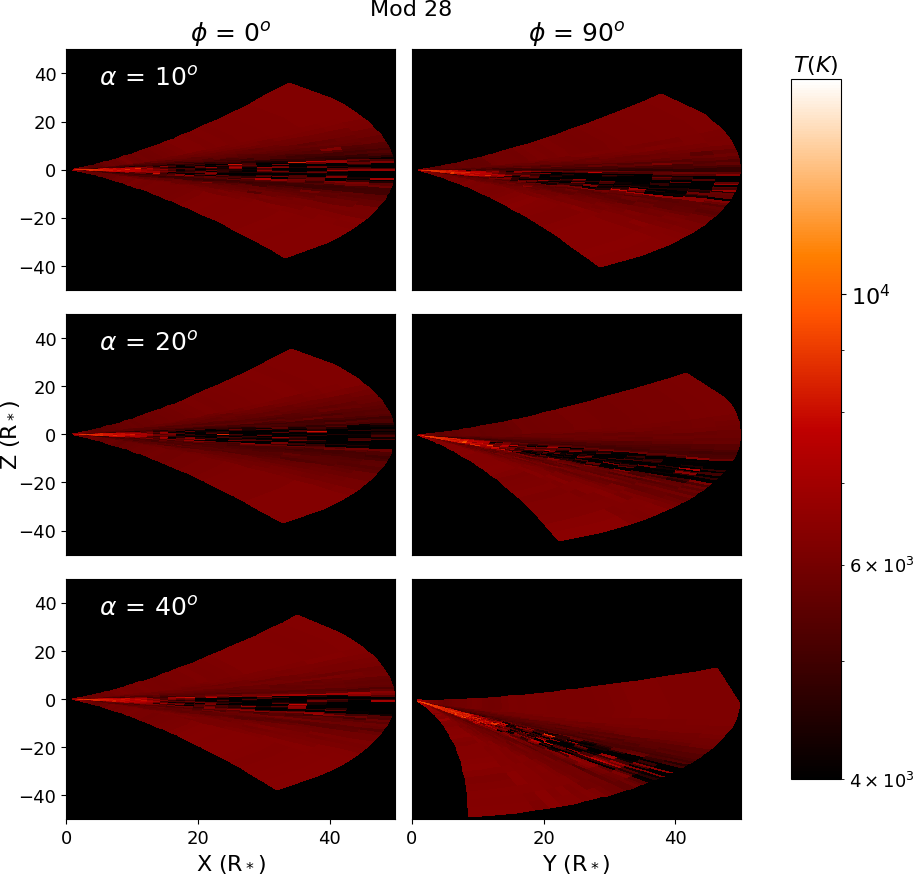}
        \vspace{3pt}
    \end{subfigure} 
    \hspace{5pt}
    \begin{subfigure}[b]{0.4\textwidth}
        \centering
        \includegraphics[width=\textwidth]{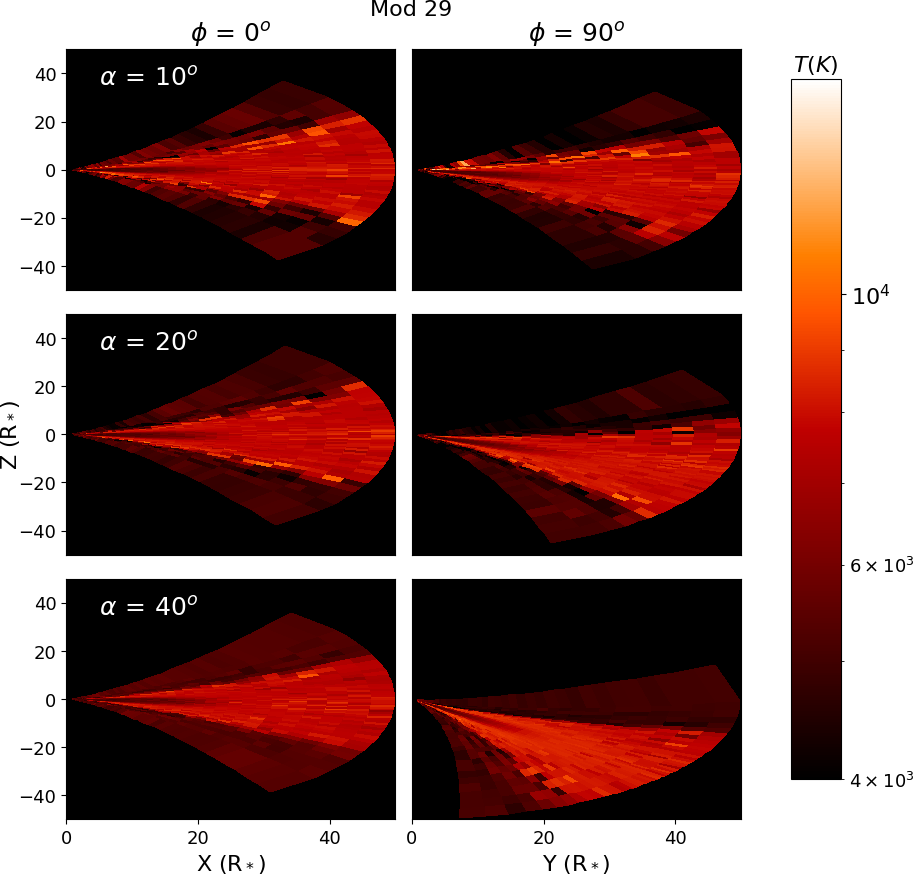}
        \vspace{3pt}
    \end{subfigure} 
    \begin{subfigure}[b]{0.4\textwidth}
        \centering
        \includegraphics[width=\textwidth]{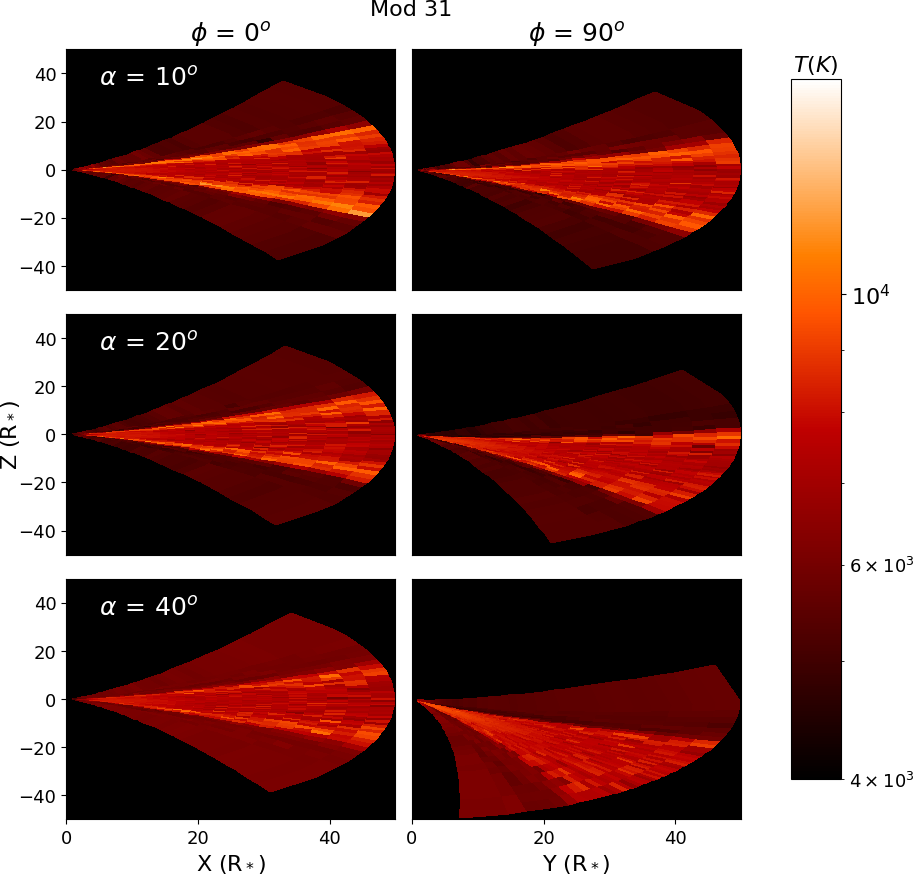}
    \end{subfigure} 
    \hspace{5pt}
    \begin{subfigure}[b]{0.4\textwidth}
        \centering
        \includegraphics[width=\textwidth]{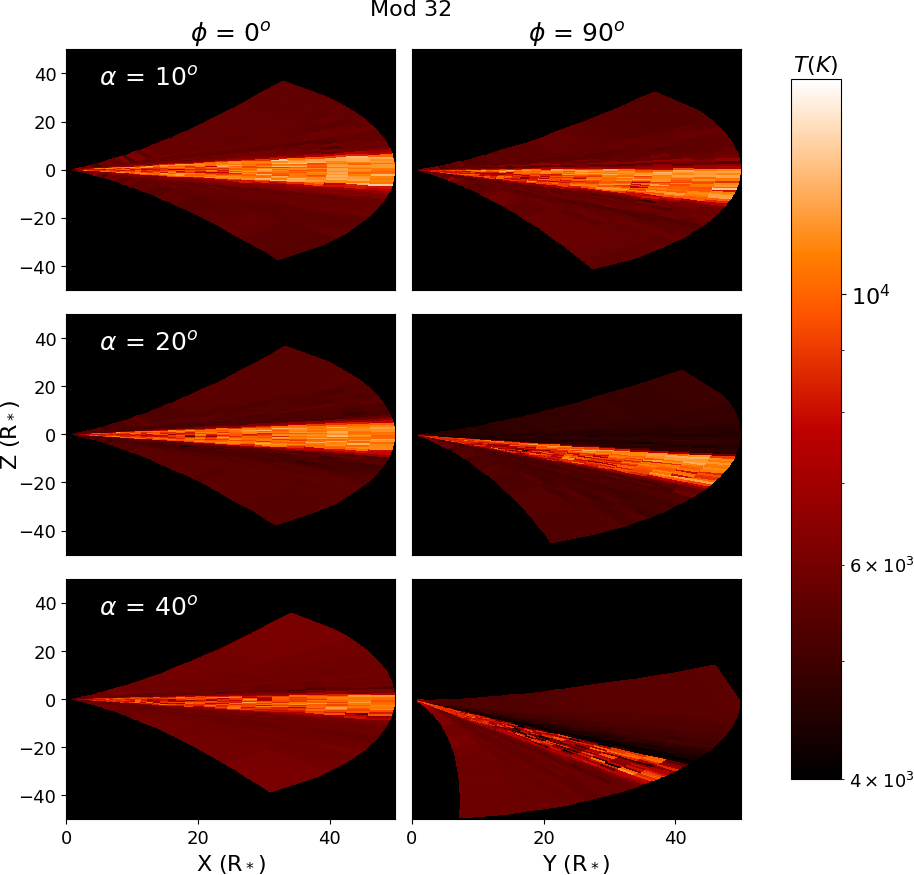}
    \end{subfigure} 
    \caption{Same as Figure \ref{fig:mod33_tilt_temps}, but for models 26-29, 31 and 32}
    \label{fig:mod58-64_tilt_temps}
\end{figure*}

\section{Midplane Temperatures of Tilted vs Non-Tilted Discs}
\label{sec:midplane_temps}

\begin{figure*}
    \centering
    \includegraphics[scale = 0.3]{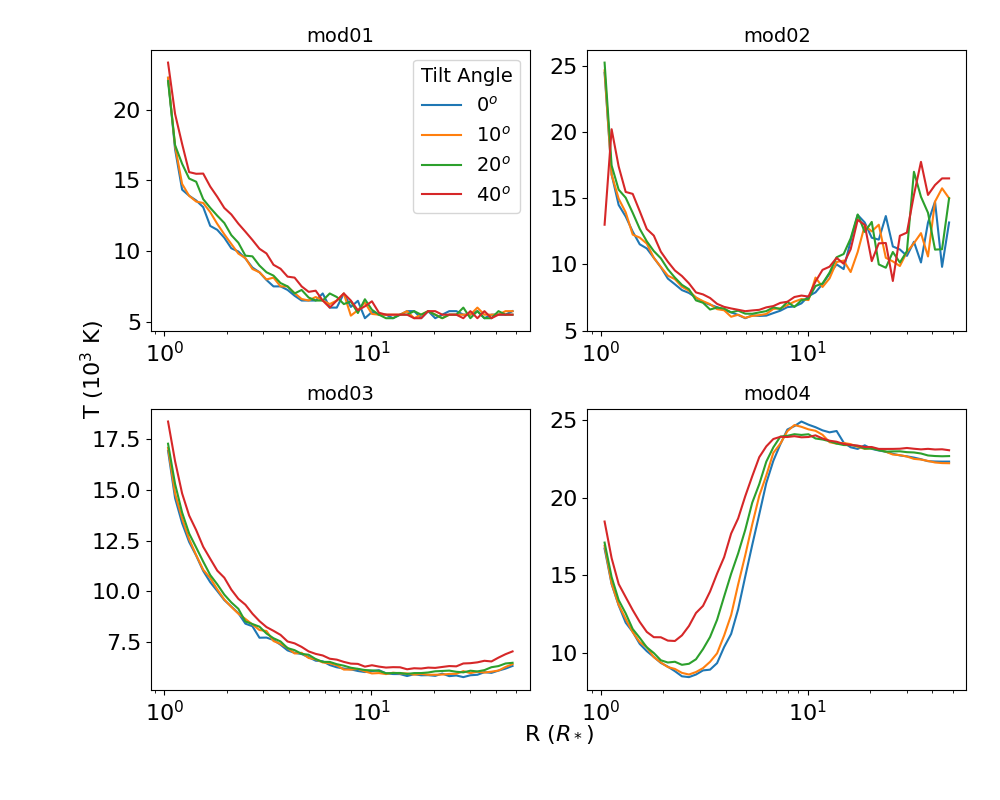}
    \includegraphics[scale = 0.3]{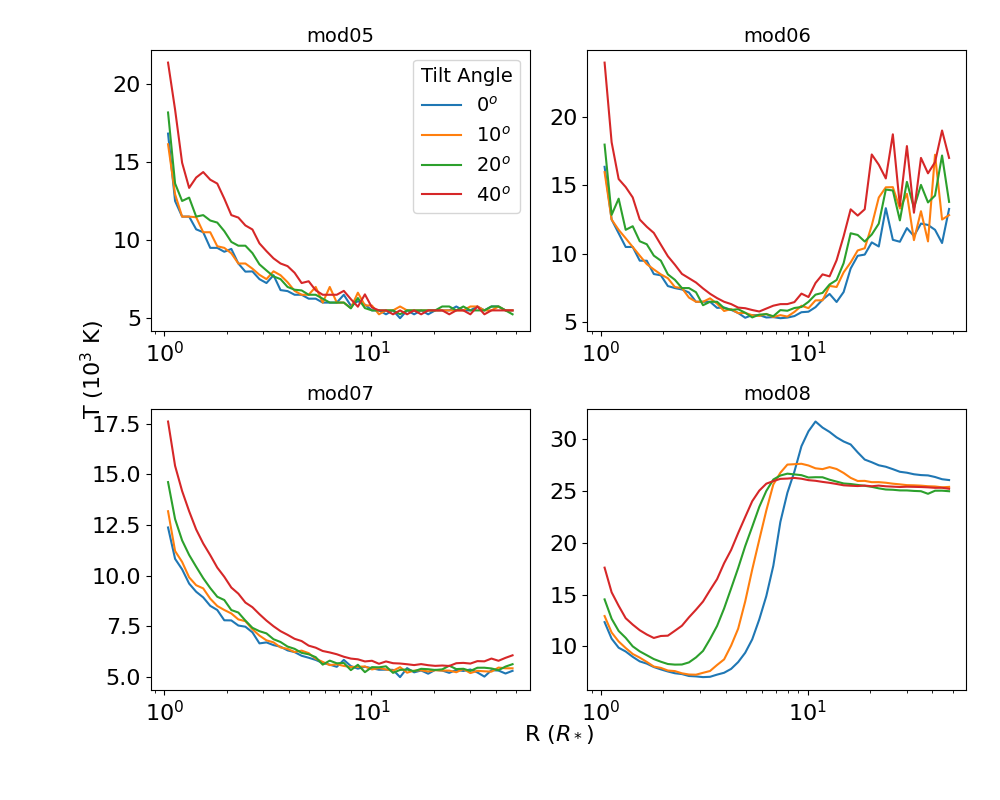}
    \includegraphics[scale = 0.3]{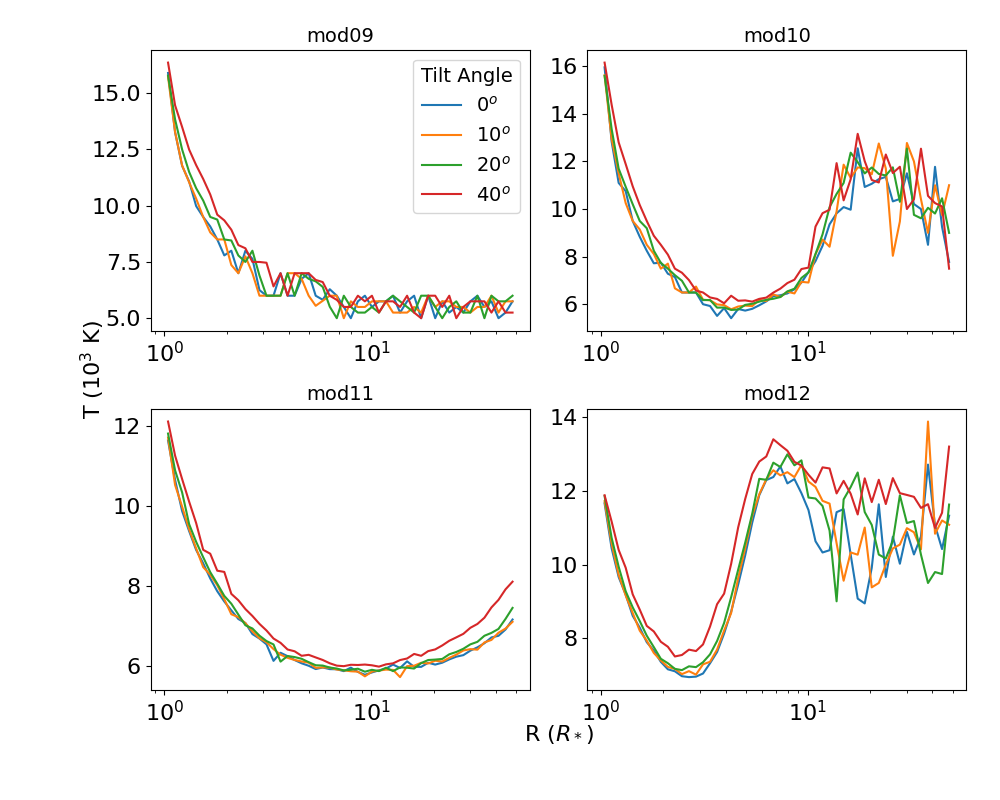}
    \includegraphics[scale = 0.3]{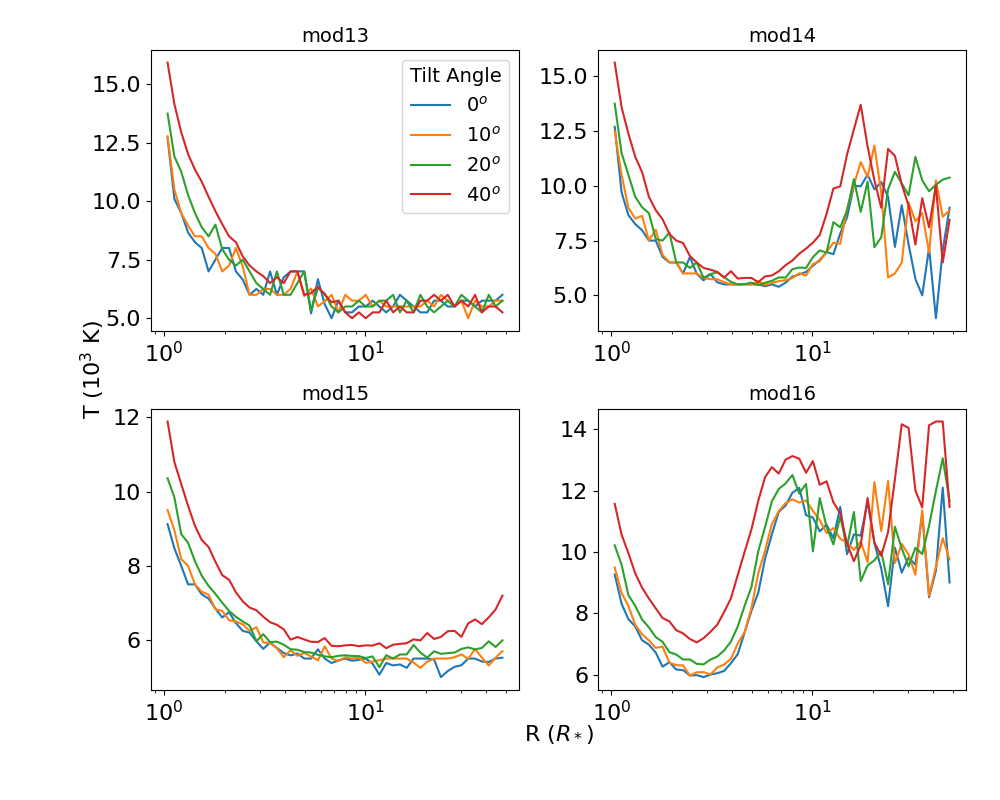}
    \caption{Temperature vs. radius at the disc midplane for all models with a B0 or B2 central star, in the direction $\phi\,=\,\ang{90}$. The four lines are for four different tilt angles as indicated by the legend.}
    \label{fig:early_midplane_temps}
\end{figure*}

\begin{figure*}
    \centering
    \includegraphics[scale = 0.3]{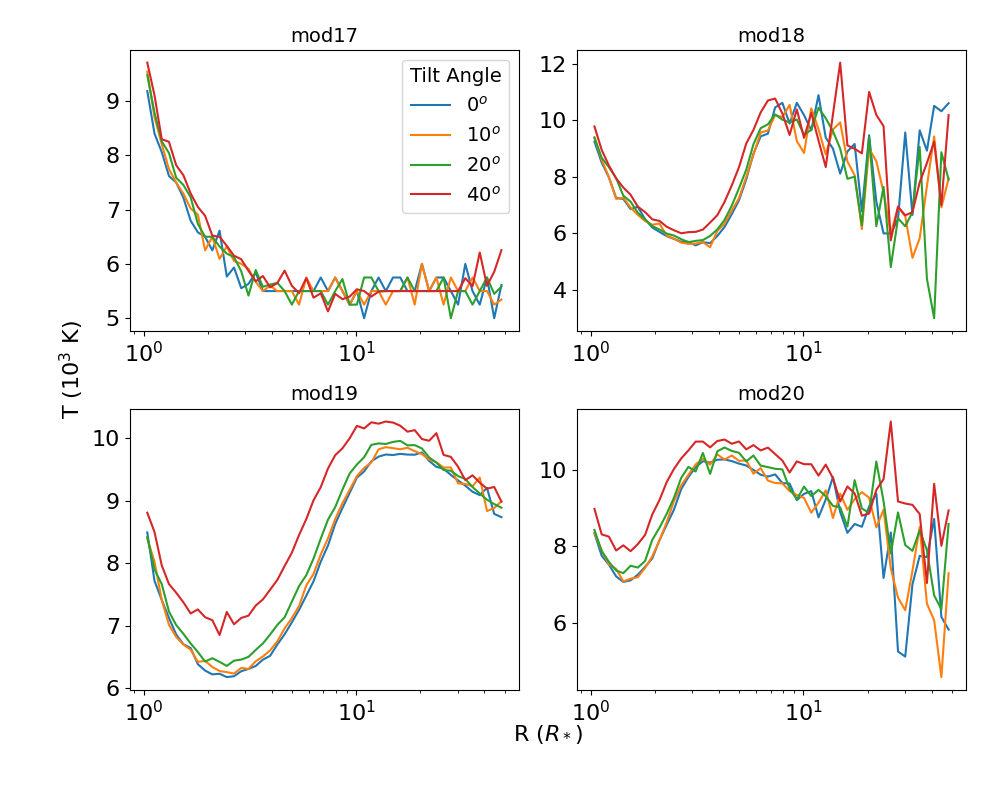}
    \includegraphics[scale = 0.3]{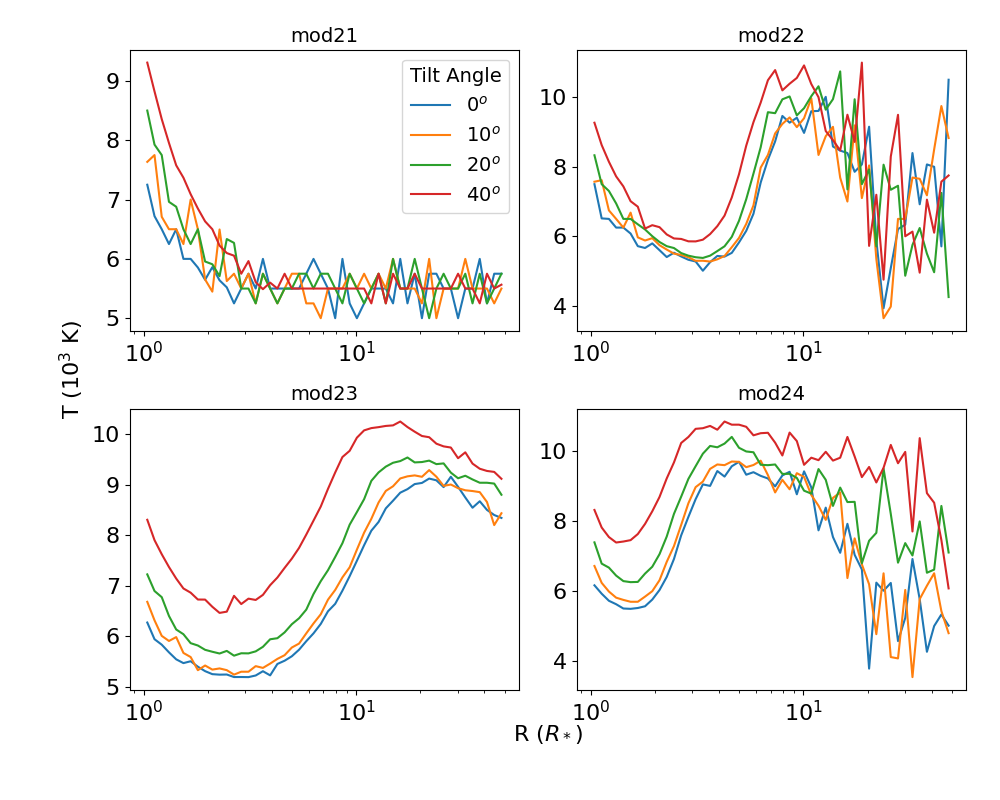}
    \includegraphics[scale = 0.3]{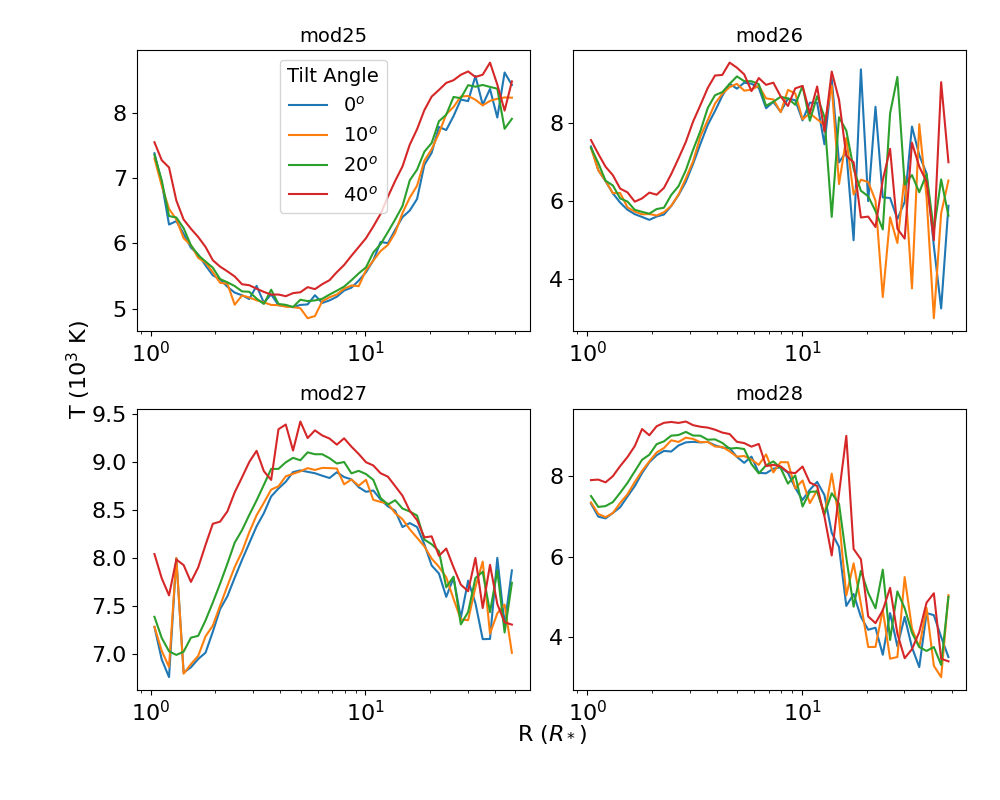}
    \includegraphics[scale = 0.3]{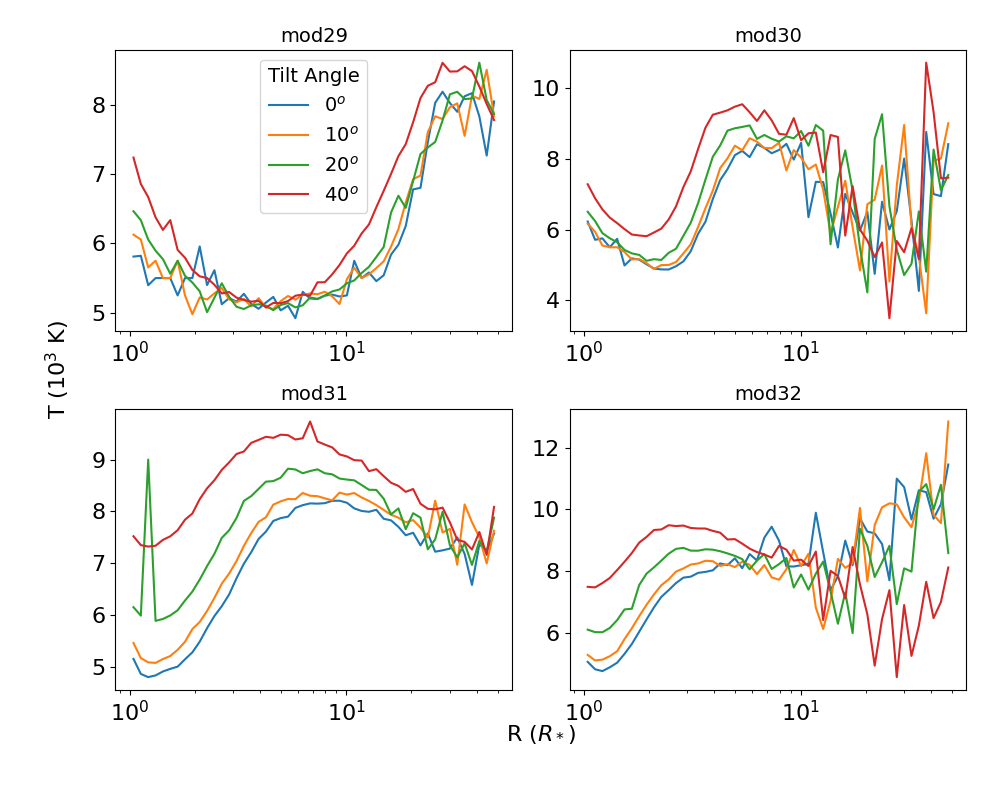}
    \caption{Same as Figure \ref{fig:early_midplane_temps} but for our B5 and B8 models.}
    \label{fig:late_midplane_temps}
\end{figure*}

\end{appendices}

\bsp	
\label{lastpage}
\end{document}